\documentclass[superscriptaddress,nofootinbib,amsmath,amssymb,aps,pra,twocolumn]{revtex4-2}

\usepackage{amsmath}
\usepackage{amssymb}
\usepackage{amsfonts}
\usepackage{bm} 
\usepackage{bbm}
\usepackage{braket}
\usepackage{color}
\usepackage{comment}
\usepackage{dcolumn} 
\usepackage{dsfont}
\usepackage{enumerate}
\usepackage{epsfig}
\usepackage{esint}
\usepackage[T1]{fontenc}
\usepackage{framed}
\usepackage{gensymb}
\usepackage{graphicx} 
\usepackage[colorlinks,linkcolor=blue,citecolor=blue,urlcolor=blue,hyperindex,driverfallback=dvipdfm]{hyperref}
\usepackage{indentfirst}
\usepackage{lmodern}
\usepackage{mathrsfs}
\usepackage{mathtools}
\usepackage{multirow}
\usepackage{psfrag}
\usepackage{pst-all}
\usepackage{soul}
\usepackage{xcolor}
\usepackage{xspace}
\usepackage{orcidlink}

\newcommand{\abs}[1] {\mathopen{}\left|#1\right|\mathclose{}}

\newcommand{\ccpar}[1] {\mathopen{}\left(#1\right)\mathclose{}}
\newcommand{\sqpar}[1] {\mathopen{}\left[#1\right]\mathclose{}}

\def\ii{{\rm i}}  \def\ee{{\rm e}}
  \def\kB{{k_{\rm B}}}
  \def\Imm{{\rm Im}}
\newcommand{\pd}[2] {\mathopen{}\frac{\partial#1}{\partial#2}\mathclose{}}

\def\rb{{\bf r}}  \def\Rb{{\bf R}}
    
\def\xx{\hat{\bf x}}    
    
\def\Eb{{\bf E}}  \def\pb{{\bf p}}
\def\EF{{E_{\rm F}}}    
  \def\vep{\varepsilon}
\def\ww{\omega}
\def\Hm{\mathcal{H}}

\begin{document}

\title{Nonlocal and nonlinear plasmonics in atomically thin heterostructures}

\author{Line Jelver\,\orcidlink{0000-0001-5503-5604}}
\affiliation{POLIMA---Center for Polariton-driven Light--Matter Interactions, University of Southern Denmark, Campusvej 55, DK-5230 Odense M, Denmark}

\author{Joel~D.~Cox\,\orcidlink{0000-0002-5954-6038}}
\email[Corresponding author: ]{cox@mci.sdu.dk}
\affiliation{POLIMA---Center for Polariton-driven Light--Matter Interactions, University of Southern Denmark, Campusvej 55, DK-5230 Odense M, Denmark}
\affiliation{Danish Institute for Advanced Study, University of Southern Denmark, Campusvej 55, DK-5230 Odense M, Denmark}

\begin{abstract}
Plasmons in atomically thin materials offer a compelling route to trigger nonlinear light-matter interactions through extreme optical confinement in the two-dimensional (2D) limit. However, optical nonlocality in plasmons is typically associated with losses in the linear response regime. Here, we show that nonlocal effects mediate strong plasmon-assisted optical nonlinearity in electrically reconfigurable 2D heterostructures. Using atomistic simulations that capture quantum finite-size and nonlocal effects in the nonlinear plasmonic response of graphene and phosphorene nanoribbon dimers, we reveal how symmetry and inter-ribbon coupling shape harmonic generation processes in perturbative and high-harmonic regimes. Independent tuning of geometry and carrier density in nanoribbon heterostructures is shown to induce inter-ribbon plasmon hybridization, impacting inversion symmetry governing even-ordered nonlinear processes like second-harmonic generation. These results reveal design principles for active and passive tuning of nonlinear plasmonic effects and enable selective enhancement of specific harmonic processes, establishing 2D heterostructures as a versatile platform for nonlinear nanophotonics.
\end{abstract}

\maketitle

\section{Introduction}

The diverse optical and electronic properties of two-dimensional (2D) materials hold significant potential for advancing compact and efficient photonic technologies~\cite{sun2016optical,reserbat2021quantum,ma2024inplane,sortino2025light,garciadeabajo2025roadmap}. Besides offering unprecedented tunability through passive means, such as stacking, patterning, and heterostructuring, the electronic properties of 2D van der Waals materials exhibit extreme sensitivity to external stimuli, including static electric and magnetic fields, strain, and chemical environments, enabling dynamic control of their optical response~\cite{fei2012gatetuning,chen2012optical,thongrattanasiri2013optical,rodrigo2015midinfrared,hafez2018extremely,cox2018transient,dias2020thermal,zeng2023nonlinear,ninhos2024tunable}. These features render 2D materials a unique platform upon which to engineer nonlinear light–matter interactions---depending crucially on electronic band structure and electromagnetic field strength at multiple optical frequencies---for applications that leverage frequency conversion, ultrafast optical modulation, and harmonic generation~\cite{liu2017highharmonic,kundys2018nonlinear,baudisch2018ultrafast,autere2018optical,alonsocalafell2021giant,li2022nonlinear,zhang2022chirality}.

Although the inherently low light-matter interaction volumes associated with the atomic-scale thickness of 2D materials impedes their nonlinear optical response, this limitation can be overcome by polaritons~\cite{basov2016polaritons,low2017polaritons}, hybrid quasiparticles arising from the coupling of light with collective matter excitations such as plasmons (free electron oscillations in metals), phonons (lattice vibrations), or excitons (electron-hole pairs in semiconductors)~\cite{basov2021polariton}. Polaritonic modes can concentrate electromagnetic fields on subwavelength dimensions~\cite{thongrattanasiri2013optical,garciadeabajo2014graphene,zhang2021interface}, significantly enhancing nonlinear optical phenomena. In particular, the extreme near-field confinement associated with 2D plasmons can produce large in-plane electric field gradients that trigger nonlocal light-matter interactions~\cite{menabde2022image,monticone2025nonlocality}. Intriguingly, the breaking of spatial inversion symmetry associated with plasmon-driven optical nonlocality permits even-ordered nonlinear processes to occur in centrosymmetric 2D materials like graphene~\cite{manzoni2015second,constant2016alloptical}. 
The patterning of 2D materials into nanoislands or nanoribbons offers the means to couple far-field illumination with localized plasmons that shape the optical near-fields to enhance both even- and odd-ordered nonlinear optical phenomena~\cite{cox2019nonlinear}.

The strong optical field confinement associated with localized plasmons in patterned 2D materials can be further intensified by engineering ensembles of closely spaced 2D nanostructures featuring narrow gaps or hotspots~\cite{christensen2012graphene,thongrattanasiri2013optical,silveiro2015quantum,rodrigo2017double}. Thus far, most studies have focused on dimers of identical units in the linear response regime. In configurations of strongly interacting 2D nanostructures, the independent tuning of plasmon resonances in each constituent offers broad control over the intensity and spatial distribution of concentrated near-fields that drive nonlinear light-matter interactions. This concept was recently explored in electrically doped graphene nanoribbon dimers and trimers, where plasmon hybridization was predicted in semiclassical simulations to produce large and actively tunable second-harmonic generation (SHG) in inversion-symmetric morphologies containing units with distinct electrical doping levels~\cite{rasmussen2023nonlocal}. However, the interplay of quantum finite-size and nonlocal effects in the nonlinear plasmonic response of atomically thin heterostructures remains unexplored.

Here we reveal that the nonlocal hybridization of plasmons in mesoscale 2D heterostructures can be engineered to drive an intense and highly tunable nonlinear optical response. Our investigations focus on electrically doped nanoribbons of graphene, characterized in the bulk by a conical electronic dispersion relation that imposes an isotropic and intrinsically anharmonic free electron response~\cite{cox2017plasmonassisted}, and phosphorene, exhibiting an electrical band gap and pronounced in-plane anisotropy that supports polarization sensitive nonlinear optical processes~\cite{xia2014rediscovering,qiao2014highmobility,yang2015optical,nourbakhsh2016excitons,carvalho2016phosphorene,neupane2022bending,chaves2025nonlocal}. To capture nonlocal effects in the nonlinear optical response that become important on few-nanometer length scales, we employ atomistic simulations based on first-principles electronic structure calculations of nanoribbons, combined with a single-particle density matrix formalism to compute their optical response self-consistently~\cite{cox2016quantum,jelver2023nonlinear,jelver2024nonlinear}. Focusing on concerted nonlocal and nonlinear light-matter interactions driven by plasmons in strongly interacting dimers formed by independently tunable nanoribbons, our results show that combined nanoribbons with complementary properties enhance the magnitude and tunability of the resulting nonlinear optical response. By engineering nanoribbon dimer geometries and charge carrier doping levels, plasmon-driven even-ordered nonlinear optical processes can occur in inversion-symmetric morphologies, both for perturbative processes, including SHG, and extreme nonlinear optical phenomena, such as high-harmonic generation (HHG). Our findings open paths to design nonlinear nanophotonic devices that can efficiently control light by light itself at the nanoscale.

\section{Results and discussion}

\begin{figure*}[t]
    \centering
    \includegraphics[width=1\textwidth]{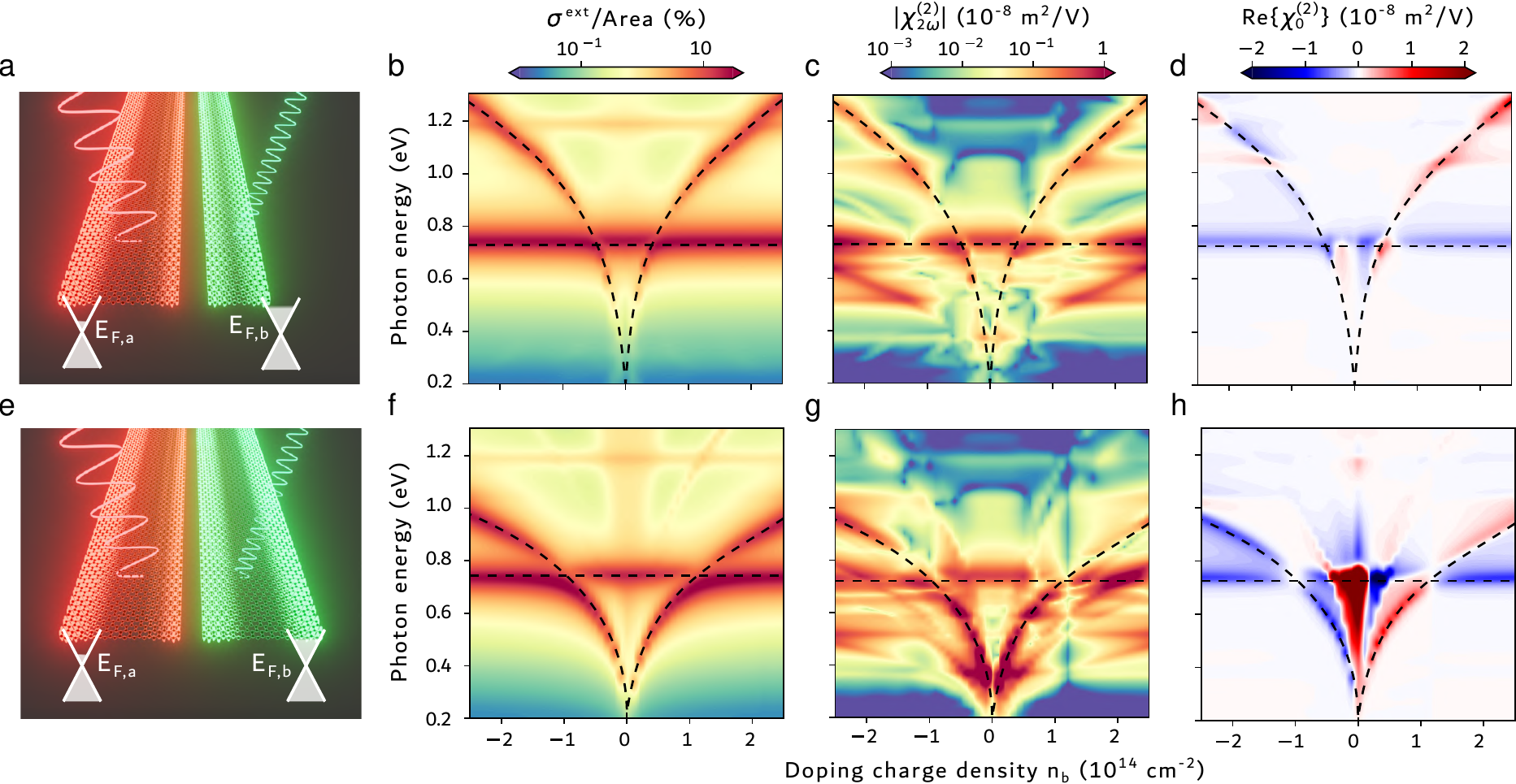}
    \caption{\textbf{Active tuning of linear and second-order optical processes in mesoscale co-planar graphene nanoribbon dimers.} (a) Illustration of second harmonic generation in a co-planar graphene nanoribbon (GNR) dimer comprised of ribbons with unequal widths and independent doping charge carrier concentrations (indicated by the Fermi energies $E_{\rm{F,a}}$ and $E_{\rm{F,b}}$). (b) The extinction cross-section $\sigma^{\rm ext}$, calculated from the linear optical response to normally impinging plane-wave illumination polarized along the ribbon confinement direction, of a dimer formed by co-planar armchair-edge terminated GNRs of width $W_{\rm a}\approx10$\,nm and $W_{\rm b}\approx5$\,nm with a 2\,nm edge-to-edge separation. The first ribbon has a fixed electron doping density $n_{\rm a}=1.2\times10^{14}$\,cm$^2$, corresponding to a plasmon resonance energy $\hbar\omega=0.75$\,eV for the isolated ribbon (indicated by the horizontal dashed line), while the doping density $n_{\rm b}$ of the second ribbon varies continuously from hole doping ($n_{\rm b}<0$) to electron doping ($n_{\rm b}>0$), with its plasmon resonance (in isolation) indicated by the dashed curves. Panels (c) and (d) show the nonlinear optical susceptibilities $\chi_{2\ww}^{(2)}$ and $\chi_0^{(2)}$ associated with second-harmonic generation and optical rectification, respectively, for the system considered in (a). Panels (e-h) show analogous results as panels (a-d) when the dimer is comprised of identical armchair edge-terminated GNRs of width $W_{\rm a}=W_\mathrm{b}\approx10$\,nm and 2\,nm edge-to-edge separation. All results are obtained by assuming a phenomenological inelastic scattering rate $\gamma$ corresponding to $\hbar\gamma=50$\,meV.
    }
    \label{fig:fig1}
\end{figure*}

We apply the atomistic simulation framework detailed in Methods to study nonlinear optical phenomena driven by localized plasmons in dimers of closely spaced graphene nanoribbons (GNRs) or phosphorene nanoribbons (PNRs) with up to $W\approx10$\,nm width, a length scale for which quantum confinement effects in the electronic band structure become important ~\cite{cox2019nonlinear}. We first discuss results obtained using perturbation theory that quantify the response of highly doped GNR dimers to monochromatic plane-wave illumination, polarized along the ribbon confinement direction, at specific orders in the electric field strength. In particular, we examine how spatial gradients in the near-fields break inversion symmetry to enable a dipolar second-order response. We then perform time-domain calculations of GNR and PNR dimers, focusing on HHG produced by intense ultrashort optical pulses and the role of near-field interactions.

\subsection{Actively tuning plasmon hybridization and symmetry in the linear and nonlinear optical response}

The possibility to passively and actively control inversion symmetry in GNR dimers motivates explorations of even-ordered nonlinear optical processes enabled by symmetry breaking~\cite{boyd2020nonlinear}. In Fig.~\ref{fig:fig1}a, we schematically illustrate the concept of plasmon-driven SHG by an asymmetric GNR dimer, comprised of ribbons with unequal widths and independent charge carrier doping levels. Figs.~\ref{fig:fig1}b-d show perturbative optical response simulations of such a dimer, formed by co-planar GNRs with widths $W_{\rm a}\approx10$\,nm and $W_{\rm b}\approx5$\,nm, 2\,nm edge-to-edge separation, and armchair edge terminations. We consider a situation where the carrier density $n_{\rm a}=1.2\times10^{14}$\,cm$^{-2}$ in the wide GNR is fixed and the density $n_{\rm b}$ in the narrow GNR is continuously varied from hole ($n_{\rm b}<0$) to electron ($n_{\rm b}>0$) doping. The linear optical response, characterized by the extinction cross section in Fig.~\ref{fig:fig1}b, features a fixed plasmon resonance at $\hbar\omega = 0.75$\,eV supported by the wide ribbon (dashed line) and a resonance scaling with $n_{\rm b}^{1/4}/\sqrt{W_{\rm b}}$ (dashed curve) in the narrow ribbon. Notably, the linear response is approximately symmetric with respect to electron/hole doping in the narrow GNR, exhibiting only minor deviations due to the asymmetry of electronic bands around the Fermi level in undoped ribbons (see SI). The SHG response in Fig.~\ref{fig:fig1}c displays a much richer spectral dependence compared to that of the extinction, with large susceptibilities appearing when the fundamental excitation frequency or its second harmonic coincide with a plasmon resonance in either GNR. A comparably large susceptibility associated with optical rectification (OR), presented in Fig.~\ref{fig:fig1}d and quantifying the static dipole moment induced in the dimer by the second-order response, follows the plasmon features at the fundamental frequency and exhibits a directional dependence with doping charge polarity.

We study active tuning of symmetry in a dimer of co-planar GNRs with equal width. These exhibit inversion symmetry in the plasmon confinement direction that is broken by electrically tuning each ribbon independently, as illustrated schematically in Fig.~\ref{fig:fig1}e. In Figs.~\ref{fig:fig1}f-h we present simulations of the linear and second-order optical response in a dimer formed by $W_{\rm a}=W_{\rm b}\approx10$\,nm armchair edge-terminated GNRs with 2\,nm edge-to-edge separation, once again by fixing the doping charge density $n_{\rm a}=1.2\times10^{14}$\,cm$^{-2}$ in the first ribbon and varying the density $n_{\rm b}$ in the second ribbon. The linear extinction cross section in Fig.~\ref{fig:fig1}f features more prominent resonances at lower energies compared to its $W_{\rm a}\neq W_{\rm b}$ counterpart in Fig.~\ref{fig:fig1}b, as well as stronger hybridization for $n_{\rm a}\approx|n_{\rm b}|$, leading to a discernible redshift of the plasmon peak at $\hbar\omega=0.75$\,eV. The SHG response of the equal-width dimer in Fig.~\ref{fig:fig1}g is also comparatively larger than that of the $W_{\rm a}\neq W_{\rm b}$ dimer (c.f., Fig.~\ref{fig:fig1}c), displaying a large plasmon-driven SHG response at lower energies when one GNR is only lightly doped. Inversion symmetry is recovered only when $n_{\rm b}\approx n_{\rm a}$, with only small differences in doping leading to a large and highly tunable SHG response. The OR susceptibility presented in Fig.~\ref{fig:fig1}h shows qualitatively similar behavior as the SHG response, and offers large tunability over the induced static dipole polarization direction.

The atomistic simulations featured in Fig.~\ref{fig:fig1} reveal that plasmon hybridization in GNR dimers can produce a large second-order nonlinear optical response that can be passively and actively tuned. These findings are in qualitative agreement with investigations based on a classical electrodynamic description of nonlocal light-matter interactions driven by free electrons in graphene, underscoring the role of inter-ribbon asymmetry and doping control in breaking inversion symmetry to enable SHG~\cite{rasmussen2023nonlocal}. In the mesoscopic systems considered here, the second-order response is particularly sensitive to quantum finite-size effects in the electronic band structure, indicated by the moderate quenching of plasmon resonances at certain photon energies that occurs independently of doping density. Incidentally, as shown in the Supplementary Information (SI), the SHG response is particularly tunable in zigzag-terminated GNRs, which we attribute to the steeper band dispersions characteristic of these ribbons.

\begin{figure*}[t]
    \centering
    \includegraphics[width=1\textwidth]{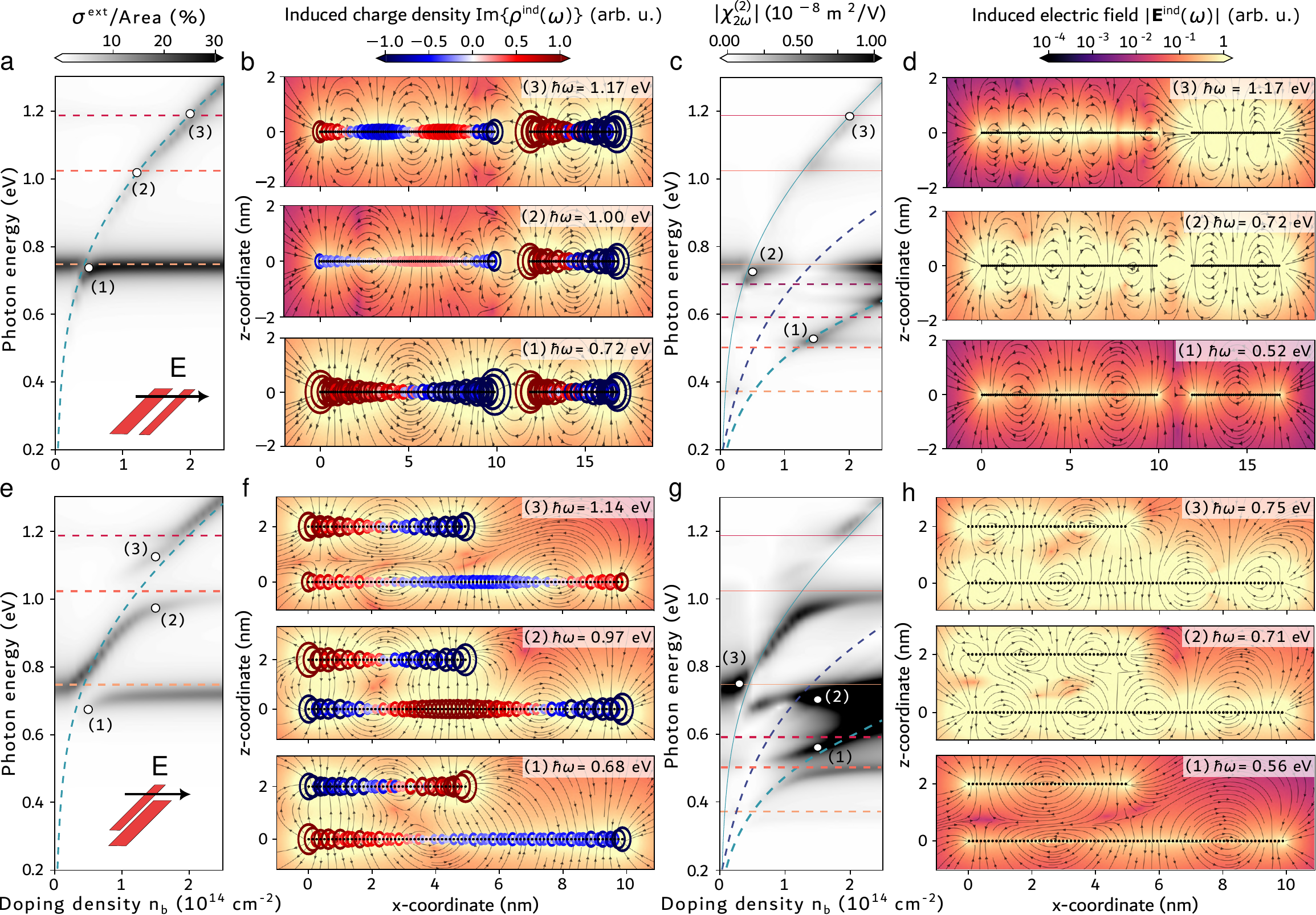}
    \caption{\textbf{Linear and nonlinear near-fields produced by hybridized plasmons in graphene nanoribbon dimers.} (a) Extinction cross-section of a dimer formed by co-planar armchair edge-terminated graphene nanoribbons (GNRs) with 2\,nm end-to-end separation, where one ribbon has width $W_{\rm a}\approx10$\,nm and a fixed doping density $n_{\rm a}=1.2\times10^{14}$\,cm$^{-2}$, and the other has width $W_{\rm b}\approx5$\,nm and tunable doping density $n_{\rm b}$. The horizontal dashed lines indicate the first three plasmon modes of the static GNR and the dashed curve shows the lowest-order dipolar mode in the tunable GNR. (b) The induced near-fields $\Eb^{(1)}$ (to linear order) of hybridized plasmon modes, i.e., at the doping densities $n_{\rm b}$ and photon energies $\hbar\omega$ indicated by the numbered points in (a), are calculated from the induced charge density $\rho_{i,l}^{\rm ind}$, the imaginary parts of which are represented by color-coordinated open circles (see first colorbar) at the atomic positions $x_{i,l}$ (marked by filled black circles). Likewise, the imaginary part of the induced field is plotted as quivers superimposing contour plots of the total field magnitude (see second colorbar). (c) The second-harmonic susceptibility for the same GNR configuration in panels (a) and (b), with dashed curves indicating where the generated frequency $2\omega$ coincides with the plasmon modes highlighted in (a) and solid curves indicating the frequency of the incoming radiation $\omega$ matching these modes. (d) The magnitude of the induced near-fields $\Eb^{(2)}$ (to second order) at the parameters corresponding to the numbered points in (c) with atomic positions marked by filled circles. Panels (e-h) show analogous results to panels (a-d) for a stacked configuration of the same GNRs with 2\,nm vertical separation and one aligned edge.
    }
    \label{fig:fig2}
\end{figure*}

\subsection{Nonlinear plasmonic near-fields in nanoribbon dimers}

To understand the optical response associated with hybridized plasmons in actively tunable GNR heterostructures, we examine the electric near-fields in co-planar and stacked dimers of closely spaced armchair-edge-terminated ribbons (separated by 2\,nm in either case) with $W_{\rm a}\approx10$\,nm and $W_{\rm b}\approx5$\,nm widths. We maintain a fixed doping charge density $n_{\rm a}$ in the former and a tunable density $n_{\rm b}$ in the latter. In Fig.~\ref{fig:fig2}a we examine the extinction spectrum of the co-planar dimer, identifying cases for which the dipolar plasmon resonance in the tunable GNR hybridizes with the dipolar (1), tripolar (2), and quadrupolar (3) plasmon modes of the static GNR at $\hbar\ww=0.72$\,eV, $1.00$\,eV, and $1.17$\,eV, respectively. These cases are featured in Fig.~\ref{fig:fig2}b, where the magnitude of the electric near-field
\begin{equation}
    \Eb(\rb) = -\nabla_\rb\sum_{i,l}\frac{\rho_{i,l}^{\rm ind}}{\abs{\rb-\Rb_{i,l}}}
\end{equation}
is plotted in the contour, superimposed by quiver plots illustrating the imaginary part of the field as well as color-coordinated circles with areas proportional to the corresponding imaginary part of the induced density $\Imm\{\rho_{i,l}^{\rm ind}\}$ at the carbon p$_z$-like Wannier orbitals $\ket{i,l}$ in structure $i$ with Wannier centers $\Rb_{i,l}=(x_{i,l},y_{i,l},z_{i,l})$ (see details in Methods). The charge density profiles in the dipolar and quadrupolar cases exhibit odd symmetry, leading to a finite net dipole moment $p_x=\sum_{i,l}x_{i,l}\rho_{i,l}^{\rm ind}$ and corresponding rise in the extinction cross section, while the tripolar mode, with zero net dipole, remains dark. The near-field maps underscore the trade-off between the large near-field intensity produced by lower-order plasmon modes and the sharper electric field gradients offered by higher-order modes.

The simulated SHG susceptibility of the co-planar dimer is shown in Fig.~\ref{fig:fig2}c, and exhibits a more complex spectral dependence on doping than the linear response featured in Fig.~\ref{fig:fig2}a. Hybridization occurs when the fundamental (solid curves) or the second harmonic (dashed curves) frequencies of the tunable and stationary plasmons in the narrow and wide ribbons, respectively, coincide. The second harmonic near-fields displayed in Fig.~\ref{fig:fig2}d show three representative cases: hybridization of dipolar plasmons in both ribbons (1), plasmon-enhanced response when the second-harmonic signal overlaps with the fundamental dipolar resonance of the wide ribbon via coupling of dipolar and tripolar modes (2), and hybridization of dipolar and quadrupolar modes (3). Compared to the linear near-field distribution featured in Fig.~\ref{fig:fig2}b, the second harmonic near-fields exhibit sharper spatial variations across the ribbon, stemming from the nonlocal character of the linear electric near-field that produces a second-order response in the bulk of centrosymmetric graphene~\cite{manzoni2015second,rasmussen2023nonlocal}.

To introduce stronger electric field gradients and enhance the associated plasmon-driven SHG, we turn to the case of stacked GNRs in Fig.~\ref{fig:fig2}e-h. The extinction spectrum in Fig.~\ref{fig:fig2}e exhibits more pronounced plasmon hybridization effects, attributed both to the larger area over which plasmonic near-fields can appreciably interact (as compared to end-to-end coupling in co-planar geometries) and the broken symmetry associated with the condition $W_{\rm a}\neq W_{\rm b}$ when the ribbon edges are aligned---here, the dipolar plasmon mode in the wider ``static'' ribbon approximately maintains its intrinsic resonance frequency after exhibiting an anticrossing behavior with the tunable plasmon mode in the narrow GNR. The associated near-fields in Fig.~\ref{fig:fig2}f are shown for situations where the plasmon in the narrow ribbon hybridizes with the dipolar (1) or the tripolar (2)-(3) modes of the wide GNR. At (1), the charge density exhibits clear dipolar character in both GNRs (albeit with charges slightly skewed relative to the GNR centers according to the dimer geometry), but the opposing polarity of these moments in each GNR lead to a diminished net dipole moment. The interaction of the dark mode of the wide GNR and the bright mode of the narrow GNR at (2) results in a pronounced plasmon-induced transparency feature where a net dipole is maintained due to the dipolar mode in the narrow ribbon and the opposing polarity creates an intense field in the gap between the ribbons. This is in contrast to the other branch of this hybridization point (3) where the polarity of the wide ribbon is opposite, resulting instead in a quenching of the inter-ribbon field.

Remarkably, the SHG susceptibility for the stacked dimer in Fig.~\ref{fig:fig2}g exhibits up to an order-of-magnitude enhancement compared with the co-planar case. The largest SHG response consistently appears when the second-harmonic signal overlaps with the fundamental resonance of the wide ribbon (3), and can be significantly modulated by tuning the doping level of the narrow GNR to hybridize with this mode. Strong SHG signals also arise when the generated second harmonic overlaps with the hybridized dipolar and tripolar plasmons in the narrow and wide ribbons, respectively (1), and when the second harmonic frequency coincides with the tripolar plasmon mode in the narrow ribbon (2). The induced second-harmonic near-field maps featured in Fig.~\ref{fig:fig2}h, corresponding to the specified photon energy and doping density in Fig.~\ref{fig:fig2}g, are characterized by rapid spatial variations along the ribbons. The most intense inter-ribbon fields occur at frequencies near the dipolar plasmon resonance in the wide ribbon, while the field profiles exhibit a high multipolar character. These results demonstrate how mode hybridization in asymmetric GNR dimers not only reshapes the linear plasmonic spectrum but also strongly amplifies nonlinear responses, highlighting the key role of active symmetry breaking in engineering second-order processes at the nanoscale.

\subsection{Tunable third-order plasmon-driven nonlinear phenomena in nanoribbon heterostructures\label{sec:III}}

\begin{figure*}[t]
    \centering
    \includegraphics[width=1\textwidth]{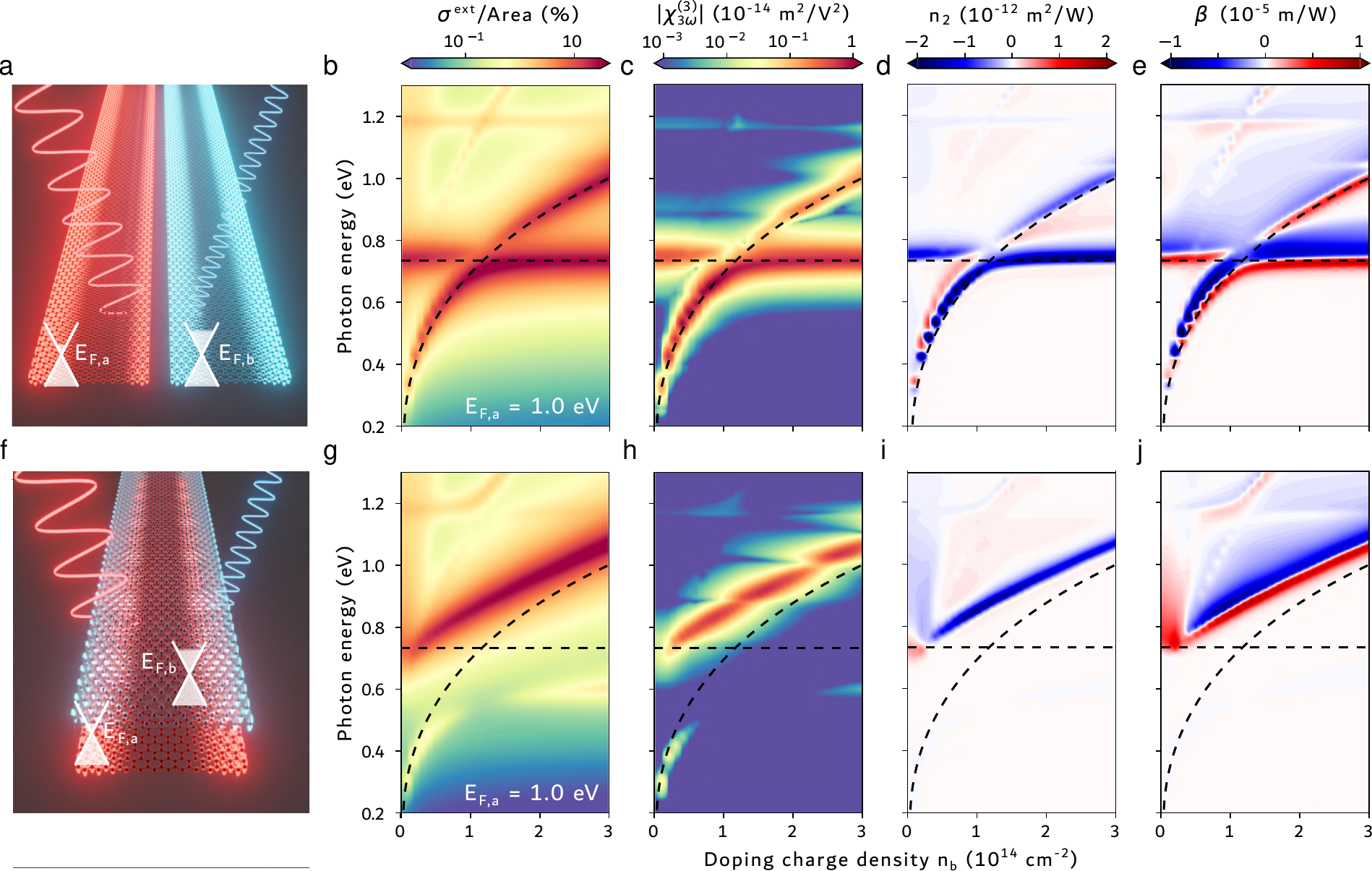}
    \caption{ \textbf{Plasmon tunability and hybridization effects in the linear and third order optical response of graphene nanoribbon dimers.} (a) Illustration of harmonic generation in co-planar graphene nanoribbon (GNR) dimers comprised of ribbons with equal width but independent doping charge carrier concentrations, indicated by the Fermi energies $E_{\rm{F,a}}$ and $E_{\rm{ F,b}}$. (b) Spectral dependence of the linear optical response for a dimer of $W\approx10$\,nm wide GNRs with edge-to-edge separation 2\,nm and armchair edge terminations, characterized by the optical extinction cross section $\sigma^{\rm ext}$ spectra per unit of the total graphene area, for a fixed carrier density $n_{\rm a}=1.2\times10^{14}$\,cm$^2$ in the first GNR and a variable density $n_{\rm b}$ in the second GNR. The horizontal dashed lines indicate the fixed plasmon resonance energy $\hbar\ww=0.75$\,eV of the first GNR in isolation while the dashed curves correspond to that of the second GNR. Panels (c-e) correspondingly show the nonlinear optical response associated with (c) third-harmonic generation (quantified by the nonlinear susceptibility $\chi_{3\omega}^{(3)}$) (d) the nonlinear refractive index $n_2$, and (e) the two-photon absorption coefficient $\beta$. Panels (f-j) show analogous results to panels (a-e) but for stacked GNR dimers separated by 2\,nm.}
    \label{fig:fig3}
\end{figure*}

Odd-ordered nonlinear optical processes occur in the bulk of centrosymmetric materials, and can thus occur in GNR dimers independently of their symmetry. We characterize the linear and third-order optical response of dimers formed by identical GNRs, with $W\approx10$\,nm width and armchair edge terminations, that are arranged in a co-planar configuration and separated by 2\,nm, as illustrated schematically in Fig.~\ref{fig:fig3}a. The linear optical response, characterized by the extinction cross section, is presented in Fig.~\ref{fig:fig3}b as the doping charge carrier density in the first ribbon is fixed to $n_{\rm a}=1.2\times10^{14}$\,cm$^{-2}$ (corresponding to a Fermi energy $\EF=1.0$\,eV) while the carrier density in the second ribbon $n_{\rm b}$ is varied continuously. The dashed horizontal lines mark the plasmon resonance at $\hbar\ww=0.75$\,eV supported by the GNR with fixed carrier density in isolation, while the plasmon resonance in the tunable GNR, indicated by the dashed curve, scales as $n_{\rm b}^{1/4}$. Resonances in the extinction generally follow the trends of isolated ribbons, deviating by a significant redshift where they would otherwise intersect due to enhanced near-field coupling. Prominent spectral features in the third-harmonic susceptibility $\chi_{3\ww}^{(3)}$ of the dimer, presented in Fig.~\ref{fig:fig3}c, closely follow the plasmon peaks in the extinction spectra, underscoring the importance of plasmonic near-field enhancement at the fundamental excitation frequency in third-harmonic generation (THG) from GNR dimers. Plasmon resonances similarly dominate the nonlinear refractive index $n_2$ and the two-photon absorption coefficient $\beta$, featured in Figs.~\ref{fig:fig3}d and \ref{fig:fig3}e, respectively, and quantified according to the prescription of Ref.~\cite{delcoso2004relation} using the simulated third-order susceptibility $\chi_\ww^{(3)}$ associated with the optical Kerr effect (see Methods). Interestingly, for a fixed photon energy, these quantities can undergo a sign change according to the charge carrier doping level.

The optical response of stacked GNRs, illustrated schematically in Fig.~\ref{fig:fig3}f, exhibits markedly different behavior than the co-planar configuration under the same conditions. Specifically, we show the extinction cross section spectra in Fig.~\ref{fig:fig3}g for the $W\approx10$\,nm wide armchair GNRs, with the fixed and varying carrier densities considered previously, separated vertically by 2\,nm. In this stacked configuration, the close proximity of atoms across the ribbons leads to a much stronger near-field interaction and a single resonance that is consistently blueshifted from that of the isolated and tunable GNR, resembling the response of a single GNR of the same width and combined doping electron density. The THG susceptibility in Fig.~\ref{fig:fig3}h is similarly dominated by this blueshifted plasmon resonance, which also displays weak dips at fixed photon energies that are attributed to electronic band structure features at the generated harmonic frequency. The nonlinear refractive index and two-photon absorption coefficient in Figs.~\ref{fig:fig3}i and \ref{fig:fig3}j, respectively, are dominated by these tunable plasmon resonances, but are less sensitive to quantum confinement effects than the THG response.

The results in Fig.~\ref{fig:fig3} indicate that independently tunable GNRs in a simple dimer configuration offer broad and active control of plasmon hybridization and the associated nonlinear response. Notably, trends in plasmon hybridization pertaining to redshifts and blueshifts in the co-planar and stacked configurations explored here are in agreement with investigations of plasmons in GNR dimers probed in the linear response regime~\cite{christensen2012graphene,silveiro2015quantum,rodrigo2017double}. In contrast to the linear response, for which the intensity of plasmonic features increases with charge carrier doping levels, plasmon-driven THG tends to decrease with carrier density, a behavior that is consistent with the third-order free electron response of graphene predicted in the Boltzmann transport equation formalism~\cite{cox2016quantum}. In contrast, the nonlinear refractive index $n_2$ and two-photon absorption coefficient $\beta$, calculated from the third-order susceptibility $\chi_\ww^{(3)}$ and generating a response at the fundamental frequency, exhibit a greater dependence on the plasmon resonance. Overall, the third-order nonlinear plasmonic response in GNR dimers presents a high sensitivity to quantum confinement effects in the electronic band structure compared to the linear response.

\subsection{High-harmonic generation in graphene–phosphorene heterostructures \label{sec:V}}

\begin{figure*}[t]
    \centering
    \includegraphics[width=1\textwidth]{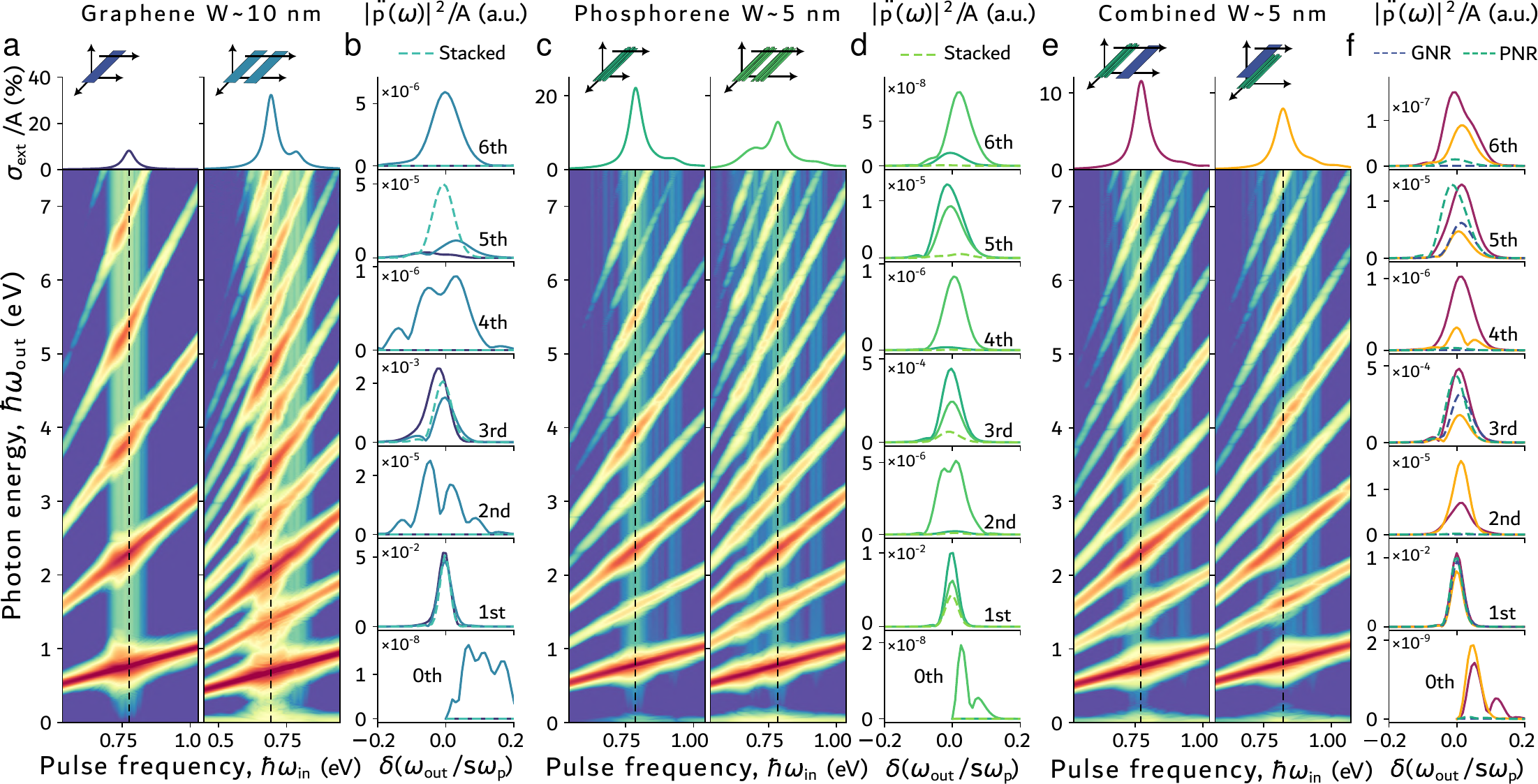}
    \caption{\textbf{High-harmonic generation in nanoribbon dimers of graphene and phosphorene.} (a) Extinction cross section (top) and harmonic spectra (bottom) for $W_\mathrm{a}=W_\mathrm{b}\approx10$\,nm GNRs. Results are shown for a single ribbon and for two co-planar zigzag-terminated ribbons with an edge-separation of 2\,nm. The isolated or first ribbon is doped at $n_{\rm a}=1.2\times10^{14}$ cm$^{-2}$, while the second ribbon is doped at $n_{\rm b}=-1.2\times10^{14}$ cm$^{-2}$, breaking inversion symmetry and aligning the plasmonic resonances of the two GNRs. (b) Comparison of optical rectification and the first six harmonic orders, obtained when the driving pulse frequency is resonant with the plasmon (dashed vertical lines in (a)), quantified by the squared modulus of the dipole acceleration per unit area, for a single ribbon, a co-planar dimer, and a stacked dimer with 2\,nm vertical spacing (dashed lines). (c-d) Results for $W_\mathrm{a}=W_\mathrm{b}\approx5$\,nm phosphorene nanoribbons (PNRs) with zigzag edge terminations. Doping densities are $n_{\rm a}=-1.5\times10^{14}$\,cm$^{-2}$ and $n_{\rm b}=2.1\times10^{14}$\,cm$^{-2}$, chosen to align the two plasmon resonance frequencies. (e) Results for $W_\mathrm{a}=W_\mathrm{b}\approx5$\,nm PNR-GNR hetero-dimers in co-planar (left panel) or stacked (right panel) geometries. The PNR is doped at $n_{\rm a}=-1.5\times10^{14}$\,cm$^{-2}$ and the GNR at $n_{\rm b}=-1.0\times10^{14}$\,cm$^{-2}$, both corresponding to $\hbar\omega_p=0.78$ eV. (f) Same as (b) and (d), but for the graphene-phosphorene dimers, with comparison to a single zigzag-terminated GNR or PNR (dashed lines).
    }
    \label{fig:fig4}
\end{figure*}

To explore the potential of graphene-based nanoribbon heterostructures for nonlinear photonics beyond the perturbative light-matter interaction regime, we investigate plasmon-assisted high-harmonic generation (HHG) in both homogeneous graphene systems and mixed-material configurations combining graphene and phosphorene. In Fig.~\ref{fig:fig4} we present the dipole acceleration spectra obtained from time-domain simulations of nanoribbon dimers excited by a Gaussian pulse with a peak intensity of $10$\,TW/m$^2$, a full width at half maximum of 100 femtoseconds, and a central photon energy $\hbar\omega_{\rm in}$ tuned around the plasmonic extinction peak. Fig.~\ref{fig:fig4}a compares the optical response of a single (left panel) $\approx10$\,nm wide zigzag edge-terminated GNR and a dimer (right panel) formed by two of these GNRs arranged in a co-planar configuration with 2\,nm edge-to-edge separation. The doping levels in the dimer are selected to break inversion symmetry while matching the energies of the first-order plasmonic modes in each ribbon: the left is fixed at $n_{\rm a}=1.2\times10^{14}$ cm$^{-2}$, and the right ribbon is doped at $n_{\rm b}=-1.2\times10^{14}$ cm$^{-2}$ ($\EF\pm1.0$\,eV). The top panels in Fig.~\ref{fig:fig4}a display the main peaks of the extinction cross section associated with the interacting plasmonic modes, while the bottom panels show the corresponding high-harmonic spectra, quantified by the square modulus of the dipole acceleration, which is proportional to the far-field power spectrum. Plasmonic enhancement of HHG is evident, when the frequency of the incoming light coincides with the main extinction peak (vertical dashed lines). Additionally, strong even-order harmonics emerge due to symmetry breaking in the dimer. In Fig.~\ref{fig:fig4}b we compare the intensities of optical rectification and the first six harmonic orders for both configurations, as well as for a stacked dimer with 2\,nm vertical separation composed of the same two GNRs as in the co-planar geometry. While only the co-planar arrangement yields even-order responses, the trends in the odd-order harmonics are less systematic. The first- and third-order peaks are generally weaker than in the isolated AC ribbon, whereas the fifth order is substantially stronger in the dimers, especially for the stacked configuration.

Highly doped phosphorene nanoribbons (PNRs) present an alternative platform to GNRs for nonlinear plasmonics, exhibiting a large band gap that limits interband quenching of HHG~\cite{jelver2023nonlinear}. Figs.~\ref{fig:fig4}c–d show analogous results to Figs,~\ref{fig:fig4}a-b for $W_\mathrm{a}=W_\mathrm{b} \approx 5$\,nm PNRs. Here we consider carrier doping densities of $n_{\rm a}=-1.5\times10^{14}$\,cm$^{-2}$ and $n_{\rm b}=2.1\times10^{14}$\,cm$^{-2}$, yielding a main extinction peak at $\hbar\omega = 0.78$\,eV in isolated ribbons. Comparing the response of isolated, co-planar, and stacked configurations in Fig.~\ref{fig:fig4}d, we find a weak even-ordered response in the isolated PNR associated with edge-induced symmetry breaking that becomes significantly larger in the co-planar dimers, where symmetry is broken across the entire ribbon width by the unequal doping levels. In contrast, the uneven response is most efficient per unit area in the isolated ribbon compared to the dimer configurations. Specifically, the stacked dimers exhibit a consistently lower response across all response orders shown here.

To explore possible synergies in the nonlocal and nonlinear response of mixed-material configurations, we simulate HHG in dimers formed by combining hole-doped $W_\mathrm{a}\approx 5$\,nm wide PNRs with $W_{\rm b}\approx5$\,nm GNRs, as illustrated in Fig.~\ref{fig:fig4}e. Specifically, we pair ZZ-terminated PNRs and GNRs in co-planar (left panel) and stacked (right panel) configurations, selecting the PNR doping level to match that of the isolated ribbon in Fig.~\ref{fig:fig4}c, while the GNR is doped to $n_{\rm b}=-1.0\times10^{14}$\,cm$^{-2}$ such that its main extinction peak coincides with that of the PNR. Fig.~\ref{fig:fig4}f compares the harmonic yields of these dimers with those of isolated graphene and phosphorene ribbons. For the even orders, a pronounced enhancement is observed relative to the isolated PNR. The second-order response is optimized in stacked systems, whereas the fourth- and sixth-order harmonics are strongest in the co-planar configuration. Odd-order processes are consistently more efficient in the co-planar geometry, although they never exceed the yields of the isolated ZZ-PNR. These results highlight the complex interplay between material, geometry, and doping in graphene–phosphorene heterostructures, and suggest that such systems provide a tunable platform where specific harmonic orders can be selectively enhanced, rather than exhibiting uniform improvements across all even or odd orders.

\section{Conclusion and Outlook}

We have systematically investigated the plasmon-driven linear and nonlinear optical response of graphene and phosphorene nanoribbon heterostructures, with particular emphasis on the role of nonlocal near-field interactions in second- and third-order processes (in the perturbative light-matter interaction regime) and (extreme) high-harmonic generation. By engineering heterostructures with different ribbon widths, varied edge terminations, and vertical or coplanar stacking geometries, we demonstrate that both intense second- and third-order nonlinearities can be controlled at the nanoscale through a combination of passive symmetry breaking and actively tuning charge carrier doping levels (e.g., through electrostatic gating).

Our results indicate that second-order nonlinear optical processes are more pronounced in vertically stacked configurations of heterodimers, where strong plasmonic mode hybridization enables access to otherwise dark resonances. In contrast, third-harmonic generation tends to be optimized in coplanar geometries though with modest enhancements over isolated ribbons. On the other hand, optimized mode matching and careful doping selection were shown to enhance nonlinear absorption coefficients, suggesting routes toward efficient two-photon processes. Extending our analysis to high-harmonic generation (HHG), we showed that nanoribbon dimers, both homogeneously formed from graphene or phosphorene and graphene–phosphorene hybrids, exhibit rich and highly tunable harmonic spectra. Nonlocal symmetry breaking enables strong even-order harmonics, while specific configurations outperform isolated ribbons for individual harmonic orders. Notably, the graphene–phosphorene systems revealed a complex dependence on stacking and edge termination, presenting opportunities for targeted enhancement of specific harmonics. Our findings establish nanoribbon heterostructures as a versatile platform for nonlinear nanophotonics, where structural and electrostatic degrees of freedom can be finely tuned to control the optical nonlinearities. Other degrees of freedom which might be investigated in future work are the integration of twist angles or strain fields. With continued advances in nanoscale fabrication and material assembly~\cite{watts2019production,zhao2020optical,lyu2024graphene}, atomically thin heterostructures are poised to play a key role in the development of compact and electrically tunable nonlinear optical components.

\section{Methods\label{sec:VI}}

Building on the framework introduced in Refs.~\cite{jelver2023nonlinear,jelver2024nonlinear}, our approach integrates first-principles electronic structure calculations with self-consistent optical response simulations based on the single-particle density matrix formalism. More specifically, the electronic band structure is computed using density functional theory (DFT), from which maximally localized Wannier functions (MLWFs) are constructed. The MLWFs serve as a compact and physically transparent basis for deriving tight-binding (TB) Hamiltonians and evaluating Coulomb matrix elements. These quantities provide the foundation for both perturbative and time-domain optical response simulations following Refs.~\cite{cox2016quantum,cox2017plasmonassisted} (applied to graphene nanoribbons), where electron–electron interactions and light–matter coupling are described within the quasistatic approximation. In what follows, we summarize the generalization of this atomistic framework to describe arbitrary ensembles of nanoribbons, including the graphene and phosphorene structures considered here.

\subsection{Electronic states of nanoribbons}

The electronic ground states of all nanoribbons are obtained from DFT using the Perdew–Burke–Ernzerhof (PBE) exchange–correlation functional as implemented in Quantum ESPRESSO~\cite{perdew1996generalized,giannozzi2009quantum,giannozzi2017advanced}. Norm-conserving pseudopotentials are taken from the PseudoDojo database~\cite{vansetten2018pseudodojo}, and wavefunctions are expanded in a plane-wave basis. Wannierization and the construction of the tight-binding Hamiltonians $\Hm^{\rm TB}$ are carried out using the Wannier90 code~\cite{pizzi2020wannier90}. Additional details on computational parameters and the electronic band structures are provided in the Supplementary Information.

For a system comprised of multiple structures, the eigenstates of each structure $i$, characterized by a TB Hamiltonian $\Hm_i^{\rm TB}$, are obtained by solving the eigenvalue problem
\begin{equation}
    \sum_{l'}^{L_i}\Hm_{i,ll'}^{\rm TB}a_{i,jl'}=\hbar\vep_{i,j}a_{i,jl} ,
\end{equation}
where $L_i$ denotes the number of atomic orbitals $\ket{i,l}$ at positions $\mathbf{R}_{i,l}=(x_{i,l},y_{i,l},z_{i,l})$ in structure $i$ and $a_{i,jl}$ are the expansion coefficients of the single-particle state $\ket{i,j}$, defined as
\begin{equation}
    \ket{i,j} = \sum_{l}^{L_i}a_{i,jl}\ket{i,l} ,
\end{equation}
with energy $\hbar\vep_{i,j}$. The equilibrium state of structure $i$ is then conveniently defined according to a density matrix
\begin{equation}
    \rho_i^{(0)} = \sum_{j}^{L_i}f_{i,j}\ket{i,j}\bra{i,j} ,
\end{equation}
where $f_{i,j}=\left[\ee^{(\hbar\vep_{i,j}-\mu_i)/\kB T_i)}+1\right]^{-1}$ are Fermi-Dirac occupation factors depending on the chemical potential $\mu_i$ and electronic temperature $T_i$ in the structure.

For a one-dimensional nanoribbon, the index $j$ of state $\ket{i,j}$ denoted above corresponds to multiplexed electronic Bloch states $j\to \{j,k\}$ indexed by $L_i$ bands $j$ and Bloch wave vector $k$. The carrier density $n$ and the Fermi energy $\EF$ are then related by
\begin{equation}
    n_i = \frac{b_i}{2\pi}\sum_j \int_{-\pi/b_i}^{\pi/b_i}{\rm d}k f_{i,\{j,k\}} ,
\end{equation}
where $b_i$ denotes the length of the unit cell in structure $i$ (containing $L_i$ orbitals) in the direction of translational invariance.

\subsection{Optical response simulations}

For a system of $N$ structures, we describe the time-evolution of the single-particle density matrix $\rho_i$ in structure $i$ by the modified Liouville-von Neumann equation
\begin{equation} \label{eq:rho_eom}
    \pd{\rho_{i}}{t} = -\frac{\ii}{\hbar}\sqpar{\Hm_i^{\rm TB}-e\phi_i,\rho_i} - \frac{\gamma_i}{2}\ccpar{\rho_i-\rho_i^{(0)}} ,
\end{equation}
where $\Hm_i^{\rm TB}$ is the TB Hamiltonian, $e$ denotes the elementary charge, $\phi_i$ is the total electrostatic potential, and $\gamma_i$ is the phenomenological rate at which the system relaxes to the equilibrium state via inelastic scattering. In the above equation of motion, the coupling of structure $i$ (with $L_i$ atomic orbitals) to an external plane-wave electric field $\Eb^{\rm ext}$ and the induced near-fields enters through the potential
\begin{equation} \label{eq:phi_i}
    \phi_{i,l} = -\Rb_{i,l}\cdot\Eb^{\rm ext} + \sum_{i'}^N\sum_{l'}^{L_{i'}}v_{ii',ll'}\rho_{i',l'}^{\rm ind} ,
\end{equation}
where $v_{ii',ll'}$ quantifies the spatial dependence of the Coulomb interaction between charges in orbital $l$ of structure $i$ and the induced charge $\rho_{i,l}^{\rm ind}=-2e\ccpar{\rho_{i,ll}-\rho_{i,ll}^{(0)}}$ in orbital $l'$ of structure $i'$ (the prefactor of 2 accounting for spin degeneracy), calculated from the density matrix elements $\rho_{i,ll'}$. Note that the above prescription assumes that the constituent nanoribbons are separated sufficiently to prevent quantum tunneling, such that the Coulomb interaction constitutes the sole mechanism of inter-structure coupling.

To simulate the optical response of nanoribbon heterostructures to an arbitrary time-dependent external field $\Eb^{\rm ext}(t)$, direct numerical integration of the system of equations represented by Eq.~\eqref{eq:rho_eom} yields the density matrices $\rho_i(t)$, from which we compute the induced dipole moment
\begin{equation}
    \pb(t)=\sum_i^N\sum_l^{L_i}\Rb_{i,l}\rho_{i,l}^{\rm ind}(t) .
\end{equation}
In this work, we employ time-domain simulations to investigate the optical response of nanoribbon heterostructures illuminated by intense ultrashort pulses, whereby Fourier transformation of the induced dipole acceleration provides access to the high-harmonic spectrum~\cite{devega2020strong}.

In a complementary approach, we solve Eq.~\eqref{eq:rho_eom} perturbatively by considering monochromatic fields $\Eb^{\rm ext}\ee^{-\ii\omega t}+{\rm c.c.}$ in the frequency domain, which leads to self-consistent expressions for the induced charge density components $\rho_i^{ns}$ in structure $i$, to order $n$ in the electric field oscillating with harmonic $s$ of its fundamental frequency, as 
\begin{equation} \label{eq:rho_nl}
    \rho_{i,l}^{ns} = \sum_{l'}^{L_{i}}\chi_{i,ll'}^{0}(s\omega)\phi_{i,l'}^{ns} + \tilde{\rho}_{i,l}^{\,ns} ,
\end{equation}
where
\begin{equation}
    \chi_{i,ll'}^0(s\omega) = -\frac{2e}{\hbar}\sum_{jj'}\ccpar{f_{i,j'}-f_{i,j}}\frac{a_{i,jl}a_{i,j'l}^*a_{i,jl'}^*a_{i,j'l'}}{s\omega - \ccpar{\vep_{i,j}-\vep_{i,j'}} + \ii\gamma/2}
\end{equation}
is the non-interacting RPA susceptibility,
\begin{equation}
    \phi_{i,l}^{ns} = -\Rb_{i,l}\cdot\Eb^{\rm ext}\delta_{n,1} + \sum_{i'}^N\sum_{l'}^{L_{i'}}v_{ii',ll'}\rho_{i',l'}^{ns}
\end{equation}
is the self-consistent potential, analogous to that in Eq.~\eqref{eq:phi_i}, and $\tilde{\rho}_{i,l}^{\,ns}$ is a nonlinear source term that is non-zero for $n>1$ and is constructed recursively from lower-order terms. In practice, the system of equations represented by Eq.~\eqref{eq:rho_nl} is solved self-consistently through the inversion of block matrices.

The perturbative formalism described above is used to quantify the linear extinction cross sections and nonlinear optical susceptibilities of graphene nanoribbon heterostructures, illuminated by normally incident fields polarized along the ribbon confinement direction $\xx$, from their per unit length dipolar polarizability
\begin{equation}
    \alpha_{s\omega}^{(n)} = \sum_{i}^N\sum_{l}^{L_i}\frac{x_{i,l}\rho_{i,l}^{ns}}{(E^{\rm ext})^n} .
\end{equation}
Specifically, the linear extinction cross section is $\sigma^{\rm ext}=(4\pi\ww/c)\Imm\{\alpha_\ww^{(1)}\}$ and the nonlinear susceptibilities are obtained as $\chi_{s\ww}^{(n)}=\alpha_{s\ww}^{(n)}/(d_{\rm gr}\sum_i W_i)$, the latter quantity expressed in terms of the ribbon widths $W_i$ and the effective graphene thickness, estimated as $d_{\rm gr}=0.33$\,nm, corresponding to the interlayer spacing in graphite. Incidentally, we can compute the induced field associated with the charge density $\rho^{ns}$ according to
\begin{equation}
    \mathbf{E}^{ns}(\rb) = \sum_i^N \sum_l^{L_i} \rho_{i,l}^{ns} \frac{\rb-\Rb_{i,l}}{\left\vert\rb-\Rb_{i,l}\right\vert^3} ,
\end{equation}
where $\rb$ is an arbitrary position.

\section{Acknowledgments}

L.~J. and J.~D.~C. acknowledge support from Independent Research Fund Denmark (grant no. 0165-00051B).
The Center for Polariton-driven Light--Matter Interactions (POLIMA) is funded by the Danish National Research Foundation (Project No.~DNRF165).
Part of the computation in this project was performed on the DeiC Large Memory HPC system managed by the eScience Center at the University of Southern Denmark.

\section{Author contributions}

J.~D.~C. conceived the project and developed the theory.
L.~J. performed the numerical simulations and drafted the manuscript.
All authors contributed to the analysis of the results and writing of the manuscript.


\begin{thebibliography}{57}%
\makeatletter
\providecommand \@ifxundefined [1]{%
 \@ifx{#1\undefined}
}%
\providecommand \@ifnum [1]{%
 \ifnum #1\expandafter \@firstoftwo
 \else \expandafter \@secondoftwo
 \fi
}%
\providecommand \@ifx [1]{%
 \ifx #1\expandafter \@firstoftwo
 \else \expandafter \@secondoftwo
 \fi
}%
\providecommand \natexlab [1]{#1}%
\providecommand \enquote  [1]{``#1''}%
\providecommand \bibnamefont  [1]{#1}%
\providecommand \bibfnamefont [1]{#1}%
\providecommand \citenamefont [1]{#1}%
\providecommand \href@noop [0]{\@secondoftwo}%
\providecommand \href [0]{\begingroup \@sanitize@url \@href}%
\providecommand \@href[1]{\@@startlink{#1}\@@href}%
\providecommand \@@href[1]{\endgroup#1\@@endlink}%
\providecommand \@sanitize@url [0]{\catcode `\\12\catcode `\$12\catcode `\&12\catcode `\#12\catcode `\^12\catcode `\_12\catcode `\%12\relax}%
\providecommand \@@startlink[1]{}%
\providecommand \@@endlink[0]{}%
\providecommand \url  [0]{\begingroup\@sanitize@url \@url }%
\providecommand \@url [1]{\endgroup\@href {#1}{\urlprefix }}%
\providecommand \urlprefix  [0]{URL }%
\providecommand \Eprint [0]{\href }%
\providecommand \doibase [0]{https://doi.org/}%
\providecommand \selectlanguage [0]{\@gobble}%
\providecommand \bibinfo  [0]{\@secondoftwo}%
\providecommand \bibfield  [0]{\@secondoftwo}%
\providecommand \translation [1]{[#1]}%
\providecommand \BibitemOpen [0]{}%
\providecommand \bibitemStop [0]{}%
\providecommand \bibitemNoStop [0]{.\EOS\space}%
\providecommand \EOS [0]{\spacefactor3000\relax}%
\providecommand \BibitemShut  [1]{\csname bibitem#1\endcsname}%
\let\auto@bib@innerbib\@empty
\bibitem [{\citenamefont {Sun}\ \emph {et~al.}(2016)\citenamefont {Sun}, \citenamefont {Martinez},\ and\ \citenamefont {Wang}}]{sun2016optical}%
  \BibitemOpen
  \bibfield  {author} {\bibinfo {author} {\bibfnamefont {Z.}~\bibnamefont {Sun}}, \bibinfo {author} {\bibfnamefont {A.}~\bibnamefont {Martinez}},\ and\ \bibinfo {author} {\bibfnamefont {F.}~\bibnamefont {Wang}},\ }\bibfield  {title} {\bibinfo {title} {Optical modulators with 2d layered materials},\ }\href {https://doi.org/10.1038/nphoton.2016.15} {\bibfield  {journal} {\bibinfo  {journal} {Nat.\ Photonics}\ }\textbf {\bibinfo {volume} {10}},\ \bibinfo {pages} {227} (\bibinfo {year} {2016})}\BibitemShut {NoStop}%
\bibitem [{\citenamefont {Reserbat-Plantey}\ \emph {et~al.}(2021)\citenamefont {Reserbat-Plantey}, \citenamefont {Epstein}, \citenamefont {Torre}, \citenamefont {Costa}, \citenamefont {Gon{\c{c}}alves}, \citenamefont {Mortensen}, \citenamefont {Polini}, \citenamefont {Song}, \citenamefont {Peres},\ and\ \citenamefont {Koppens}}]{reserbat2021quantum}%
  \BibitemOpen
  \bibfield  {author} {\bibinfo {author} {\bibfnamefont {A.}~\bibnamefont {Reserbat-Plantey}}, \bibinfo {author} {\bibfnamefont {I.}~\bibnamefont {Epstein}}, \bibinfo {author} {\bibfnamefont {I.}~\bibnamefont {Torre}}, \bibinfo {author} {\bibfnamefont {A.~T.}\ \bibnamefont {Costa}}, \bibinfo {author} {\bibfnamefont {P.~A.~D.}\ \bibnamefont {Gon{\c{c}}alves}}, \bibinfo {author} {\bibfnamefont {N.~A.}\ \bibnamefont {Mortensen}}, \bibinfo {author} {\bibfnamefont {M.}~\bibnamefont {Polini}}, \bibinfo {author} {\bibfnamefont {J.~C.~W.}\ \bibnamefont {Song}}, \bibinfo {author} {\bibfnamefont {N.~M.~R.}\ \bibnamefont {Peres}},\ and\ \bibinfo {author} {\bibfnamefont {F.~H.~L.}\ \bibnamefont {Koppens}},\ }\bibfield  {title} {\bibinfo {title} {Quantum nanophotonics in two-dimensional materials},\ }\href {https://doi.org/10.1021/acsphotonics.0c01224} {\bibfield  {journal} {\bibinfo  {journal} {ACS Photonics}\ }\textbf {\bibinfo {volume} {8}},\ \bibinfo {pages} {85} (\bibinfo {year} {2021})}\BibitemShut {NoStop}%
\bibitem [{\citenamefont {Ma}\ \emph {et~al.}(2024)\citenamefont {Ma}, \citenamefont {Zhong}, \citenamefont {Dai},\ and\ \citenamefont {Ou}}]{ma2024inplane}%
  \BibitemOpen
  \bibfield  {author} {\bibinfo {author} {\bibfnamefont {Y.}~\bibnamefont {Ma}}, \bibinfo {author} {\bibfnamefont {G.}~\bibnamefont {Zhong}}, \bibinfo {author} {\bibfnamefont {Z.}~\bibnamefont {Dai}},\ and\ \bibinfo {author} {\bibfnamefont {Q.}~\bibnamefont {Ou}},\ }\bibfield  {title} {\bibinfo {title} {In-plane hyperbolic phonon polaritons: materials, properties, and nanophotonic devices},\ }\href {https://doi.org/10.1038/s44310-024-00019-4} {\bibfield  {journal} {\bibinfo  {journal} {npj Nanophotonics}\ }\textbf {\bibinfo {volume} {1}},\ \bibinfo {pages} {25} (\bibinfo {year} {2024})}\BibitemShut {NoStop}%
\bibitem [{\citenamefont {Sortino}\ \emph {et~al.}(2025)\citenamefont {Sortino}, \citenamefont {Guimar{\~a}es}, \citenamefont {Molina-S{\'a}nchez}, \citenamefont {Quan}, \citenamefont {Garoli},\ and\ \citenamefont {Maccaferri}}]{sortino2025light}%
  \BibitemOpen
  \bibfield  {author} {\bibinfo {author} {\bibfnamefont {L.}~\bibnamefont {Sortino}}, \bibinfo {author} {\bibfnamefont {M.~H.~D.}\ \bibnamefont {Guimar{\~a}es}}, \bibinfo {author} {\bibfnamefont {A.}~\bibnamefont {Molina-S{\'a}nchez}}, \bibinfo {author} {\bibfnamefont {J.}~\bibnamefont {Quan}}, \bibinfo {author} {\bibfnamefont {D.}~\bibnamefont {Garoli}},\ and\ \bibinfo {author} {\bibfnamefont {N.}~\bibnamefont {Maccaferri}},\ }\bibfield  {title} {\bibinfo {title} {Light-matter interactions in layered materials and heterostructures: from moir{\'e} physics and magneto-optical effects to ultrafast dynamics and hybrid meta-photonics},\ }\href {https://doi.org/10.1088/2053-1583/adc4f5} {\bibfield  {journal} {\bibinfo  {journal} {2D Mater.}\ }\textbf {\bibinfo {volume} {12}},\ \bibinfo {pages} {033003} (\bibinfo {year} {2025})}\BibitemShut {NoStop}%
\bibitem [{\citenamefont {{Garc\'{\i}a de Abajo}}\ \emph {et~al.}(2025)\citenamefont {{Garc\'{\i}a de Abajo}}, \citenamefont {Basov}, \citenamefont {Koppens}, \citenamefont {Orsini}, \citenamefont {Ceccanti}, \citenamefont {Castilla}, \citenamefont {Cavicchi}, \citenamefont {Polini}, \citenamefont {Gon\c{c}alves}, \citenamefont {Costa}, \citenamefont {Peres}, \citenamefont {Mortensen}, \citenamefont {Bharadwaj}, \citenamefont {Jacob}, \citenamefont {Schuck}, \citenamefont {Pasupathy}, \citenamefont {Delor}, \citenamefont {Liu}, \citenamefont {Mugarza}, \citenamefont {Merino}, \citenamefont {Cuxart}, \citenamefont {Ch\'{a}vez-Angel}, \citenamefont {Svec}, \citenamefont {Tizei}, \citenamefont {Dirnberger}, \citenamefont {Deng}, \citenamefont {Schneider}, \citenamefont {Menon}, \citenamefont {Deilmann}, \citenamefont {Chernikov}, \citenamefont {Thygesen}, \citenamefont {Abate}, \citenamefont {Terrones}, \citenamefont {Sangwan}, \citenamefont {Hersam}, \citenamefont {Yu}, \citenamefont {Chen}, \citenamefont
  {Heinz}, \citenamefont {Murthy}, \citenamefont {Kroner}, \citenamefont {Smolenski}, \citenamefont {Thureja}, \citenamefont {Chervy}, \citenamefont {Genco}, \citenamefont {Trovatello}, \citenamefont {Cerullo}, \citenamefont {Conte}, \citenamefont {Timmer}, \citenamefont {Sio}, \citenamefont {Lienau}, \citenamefont {Shang}, \citenamefont {Hong}, \citenamefont {Liu}, \citenamefont {Sun}, \citenamefont {Rozema}, \citenamefont {Walther}, \citenamefont {Al\`{u}}, \citenamefont {Cotrufo}, \citenamefont {Queiroz}, \citenamefont {Zhu}, \citenamefont {Cox}, \citenamefont {Dias}, \citenamefont {{Rodr\'{\i}guez Echarri}}, \citenamefont {Iyikanat}, \citenamefont {Marini}, \citenamefont {Herrmann}, \citenamefont {Tornow}, \citenamefont {Klimmer}, \citenamefont {Wilhelm}, \citenamefont {Soavi}, \citenamefont {Sun}, \citenamefont {Wu}, \citenamefont {Xiong}, \citenamefont {Matsyshyn}, \citenamefont {Kumar}, \citenamefont {Song}, \citenamefont {Bucher}, \citenamefont {Gorlach}, \citenamefont {Tsesses}, \citenamefont
  {Kaminer}, \citenamefont {Schwab}, \citenamefont {Mangold}, \citenamefont {Giessen}, \citenamefont {S\'anchez}, \citenamefont {Efetov}, \citenamefont {Low}, \citenamefont {G\'omez-Santos}, \citenamefont {Stauber}, \citenamefont {\'Alvarez-P\'erez}, \citenamefont {Duan}, \citenamefont {Mart\'{\i}n-Moreno}, \citenamefont {Paarmann}, \citenamefont {Caldwell}, \citenamefont {Nikitin}, \citenamefont {Alonso-Gonz\'alez}, \citenamefont {Mueller}, \citenamefont {Volkov}, \citenamefont {Jariwala}, \citenamefont {Shegai}, \citenamefont {van~de Groep}, \citenamefont {Boltasseva}, \citenamefont {Bondarev}, \citenamefont {Shalaev}, \citenamefont {Simon}, \citenamefont {Fruhling}, \citenamefont {Shen}, \citenamefont {Novko}, \citenamefont {Tan}, \citenamefont {Wang}, \citenamefont {Petek}, \citenamefont {Mkhitaryan}, \citenamefont {Yu}, \citenamefont {Manjavacas}, \citenamefont {Ortega}, \citenamefont {Cheng}, \citenamefont {Tian}, \citenamefont {Mao}, \citenamefont {Thourhout}, \citenamefont {Gan}, \citenamefont {Dai},
  \citenamefont {Sternbach}, \citenamefont {Zhou}, \citenamefont {Hafezi}, \citenamefont {Litvinov}, \citenamefont {Grzeszczyk}, \citenamefont {Novoselov}, \citenamefont {Koperski}, \citenamefont {Papadopoulos}, \citenamefont {Novotny}, \citenamefont {Viti}, \citenamefont {Vitiello}, \citenamefont {Cottam}, \citenamefont {Dewes}, \citenamefont {Makarovsky}, \citenamefont {Patan\`{e}}, \citenamefont {Song}, \citenamefont {Cai}, \citenamefont {Chen}, \citenamefont {Naveh}, \citenamefont {Jang}, \citenamefont {Park}, \citenamefont {Xia}, \citenamefont {Jenke}, \citenamefont {Bajo}, \citenamefont {Braun}, \citenamefont {Burch}, \citenamefont {Zhao},\ and\ \citenamefont {Xu}}]{garciadeabajo2025roadmap}%
  \BibitemOpen
  \bibfield  {author} {\bibinfo {author} {\bibfnamefont {F.~J.}\ \bibnamefont {{Garc\'{\i}a de Abajo}}}, \bibinfo {author} {\bibfnamefont {D.~N.}\ \bibnamefont {Basov}}, \bibinfo {author} {\bibfnamefont {F.~H.~L.}\ \bibnamefont {Koppens}}, \bibinfo {author} {\bibfnamefont {L.}~\bibnamefont {Orsini}}, \bibinfo {author} {\bibfnamefont {M.}~\bibnamefont {Ceccanti}}, \bibinfo {author} {\bibfnamefont {S.}~\bibnamefont {Castilla}}, \bibinfo {author} {\bibfnamefont {L.}~\bibnamefont {Cavicchi}}, \bibinfo {author} {\bibfnamefont {M.}~\bibnamefont {Polini}}, \bibinfo {author} {\bibfnamefont {P.~A.~D.}\ \bibnamefont {Gon\c{c}alves}}, \bibinfo {author} {\bibfnamefont {A.~T.}\ \bibnamefont {Costa}}, \bibinfo {author} {\bibfnamefont {N.~M.~R.}\ \bibnamefont {Peres}}, \bibinfo {author} {\bibfnamefont {N.~A.}\ \bibnamefont {Mortensen}}, \bibinfo {author} {\bibfnamefont {S.}~\bibnamefont {Bharadwaj}}, \bibinfo {author} {\bibfnamefont {Z.}~\bibnamefont {Jacob}}, \bibinfo {author} {\bibfnamefont {P.~J.}\ \bibnamefont
  {Schuck}}, \bibinfo {author} {\bibfnamefont {A.~N.}\ \bibnamefont {Pasupathy}}, \bibinfo {author} {\bibfnamefont {M.}~\bibnamefont {Delor}}, \bibinfo {author} {\bibfnamefont {M.~K.}\ \bibnamefont {Liu}}, \bibinfo {author} {\bibfnamefont {A.}~\bibnamefont {Mugarza}}, \bibinfo {author} {\bibfnamefont {P.}~\bibnamefont {Merino}}, \bibinfo {author} {\bibfnamefont {M.~G.}\ \bibnamefont {Cuxart}}, \bibinfo {author} {\bibfnamefont {E.}~\bibnamefont {Ch\'{a}vez-Angel}}, \bibinfo {author} {\bibfnamefont {M.}~\bibnamefont {Svec}}, \bibinfo {author} {\bibfnamefont {L.~H.~G.}\ \bibnamefont {Tizei}}, \bibinfo {author} {\bibfnamefont {F.}~\bibnamefont {Dirnberger}}, \bibinfo {author} {\bibfnamefont {H.}~\bibnamefont {Deng}}, \bibinfo {author} {\bibfnamefont {C.}~\bibnamefont {Schneider}}, \bibinfo {author} {\bibfnamefont {V.}~\bibnamefont {Menon}}, \bibinfo {author} {\bibfnamefont {T.}~\bibnamefont {Deilmann}}, \bibinfo {author} {\bibfnamefont {A.}~\bibnamefont {Chernikov}}, \bibinfo {author} {\bibfnamefont {K.~S.}\
  \bibnamefont {Thygesen}}, \bibinfo {author} {\bibfnamefont {Y.}~\bibnamefont {Abate}}, \bibinfo {author} {\bibfnamefont {M.}~\bibnamefont {Terrones}}, \bibinfo {author} {\bibfnamefont {V.~K.}\ \bibnamefont {Sangwan}}, \bibinfo {author} {\bibfnamefont {M.~C.}\ \bibnamefont {Hersam}}, \bibinfo {author} {\bibfnamefont {L.}~\bibnamefont {Yu}}, \bibinfo {author} {\bibfnamefont {X.}~\bibnamefont {Chen}}, \bibinfo {author} {\bibfnamefont {T.~F.}\ \bibnamefont {Heinz}}, \bibinfo {author} {\bibfnamefont {P.}~\bibnamefont {Murthy}}, \bibinfo {author} {\bibfnamefont {M.}~\bibnamefont {Kroner}}, \bibinfo {author} {\bibfnamefont {T.}~\bibnamefont {Smolenski}}, \bibinfo {author} {\bibfnamefont {D.}~\bibnamefont {Thureja}}, \bibinfo {author} {\bibfnamefont {T.}~\bibnamefont {Chervy}}, \bibinfo {author} {\bibfnamefont {A.}~\bibnamefont {Genco}}, \bibinfo {author} {\bibfnamefont {C.}~\bibnamefont {Trovatello}}, \bibinfo {author} {\bibfnamefont {G.}~\bibnamefont {Cerullo}}, \bibinfo {author} {\bibfnamefont {S.~D.}\
  \bibnamefont {Conte}}, \bibinfo {author} {\bibfnamefont {D.}~\bibnamefont {Timmer}}, \bibinfo {author} {\bibfnamefont {A.~D.}\ \bibnamefont {Sio}}, \bibinfo {author} {\bibfnamefont {C.}~\bibnamefont {Lienau}}, \bibinfo {author} {\bibfnamefont {N.}~\bibnamefont {Shang}}, \bibinfo {author} {\bibfnamefont {H.}~\bibnamefont {Hong}}, \bibinfo {author} {\bibfnamefont {K.}~\bibnamefont {Liu}}, \bibinfo {author} {\bibfnamefont {Z.}~\bibnamefont {Sun}}, \bibinfo {author} {\bibfnamefont {L.~A.}\ \bibnamefont {Rozema}}, \bibinfo {author} {\bibfnamefont {P.}~\bibnamefont {Walther}}, \bibinfo {author} {\bibfnamefont {A.}~\bibnamefont {Al\`{u}}}, \bibinfo {author} {\bibfnamefont {M.}~\bibnamefont {Cotrufo}}, \bibinfo {author} {\bibfnamefont {R.}~\bibnamefont {Queiroz}}, \bibinfo {author} {\bibfnamefont {X.-Y.}\ \bibnamefont {Zhu}}, \bibinfo {author} {\bibfnamefont {J.~D.}\ \bibnamefont {Cox}}, \bibinfo {author} {\bibfnamefont {E.~J.~C.}\ \bibnamefont {Dias}}, \bibinfo {author} {\bibfnamefont {A.}~\bibnamefont
  {{Rodr\'{\i}guez Echarri}}}, \bibinfo {author} {\bibfnamefont {F.}~\bibnamefont {Iyikanat}}, \bibinfo {author} {\bibfnamefont {A.}~\bibnamefont {Marini}}, \bibinfo {author} {\bibfnamefont {P.}~\bibnamefont {Herrmann}}, \bibinfo {author} {\bibfnamefont {N.}~\bibnamefont {Tornow}}, \bibinfo {author} {\bibfnamefont {S.}~\bibnamefont {Klimmer}}, \bibinfo {author} {\bibfnamefont {J.}~\bibnamefont {Wilhelm}}, \bibinfo {author} {\bibfnamefont {G.}~\bibnamefont {Soavi}}, \bibinfo {author} {\bibfnamefont {Z.}~\bibnamefont {Sun}}, \bibinfo {author} {\bibfnamefont {S.}~\bibnamefont {Wu}}, \bibinfo {author} {\bibfnamefont {Y.}~\bibnamefont {Xiong}}, \bibinfo {author} {\bibfnamefont {O.}~\bibnamefont {Matsyshyn}}, \bibinfo {author} {\bibfnamefont {R.~K.}\ \bibnamefont {Kumar}}, \bibinfo {author} {\bibfnamefont {J.~C.~W.}\ \bibnamefont {Song}}, \bibinfo {author} {\bibfnamefont {T.}~\bibnamefont {Bucher}}, \bibinfo {author} {\bibfnamefont {A.}~\bibnamefont {Gorlach}}, \bibinfo {author} {\bibfnamefont {S.}~\bibnamefont
  {Tsesses}}, \bibinfo {author} {\bibfnamefont {I.}~\bibnamefont {Kaminer}}, \bibinfo {author} {\bibfnamefont {J.}~\bibnamefont {Schwab}}, \bibinfo {author} {\bibfnamefont {F.}~\bibnamefont {Mangold}}, \bibinfo {author} {\bibfnamefont {H.}~\bibnamefont {Giessen}}, \bibinfo {author} {\bibfnamefont {M.~S.}\ \bibnamefont {S\'anchez}}, \bibinfo {author} {\bibfnamefont {D.~K.}\ \bibnamefont {Efetov}}, \bibinfo {author} {\bibfnamefont {T.}~\bibnamefont {Low}}, \bibinfo {author} {\bibfnamefont {G.}~\bibnamefont {G\'omez-Santos}}, \bibinfo {author} {\bibfnamefont {T.}~\bibnamefont {Stauber}}, \bibinfo {author} {\bibfnamefont {G.}~\bibnamefont {\'Alvarez-P\'erez}}, \bibinfo {author} {\bibfnamefont {J.}~\bibnamefont {Duan}}, \bibinfo {author} {\bibfnamefont {L.}~\bibnamefont {Mart\'{\i}n-Moreno}}, \bibinfo {author} {\bibfnamefont {A.}~\bibnamefont {Paarmann}}, \bibinfo {author} {\bibfnamefont {J.~D.}\ \bibnamefont {Caldwell}}, \bibinfo {author} {\bibfnamefont {A.~Y.}\ \bibnamefont {Nikitin}}, \bibinfo {author}
  {\bibfnamefont {P.}~\bibnamefont {Alonso-Gonz\'alez}}, \bibinfo {author} {\bibfnamefont {N.~S.}\ \bibnamefont {Mueller}}, \bibinfo {author} {\bibfnamefont {V.}~\bibnamefont {Volkov}}, \bibinfo {author} {\bibfnamefont {D.}~\bibnamefont {Jariwala}}, \bibinfo {author} {\bibfnamefont {T.}~\bibnamefont {Shegai}}, \bibinfo {author} {\bibfnamefont {J.}~\bibnamefont {van~de Groep}}, \bibinfo {author} {\bibfnamefont {A.}~\bibnamefont {Boltasseva}}, \bibinfo {author} {\bibfnamefont {I.~V.}\ \bibnamefont {Bondarev}}, \bibinfo {author} {\bibfnamefont {V.~M.}\ \bibnamefont {Shalaev}}, \bibinfo {author} {\bibfnamefont {J.}~\bibnamefont {Simon}}, \bibinfo {author} {\bibfnamefont {C.}~\bibnamefont {Fruhling}}, \bibinfo {author} {\bibfnamefont {G.}~\bibnamefont {Shen}}, \bibinfo {author} {\bibfnamefont {D.}~\bibnamefont {Novko}}, \bibinfo {author} {\bibfnamefont {S.}~\bibnamefont {Tan}}, \bibinfo {author} {\bibfnamefont {B.}~\bibnamefont {Wang}}, \bibinfo {author} {\bibfnamefont {H.}~\bibnamefont {Petek}}, \bibinfo {author}
  {\bibfnamefont {V.}~\bibnamefont {Mkhitaryan}}, \bibinfo {author} {\bibfnamefont {R.}~\bibnamefont {Yu}}, \bibinfo {author} {\bibfnamefont {A.}~\bibnamefont {Manjavacas}}, \bibinfo {author} {\bibfnamefont {J.~E.}\ \bibnamefont {Ortega}}, \bibinfo {author} {\bibfnamefont {X.}~\bibnamefont {Cheng}}, \bibinfo {author} {\bibfnamefont {R.}~\bibnamefont {Tian}}, \bibinfo {author} {\bibfnamefont {D.}~\bibnamefont {Mao}}, \bibinfo {author} {\bibfnamefont {D.~V.}\ \bibnamefont {Thourhout}}, \bibinfo {author} {\bibfnamefont {X.}~\bibnamefont {Gan}}, \bibinfo {author} {\bibfnamefont {Q.}~\bibnamefont {Dai}}, \bibinfo {author} {\bibfnamefont {A.}~\bibnamefont {Sternbach}}, \bibinfo {author} {\bibfnamefont {Y.}~\bibnamefont {Zhou}}, \bibinfo {author} {\bibfnamefont {M.}~\bibnamefont {Hafezi}}, \bibinfo {author} {\bibfnamefont {D.}~\bibnamefont {Litvinov}}, \bibinfo {author} {\bibfnamefont {M.}~\bibnamefont {Grzeszczyk}}, \bibinfo {author} {\bibfnamefont {K.~S.}\ \bibnamefont {Novoselov}}, \bibinfo {author}
  {\bibfnamefont {M.}~\bibnamefont {Koperski}}, \bibinfo {author} {\bibfnamefont {S.}~\bibnamefont {Papadopoulos}}, \bibinfo {author} {\bibfnamefont {L.}~\bibnamefont {Novotny}}, \bibinfo {author} {\bibfnamefont {L.}~\bibnamefont {Viti}}, \bibinfo {author} {\bibfnamefont {M.~S.}\ \bibnamefont {Vitiello}}, \bibinfo {author} {\bibfnamefont {N.~D.}\ \bibnamefont {Cottam}}, \bibinfo {author} {\bibfnamefont {B.~T.}\ \bibnamefont {Dewes}}, \bibinfo {author} {\bibfnamefont {O.}~\bibnamefont {Makarovsky}}, \bibinfo {author} {\bibfnamefont {A.}~\bibnamefont {Patan\`{e}}}, \bibinfo {author} {\bibfnamefont {Y.}~\bibnamefont {Song}}, \bibinfo {author} {\bibfnamefont {M.}~\bibnamefont {Cai}}, \bibinfo {author} {\bibfnamefont {J.}~\bibnamefont {Chen}}, \bibinfo {author} {\bibfnamefont {D.}~\bibnamefont {Naveh}}, \bibinfo {author} {\bibfnamefont {H.}~\bibnamefont {Jang}}, \bibinfo {author} {\bibfnamefont {S.}~\bibnamefont {Park}}, \bibinfo {author} {\bibfnamefont {F.}~\bibnamefont {Xia}}, \bibinfo {author} {\bibfnamefont
  {P.~K.}\ \bibnamefont {Jenke}}, \bibinfo {author} {\bibfnamefont {J.}~\bibnamefont {Bajo}}, \bibinfo {author} {\bibfnamefont {B.}~\bibnamefont {Braun}}, \bibinfo {author} {\bibfnamefont {K.~S.}\ \bibnamefont {Burch}}, \bibinfo {author} {\bibfnamefont {L.}~\bibnamefont {Zhao}},\ and\ \bibinfo {author} {\bibfnamefont {X.}~\bibnamefont {Xu}},\ }\bibfield  {title} {\bibinfo {title} {Roadmap for photonics with 2d materials},\ }\href {https://doi.org/10.1021/acsphotonics.5c00353} {\bibfield  {journal} {\bibinfo  {journal} {ACS Photonics}\ }\textbf {\bibinfo {volume} {12}},\ \bibinfo {pages} {3961} (\bibinfo {year} {2025})}\BibitemShut {NoStop}%
\bibitem [{\citenamefont {Fei}\ \emph {et~al.}(2012)\citenamefont {Fei}, \citenamefont {Rodin}, \citenamefont {Andreev}, \citenamefont {Bao}, \citenamefont {McLeod}, \citenamefont {Wagner}, \citenamefont {Zhang}, \citenamefont {Zhao}, \citenamefont {Thiemens}, \citenamefont {Dominguez}, \citenamefont {Fogler}, \citenamefont {Neto}, \citenamefont {Lau}, \citenamefont {Leilmann},\ and\ \citenamefont {Basov}}]{fei2012gatetuning}%
  \BibitemOpen
  \bibfield  {author} {\bibinfo {author} {\bibfnamefont {Z.}~\bibnamefont {Fei}}, \bibinfo {author} {\bibfnamefont {A.~S.}\ \bibnamefont {Rodin}}, \bibinfo {author} {\bibfnamefont {G.~O.}\ \bibnamefont {Andreev}}, \bibinfo {author} {\bibfnamefont {W.}~\bibnamefont {Bao}}, \bibinfo {author} {\bibfnamefont {A.~S.}\ \bibnamefont {McLeod}}, \bibinfo {author} {\bibfnamefont {M.}~\bibnamefont {Wagner}}, \bibinfo {author} {\bibfnamefont {L.~M.}\ \bibnamefont {Zhang}}, \bibinfo {author} {\bibfnamefont {Z.}~\bibnamefont {Zhao}}, \bibinfo {author} {\bibfnamefont {M.}~\bibnamefont {Thiemens}}, \bibinfo {author} {\bibfnamefont {G.}~\bibnamefont {Dominguez}}, \bibinfo {author} {\bibfnamefont {M.~M.}\ \bibnamefont {Fogler}}, \bibinfo {author} {\bibfnamefont {A.~H.~C.}\ \bibnamefont {Neto}}, \bibinfo {author} {\bibfnamefont {C.~N.}\ \bibnamefont {Lau}}, \bibinfo {author} {\bibfnamefont {F.}~\bibnamefont {Leilmann}},\ and\ \bibinfo {author} {\bibfnamefont {D.~N.}\ \bibnamefont {Basov}},\ }\bibfield  {title} {\bibinfo {title}
  {Gate-tuning of graphene plasmons revealed by infrared nano-imaging},\ }\href {https://doi.org/10.1038/nature11253} {\bibfield  {journal} {\bibinfo  {journal} {Nature}\ }\textbf {\bibinfo {volume} {487}},\ \bibinfo {pages} {82} (\bibinfo {year} {2012})}\BibitemShut {NoStop}%
\bibitem [{\citenamefont {Chen}\ \emph {et~al.}(2012)\citenamefont {Chen}, \citenamefont {Badioli}, \citenamefont {Alonso-Gonz{\'a}lez}, \citenamefont {Thongrattanasiri}, \citenamefont {Huth}, \citenamefont {Osmond}, \citenamefont {Spasenovi{\'c}}, \citenamefont {Centeno}, \citenamefont {Pesquera}, \citenamefont {Godignon}, \citenamefont {Elorza}, \citenamefont {Camara}, \citenamefont {{Garc\'{\i}a de Abajo}}, \citenamefont {Hillenbrand},\ and\ \citenamefont {Koppens}}]{chen2012optical}%
  \BibitemOpen
  \bibfield  {author} {\bibinfo {author} {\bibfnamefont {J.}~\bibnamefont {Chen}}, \bibinfo {author} {\bibfnamefont {M.}~\bibnamefont {Badioli}}, \bibinfo {author} {\bibfnamefont {P.}~\bibnamefont {Alonso-Gonz{\'a}lez}}, \bibinfo {author} {\bibfnamefont {S.}~\bibnamefont {Thongrattanasiri}}, \bibinfo {author} {\bibfnamefont {F.}~\bibnamefont {Huth}}, \bibinfo {author} {\bibfnamefont {J.}~\bibnamefont {Osmond}}, \bibinfo {author} {\bibfnamefont {M.}~\bibnamefont {Spasenovi{\'c}}}, \bibinfo {author} {\bibfnamefont {A.}~\bibnamefont {Centeno}}, \bibinfo {author} {\bibfnamefont {A.}~\bibnamefont {Pesquera}}, \bibinfo {author} {\bibfnamefont {P.}~\bibnamefont {Godignon}}, \bibinfo {author} {\bibfnamefont {A.~Z.}\ \bibnamefont {Elorza}}, \bibinfo {author} {\bibfnamefont {N.}~\bibnamefont {Camara}}, \bibinfo {author} {\bibfnamefont {F.~J.}\ \bibnamefont {{Garc\'{\i}a de Abajo}}}, \bibinfo {author} {\bibfnamefont {R.}~\bibnamefont {Hillenbrand}},\ and\ \bibinfo {author} {\bibfnamefont {F.~H.~L.}\ \bibnamefont
  {Koppens}},\ }\bibfield  {title} {\bibinfo {title} {Optical nano-imaging of gate-tunable graphene plasmons},\ }\href {https://doi.org/10.1038/nature11254} {\bibfield  {journal} {\bibinfo  {journal} {Nature}\ }\textbf {\bibinfo {volume} {487}},\ \bibinfo {pages} {77} (\bibinfo {year} {2012})}\BibitemShut {NoStop}%
\bibitem [{\citenamefont {Thongrattanasiri}\ and\ \citenamefont {{Garc\'{\i}a de Abajo}}(2013)}]{thongrattanasiri2013optical}%
  \BibitemOpen
  \bibfield  {author} {\bibinfo {author} {\bibfnamefont {S.}~\bibnamefont {Thongrattanasiri}}\ and\ \bibinfo {author} {\bibfnamefont {F.~J.}\ \bibnamefont {{Garc\'{\i}a de Abajo}}},\ }\bibfield  {title} {\bibinfo {title} {Optical field enhancement by strong plasmon interaction in graphene nanostructures},\ }\href {https://doi.org/10.1103/PhysRevLett.110.187401} {\bibfield  {journal} {\bibinfo  {journal} {Phys.\ Rev.\ Lett.}\ }\textbf {\bibinfo {volume} {110}},\ \bibinfo {pages} {187401} (\bibinfo {year} {2013})}\BibitemShut {NoStop}%
\bibitem [{\citenamefont {Rodrigo}\ \emph {et~al.}(2015)\citenamefont {Rodrigo}, \citenamefont {Limaj}, \citenamefont {Janner}, \citenamefont {Etezadi}, \citenamefont {{Garc\'{\i}a de Abajo}}, \citenamefont {Pruneri},\ and\ \citenamefont {Altug}}]{rodrigo2015midinfrared}%
  \BibitemOpen
  \bibfield  {author} {\bibinfo {author} {\bibfnamefont {D.}~\bibnamefont {Rodrigo}}, \bibinfo {author} {\bibfnamefont {O.}~\bibnamefont {Limaj}}, \bibinfo {author} {\bibfnamefont {D.}~\bibnamefont {Janner}}, \bibinfo {author} {\bibfnamefont {D.}~\bibnamefont {Etezadi}}, \bibinfo {author} {\bibfnamefont {F.~J.}\ \bibnamefont {{Garc\'{\i}a de Abajo}}}, \bibinfo {author} {\bibfnamefont {V.}~\bibnamefont {Pruneri}},\ and\ \bibinfo {author} {\bibfnamefont {H.}~\bibnamefont {Altug}},\ }\bibfield  {title} {\bibinfo {title} {Mid-infrared plasmonic biosensing with graphene},\ }\href {https://doi.org/10.1126/science.aab2051} {\bibfield  {journal} {\bibinfo  {journal} {Science}\ }\textbf {\bibinfo {volume} {349}},\ \bibinfo {pages} {165} (\bibinfo {year} {2015})}\BibitemShut {NoStop}%
\bibitem [{\citenamefont {Hafez}\ \emph {et~al.}(2018)\citenamefont {Hafez}, \citenamefont {Kovalev}, \citenamefont {Deinert}, \citenamefont {Mics}, \citenamefont {Green}, \citenamefont {Awari}, \citenamefont {Chen}, \citenamefont {Germanskiy}, \citenamefont {Lehnert}, \citenamefont {Teichert}, \citenamefont {Wang}, \citenamefont {Tielrooij}, \citenamefont {Liu}, \citenamefont {Chen}, \citenamefont {Narita}, \citenamefont {M{\"u}llen}, \citenamefont {Bonn}, \citenamefont {Gensch},\ and\ \citenamefont {Turchinovich}}]{hafez2018extremely}%
  \BibitemOpen
  \bibfield  {author} {\bibinfo {author} {\bibfnamefont {H.~A.}\ \bibnamefont {Hafez}}, \bibinfo {author} {\bibfnamefont {S.}~\bibnamefont {Kovalev}}, \bibinfo {author} {\bibfnamefont {J.-C.}\ \bibnamefont {Deinert}}, \bibinfo {author} {\bibfnamefont {Z.}~\bibnamefont {Mics}}, \bibinfo {author} {\bibfnamefont {B.}~\bibnamefont {Green}}, \bibinfo {author} {\bibfnamefont {N.}~\bibnamefont {Awari}}, \bibinfo {author} {\bibfnamefont {M.}~\bibnamefont {Chen}}, \bibinfo {author} {\bibfnamefont {S.}~\bibnamefont {Germanskiy}}, \bibinfo {author} {\bibfnamefont {U.}~\bibnamefont {Lehnert}}, \bibinfo {author} {\bibfnamefont {J.}~\bibnamefont {Teichert}}, \bibinfo {author} {\bibfnamefont {Z.}~\bibnamefont {Wang}}, \bibinfo {author} {\bibfnamefont {K.-J.}\ \bibnamefont {Tielrooij}}, \bibinfo {author} {\bibfnamefont {Z.}~\bibnamefont {Liu}}, \bibinfo {author} {\bibfnamefont {Z.}~\bibnamefont {Chen}}, \bibinfo {author} {\bibfnamefont {A.}~\bibnamefont {Narita}}, \bibinfo {author} {\bibfnamefont {K.}~\bibnamefont
  {M{\"u}llen}}, \bibinfo {author} {\bibfnamefont {M.}~\bibnamefont {Bonn}}, \bibinfo {author} {\bibfnamefont {M.}~\bibnamefont {Gensch}},\ and\ \bibinfo {author} {\bibfnamefont {D.}~\bibnamefont {Turchinovich}},\ }\bibfield  {title} {\bibinfo {title} {Extremely efficient terahertz high-harmonic generation in graphene by hot dirac fermions},\ }\href {https://doi.org/10.1038/s41586-018-0508-1} {\bibfield  {journal} {\bibinfo  {journal} {Nature}\ }\textbf {\bibinfo {volume} {561}},\ \bibinfo {pages} {507} (\bibinfo {year} {2018})}\BibitemShut {NoStop}%
\bibitem [{\citenamefont {Cox}\ and\ \citenamefont {{Garc\'{\i}a de Abajo}}(2018)}]{cox2018transient}%
  \BibitemOpen
  \bibfield  {author} {\bibinfo {author} {\bibfnamefont {J.~D.}\ \bibnamefont {Cox}}\ and\ \bibinfo {author} {\bibfnamefont {F.~J.}\ \bibnamefont {{Garc\'{\i}a de Abajo}}},\ }\bibfield  {title} {\bibinfo {title} {Transient nonlinear plasmonics in nanostructured graphene},\ }\href {https://doi.org/10.1364/OPTICA.5.000429} {\bibfield  {journal} {\bibinfo  {journal} {Optica}\ }\textbf {\bibinfo {volume} {5}},\ \bibinfo {pages} {429} (\bibinfo {year} {2018})}\BibitemShut {NoStop}%
\bibitem [{\citenamefont {Dias}\ \emph {et~al.}(2020)\citenamefont {Dias}, \citenamefont {Yu},\ and\ \citenamefont {{Garc\'{\i}a de Abajo}}}]{dias2020thermal}%
  \BibitemOpen
  \bibfield  {author} {\bibinfo {author} {\bibfnamefont {E.~J.~C.}\ \bibnamefont {Dias}}, \bibinfo {author} {\bibfnamefont {R.}~\bibnamefont {Yu}},\ and\ \bibinfo {author} {\bibfnamefont {F.~J.}\ \bibnamefont {{Garc\'{\i}a de Abajo}}},\ }\bibfield  {title} {\bibinfo {title} {Thermal manipulation of plasmons in atomically thin films},\ }\href {https://doi.org/10.1038/s41377-020-0322-z} {\bibfield  {journal} {\bibinfo  {journal} {Light Sci.\ Appl.}\ }\textbf {\bibinfo {volume} {9}},\ \bibinfo {pages} {87} (\bibinfo {year} {2020})}\BibitemShut {NoStop}%
\bibitem [{\citenamefont {Zeng}\ \emph {et~al.}(2023)\citenamefont {Zeng}, \citenamefont {Wan}, \citenamefont {Zhao}, \citenamefont {Huang}, \citenamefont {Wang}, \citenamefont {Cheng},\ and\ \citenamefont {Jiang}}]{zeng2023nonlinear}%
  \BibitemOpen
  \bibfield  {author} {\bibinfo {author} {\bibfnamefont {X.}~\bibnamefont {Zeng}}, \bibinfo {author} {\bibfnamefont {C.}~\bibnamefont {Wan}}, \bibinfo {author} {\bibfnamefont {Z.}~\bibnamefont {Zhao}}, \bibinfo {author} {\bibfnamefont {D.}~\bibnamefont {Huang}}, \bibinfo {author} {\bibfnamefont {Z.}~\bibnamefont {Wang}}, \bibinfo {author} {\bibfnamefont {X.}~\bibnamefont {Cheng}},\ and\ \bibinfo {author} {\bibfnamefont {T.}~\bibnamefont {Jiang}},\ }\bibfield  {title} {\bibinfo {title} {Nonlinear optics of two-dimensional heterostructures},\ }\href {https://doi.org/10.1007/s11467-023-1363-6} {\bibfield  {journal} {\bibinfo  {journal} {Front.\ Phys.}\ }\textbf {\bibinfo {volume} {19}},\ \bibinfo {pages} {33301} (\bibinfo {year} {2023})}\BibitemShut {NoStop}%
\bibitem [{\citenamefont {Ninhos}\ \emph {et~al.}(2024)\citenamefont {Ninhos}, \citenamefont {Tserkezis}, \citenamefont {Mortensen},\ and\ \citenamefont {Peres}}]{ninhos2024tunable}%
  \BibitemOpen
  \bibfield  {author} {\bibinfo {author} {\bibfnamefont {P.}~\bibnamefont {Ninhos}}, \bibinfo {author} {\bibfnamefont {C.}~\bibnamefont {Tserkezis}}, \bibinfo {author} {\bibfnamefont {N.~A.}\ \bibnamefont {Mortensen}},\ and\ \bibinfo {author} {\bibfnamefont {N.~M.~R.}\ \bibnamefont {Peres}},\ }\bibfield  {title} {\bibinfo {title} {Tunable exciton polaritons in band-gap engineered hexagonal boron nitride},\ }\href {https://doi.org/10.1021/acsnano.4c07003} {\bibfield  {journal} {\bibinfo  {journal} {ACS Nano}\ }\textbf {\bibinfo {volume} {18}},\ \bibinfo {pages} {20751} (\bibinfo {year} {2024})}\BibitemShut {NoStop}%
\bibitem [{\citenamefont {Liu}\ \emph {et~al.}(2017)\citenamefont {Liu}, \citenamefont {Li}, \citenamefont {You}, \citenamefont {Ghimire}, \citenamefont {Heinz},\ and\ \citenamefont {Reis}}]{liu2017highharmonic}%
  \BibitemOpen
  \bibfield  {author} {\bibinfo {author} {\bibfnamefont {H.}~\bibnamefont {Liu}}, \bibinfo {author} {\bibfnamefont {Y.}~\bibnamefont {Li}}, \bibinfo {author} {\bibfnamefont {Y.~S.}\ \bibnamefont {You}}, \bibinfo {author} {\bibfnamefont {S.}~\bibnamefont {Ghimire}}, \bibinfo {author} {\bibfnamefont {T.~F.}\ \bibnamefont {Heinz}},\ and\ \bibinfo {author} {\bibfnamefont {D.~A.}\ \bibnamefont {Reis}},\ }\bibfield  {title} {\bibinfo {title} {High-harmonic generation from an atomically thin semiconductor},\ }\href {https://doi.org/10.1038/nphys3946} {\bibfield  {journal} {\bibinfo  {journal} {Nat.\ Phys.}\ }\textbf {\bibinfo {volume} {13}},\ \bibinfo {pages} {262} (\bibinfo {year} {2017})}\BibitemShut {NoStop}%
\bibitem [{\citenamefont {Kundys}\ \emph {et~al.}(2018)\citenamefont {Kundys}, \citenamefont {Duppen}, \citenamefont {Marshall}, \citenamefont {Rodriguez}, \citenamefont {Torre}, \citenamefont {Tomadin}, \citenamefont {Polini},\ and\ \citenamefont {Grigorenko}}]{kundys2018nonlinear}%
  \BibitemOpen
  \bibfield  {author} {\bibinfo {author} {\bibfnamefont {D.}~\bibnamefont {Kundys}}, \bibinfo {author} {\bibfnamefont {B.~V.}\ \bibnamefont {Duppen}}, \bibinfo {author} {\bibfnamefont {O.~P.}\ \bibnamefont {Marshall}}, \bibinfo {author} {\bibfnamefont {F.}~\bibnamefont {Rodriguez}}, \bibinfo {author} {\bibfnamefont {I.}~\bibnamefont {Torre}}, \bibinfo {author} {\bibfnamefont {A.}~\bibnamefont {Tomadin}}, \bibinfo {author} {\bibfnamefont {M.}~\bibnamefont {Polini}},\ and\ \bibinfo {author} {\bibfnamefont {A.~N.}\ \bibnamefont {Grigorenko}},\ }\bibfield  {title} {\bibinfo {title} {Nonlinear light mixing by graphene plasmons},\ }\href {https://doi.org/10.1021/acs.nanolett.7b04114} {\bibfield  {journal} {\bibinfo  {journal} {Nano Lett.}\ }\textbf {\bibinfo {volume} {18}},\ \bibinfo {pages} {282} (\bibinfo {year} {2018})}\BibitemShut {NoStop}%
\bibitem [{\citenamefont {Baudisch}\ \emph {et~al.}(2018)\citenamefont {Baudisch}, \citenamefont {Marini}, \citenamefont {Cox}, \citenamefont {Zhu}, \citenamefont {Silva}, \citenamefont {Teichmann}, \citenamefont {Massicotte}, \citenamefont {Koppens}, \citenamefont {Levitov}, \citenamefont {{Garc\'{\i}a de Abajo}},\ and\ \citenamefont {Biegert}}]{baudisch2018ultrafast}%
  \BibitemOpen
  \bibfield  {author} {\bibinfo {author} {\bibfnamefont {M.}~\bibnamefont {Baudisch}}, \bibinfo {author} {\bibfnamefont {A.}~\bibnamefont {Marini}}, \bibinfo {author} {\bibfnamefont {J.~D.}\ \bibnamefont {Cox}}, \bibinfo {author} {\bibfnamefont {T.}~\bibnamefont {Zhu}}, \bibinfo {author} {\bibfnamefont {F.}~\bibnamefont {Silva}}, \bibinfo {author} {\bibfnamefont {S.}~\bibnamefont {Teichmann}}, \bibinfo {author} {\bibfnamefont {M.}~\bibnamefont {Massicotte}}, \bibinfo {author} {\bibfnamefont {F.}~\bibnamefont {Koppens}}, \bibinfo {author} {\bibfnamefont {L.~S.}\ \bibnamefont {Levitov}}, \bibinfo {author} {\bibfnamefont {F.~J.}\ \bibnamefont {{Garc\'{\i}a de Abajo}}},\ and\ \bibinfo {author} {\bibfnamefont {J.}~\bibnamefont {Biegert}},\ }\bibfield  {title} {\bibinfo {title} {Ultrafast nonlinear optical response of dirac fermions in graphene},\ }\href {https://doi.org/10.1038/s41467-018-03413-7} {\bibfield  {journal} {\bibinfo  {journal} {Nat.\ Commun.}\ }\textbf {\bibinfo {volume} {9}},\ \bibinfo {pages} {1018}
  (\bibinfo {year} {2018})}\BibitemShut {NoStop}%
\bibitem [{\citenamefont {Autere}\ \emph {et~al.}(2018)\citenamefont {Autere}, \citenamefont {Jussila}, \citenamefont {Marini}, \citenamefont {Saavedra}, \citenamefont {Dai}, \citenamefont {{a}yn\ {a}tjoki}, \citenamefont {Karvonen}, \citenamefont {Yang}, \citenamefont {Amirsolaimani}, \citenamefont {Norwood}, \citenamefont {Peyghambarian}, \citenamefont {Lipsanen}, \citenamefont {Kieu}, \citenamefont {{Garc\'{\i}a de Abajo}},\ and\ \citenamefont {Sun}}]{autere2018optical}%
  \BibitemOpen
  \bibfield  {author} {\bibinfo {author} {\bibfnamefont {A.}~\bibnamefont {Autere}}, \bibinfo {author} {\bibfnamefont {H.}~\bibnamefont {Jussila}}, \bibinfo {author} {\bibfnamefont {A.}~\bibnamefont {Marini}}, \bibinfo {author} {\bibfnamefont {J.~R.~M.}\ \bibnamefont {Saavedra}}, \bibinfo {author} {\bibfnamefont {Y.}~\bibnamefont {Dai}}, \bibinfo {author} {\bibfnamefont {A.~S.}\ \bibnamefont {{a}yn\ {a}tjoki}}, \bibinfo {author} {\bibfnamefont {L.}~\bibnamefont {Karvonen}}, \bibinfo {author} {\bibfnamefont {H.}~\bibnamefont {Yang}}, \bibinfo {author} {\bibfnamefont {B.}~\bibnamefont {Amirsolaimani}}, \bibinfo {author} {\bibfnamefont {R.~A.}\ \bibnamefont {Norwood}}, \bibinfo {author} {\bibfnamefont {N.}~\bibnamefont {Peyghambarian}}, \bibinfo {author} {\bibfnamefont {H.}~\bibnamefont {Lipsanen}}, \bibinfo {author} {\bibfnamefont {K.}~\bibnamefont {Kieu}}, \bibinfo {author} {\bibfnamefont {F.~J.}\ \bibnamefont {{Garc\'{\i}a de Abajo}}},\ and\ \bibinfo {author} {\bibfnamefont {Z.}~\bibnamefont {Sun}},\
  }\bibfield  {title} {\bibinfo {title} {Optical harmonic generation in monolayer group-vi transition metal dichalcogenides},\ }\href {https://doi.org/10.1103/PhysRevB.98.115426} {\bibfield  {journal} {\bibinfo  {journal} {Phys.\ Rev.\ B}\ }\textbf {\bibinfo {volume} {98}},\ \bibinfo {pages} {115426} (\bibinfo {year} {2018})}\BibitemShut {NoStop}%
\bibitem [{\citenamefont {{Alonso Calafell}}\ \emph {et~al.}(2021)\citenamefont {{Alonso Calafell}}, \citenamefont {Rozema}, \citenamefont {Iranzo}, \citenamefont {Trenti}, \citenamefont {Jenke}, \citenamefont {Cox}, \citenamefont {Kumar}, \citenamefont {Bieliaiev}, \citenamefont {Nanot}, \citenamefont {Peng}, \citenamefont {Efetov}, \citenamefont {Hong}, \citenamefont {Kong}, \citenamefont {Englund}, \citenamefont {{Garc\'{\i}a de Abajo}}, \citenamefont {Koppens},\ and\ \citenamefont {Walther}}]{alonsocalafell2021giant}%
  \BibitemOpen
  \bibfield  {author} {\bibinfo {author} {\bibfnamefont {I.}~\bibnamefont {{Alonso Calafell}}}, \bibinfo {author} {\bibfnamefont {L.~A.}\ \bibnamefont {Rozema}}, \bibinfo {author} {\bibfnamefont {D.~A.}\ \bibnamefont {Iranzo}}, \bibinfo {author} {\bibfnamefont {A.}~\bibnamefont {Trenti}}, \bibinfo {author} {\bibfnamefont {P.~K.}\ \bibnamefont {Jenke}}, \bibinfo {author} {\bibfnamefont {J.~D.}\ \bibnamefont {Cox}}, \bibinfo {author} {\bibfnamefont {A.}~\bibnamefont {Kumar}}, \bibinfo {author} {\bibfnamefont {H.}~\bibnamefont {Bieliaiev}}, \bibinfo {author} {\bibfnamefont {S.}~\bibnamefont {Nanot}}, \bibinfo {author} {\bibfnamefont {C.}~\bibnamefont {Peng}}, \bibinfo {author} {\bibfnamefont {D.~K.}\ \bibnamefont {Efetov}}, \bibinfo {author} {\bibfnamefont {J.-Y.}\ \bibnamefont {Hong}}, \bibinfo {author} {\bibfnamefont {J.}~\bibnamefont {Kong}}, \bibinfo {author} {\bibfnamefont {D.~R.}\ \bibnamefont {Englund}}, \bibinfo {author} {\bibfnamefont {F.~J.}\ \bibnamefont {{Garc\'{\i}a de Abajo}}}, \bibinfo {author}
  {\bibfnamefont {F.~H.~L.}\ \bibnamefont {Koppens}},\ and\ \bibinfo {author} {\bibfnamefont {P.}~\bibnamefont {Walther}},\ }\bibfield  {title} {\bibinfo {title} {Giant enhancement of third-harmonic generation in graphene--metal heterostructures},\ }\href {https://doi.org/10.1038/s41565-020-00808-w} {\bibfield  {journal} {\bibinfo  {journal} {Nat.\ Nanotechnol.}\ }\textbf {\bibinfo {volume} {16}},\ \bibinfo {pages} {318} (\bibinfo {year} {2021})}\BibitemShut {NoStop}%
\bibitem [{\citenamefont {Li}\ \emph {et~al.}(2022)\citenamefont {Li}, \citenamefont {An}, \citenamefont {Lu}, \citenamefont {Wang}, \citenamefont {Chang}, \citenamefont {Tan}, \citenamefont {Guo}, \citenamefont {Xu}, \citenamefont {He}, \citenamefont {Xia}, \citenamefont {Wu}, \citenamefont {Su}, \citenamefont {Liu}, \citenamefont {Rao}, \citenamefont {Soavi},\ and\ \citenamefont {Yao}}]{li2022nonlinear}%
  \BibitemOpen
  \bibfield  {author} {\bibinfo {author} {\bibfnamefont {Y.}~\bibnamefont {Li}}, \bibinfo {author} {\bibfnamefont {N.}~\bibnamefont {An}}, \bibinfo {author} {\bibfnamefont {Z.}~\bibnamefont {Lu}}, \bibinfo {author} {\bibfnamefont {Y.}~\bibnamefont {Wang}}, \bibinfo {author} {\bibfnamefont {B.}~\bibnamefont {Chang}}, \bibinfo {author} {\bibfnamefont {T.}~\bibnamefont {Tan}}, \bibinfo {author} {\bibfnamefont {X.}~\bibnamefont {Guo}}, \bibinfo {author} {\bibfnamefont {X.}~\bibnamefont {Xu}}, \bibinfo {author} {\bibfnamefont {J.}~\bibnamefont {He}}, \bibinfo {author} {\bibfnamefont {H.}~\bibnamefont {Xia}}, \bibinfo {author} {\bibfnamefont {Z.}~\bibnamefont {Wu}}, \bibinfo {author} {\bibfnamefont {Y.}~\bibnamefont {Su}}, \bibinfo {author} {\bibfnamefont {Y.}~\bibnamefont {Liu}}, \bibinfo {author} {\bibfnamefont {Y.}~\bibnamefont {Rao}}, \bibinfo {author} {\bibfnamefont {G.}~\bibnamefont {Soavi}},\ and\ \bibinfo {author} {\bibfnamefont {B.}~\bibnamefont {Yao}},\ }\bibfield  {title} {\bibinfo {title} {Nonlinear
  co-generation of graphene plasmons for optoelectronic logic operations},\ }\href {https://doi.org/10.1038/s41467-022-30901-8} {\bibfield  {journal} {\bibinfo  {journal} {Nat.\ Commun.}\ }\textbf {\bibinfo {volume} {13}},\ \bibinfo {pages} {3138} (\bibinfo {year} {2022})}\BibitemShut {NoStop}%
\bibitem [{\citenamefont {Zhang}\ \emph {et~al.}(2022)\citenamefont {Zhang}, \citenamefont {Wang}, \citenamefont {Dai}, \citenamefont {Bai}, \citenamefont {Hu}, \citenamefont {Du}, \citenamefont {Hu}, \citenamefont {Yang}, \citenamefont {Li}, \citenamefont {Dai}, \citenamefont {Hasan},\ and\ \citenamefont {Sun}}]{zhang2022chirality}%
  \BibitemOpen
  \bibfield  {author} {\bibinfo {author} {\bibfnamefont {Y.}~\bibnamefont {Zhang}}, \bibinfo {author} {\bibfnamefont {Y.}~\bibnamefont {Wang}}, \bibinfo {author} {\bibfnamefont {Y.}~\bibnamefont {Dai}}, \bibinfo {author} {\bibfnamefont {X.}~\bibnamefont {Bai}}, \bibinfo {author} {\bibfnamefont {X.}~\bibnamefont {Hu}}, \bibinfo {author} {\bibfnamefont {L.}~\bibnamefont {Du}}, \bibinfo {author} {\bibfnamefont {H.}~\bibnamefont {Hu}}, \bibinfo {author} {\bibfnamefont {X.}~\bibnamefont {Yang}}, \bibinfo {author} {\bibfnamefont {D.}~\bibnamefont {Li}}, \bibinfo {author} {\bibfnamefont {Q.}~\bibnamefont {Dai}}, \bibinfo {author} {\bibfnamefont {T.}~\bibnamefont {Hasan}},\ and\ \bibinfo {author} {\bibfnamefont {Z.}~\bibnamefont {Sun}},\ }\bibfield  {title} {\bibinfo {title} {Chirality logic gates},\ }\href {https://doi.org/10.1126/sciadv.abq8246} {\bibfield  {journal} {\bibinfo  {journal} {Sci.\ Adv.}\ }\textbf {\bibinfo {volume} {8}},\ \bibinfo {pages} {eabq8246} (\bibinfo {year} {2022})}\BibitemShut {NoStop}%
\bibitem [{\citenamefont {Basov}\ \emph {et~al.}(2016)\citenamefont {Basov}, \citenamefont {Fogler},\ and\ \citenamefont {{Garc\'{\i}a de Abajo}}}]{basov2016polaritons}%
  \BibitemOpen
  \bibfield  {author} {\bibinfo {author} {\bibfnamefont {D.~N.}\ \bibnamefont {Basov}}, \bibinfo {author} {\bibfnamefont {M.~M.}\ \bibnamefont {Fogler}},\ and\ \bibinfo {author} {\bibfnamefont {F.~J.}\ \bibnamefont {{Garc\'{\i}a de Abajo}}},\ }\bibfield  {title} {\bibinfo {title} {Polaritons in van der waals materials},\ }\href {https://doi.org/10.1126/science.aag1992} {\bibfield  {journal} {\bibinfo  {journal} {Science}\ }\textbf {\bibinfo {volume} {354}},\ \bibinfo {pages} {aag1992} (\bibinfo {year} {2016})}\BibitemShut {NoStop}%
\bibitem [{\citenamefont {Low}\ \emph {et~al.}(2017)\citenamefont {Low}, \citenamefont {Chaves}, \citenamefont {Caldwell}, \citenamefont {Kumar}, \citenamefont {Fang}, \citenamefont {Avouris}, \citenamefont {Heinz}, \citenamefont {Guinea}, \citenamefont {Martin-Moreno},\ and\ \citenamefont {Koppens}}]{low2017polaritons}%
  \BibitemOpen
  \bibfield  {author} {\bibinfo {author} {\bibfnamefont {T.}~\bibnamefont {Low}}, \bibinfo {author} {\bibfnamefont {A.}~\bibnamefont {Chaves}}, \bibinfo {author} {\bibfnamefont {J.~D.}\ \bibnamefont {Caldwell}}, \bibinfo {author} {\bibfnamefont {A.}~\bibnamefont {Kumar}}, \bibinfo {author} {\bibfnamefont {N.~X.}\ \bibnamefont {Fang}}, \bibinfo {author} {\bibfnamefont {P.}~\bibnamefont {Avouris}}, \bibinfo {author} {\bibfnamefont {T.~F.}\ \bibnamefont {Heinz}}, \bibinfo {author} {\bibfnamefont {F.}~\bibnamefont {Guinea}}, \bibinfo {author} {\bibfnamefont {L.}~\bibnamefont {Martin-Moreno}},\ and\ \bibinfo {author} {\bibfnamefont {F.}~\bibnamefont {Koppens}},\ }\bibfield  {title} {\bibinfo {title} {Polaritons in layered two-dimensional materials},\ }\href {https://doi.org/10.1038/nmat4792} {\bibfield  {journal} {\bibinfo  {journal} {Nat.\ Mater.}\ }\textbf {\bibinfo {volume} {16}},\ \bibinfo {pages} {182} (\bibinfo {year} {2017})}\BibitemShut {NoStop}%
\bibitem [{\citenamefont {Basov}\ \emph {et~al.}(2021)\citenamefont {Basov}, \citenamefont {Asenjo-Garcia}, \citenamefont {Schuck}, \citenamefont {Zhu},\ and\ \citenamefont {Rubio}}]{basov2021polariton}%
  \BibitemOpen
  \bibfield  {author} {\bibinfo {author} {\bibfnamefont {D.~N.}\ \bibnamefont {Basov}}, \bibinfo {author} {\bibfnamefont {A.}~\bibnamefont {Asenjo-Garcia}}, \bibinfo {author} {\bibfnamefont {P.~J.}\ \bibnamefont {Schuck}}, \bibinfo {author} {\bibfnamefont {X.}~\bibnamefont {Zhu}},\ and\ \bibinfo {author} {\bibfnamefont {A.}~\bibnamefont {Rubio}},\ }\bibfield  {title} {\bibinfo {title} {Polariton panorama},\ }\href {https://doi.org/10.1515/nanoph-2020-0449} {\bibfield  {journal} {\bibinfo  {journal} {Nanophotonics}\ }\textbf {\bibinfo {volume} {10}},\ \bibinfo {pages} {549} (\bibinfo {year} {2021})}\BibitemShut {NoStop}%
\bibitem [{\citenamefont {{Garc\'{\i}a de Abajo}}(2014)}]{garciadeabajo2014graphene}%
  \BibitemOpen
  \bibfield  {author} {\bibinfo {author} {\bibfnamefont {F.~J.}\ \bibnamefont {{Garc\'{\i}a de Abajo}}},\ }\bibfield  {title} {\bibinfo {title} {Graphene plasmonics: challenges and opportunities},\ }\href {https://doi.org/10.1021/ph400147y} {\bibfield  {journal} {\bibinfo  {journal} {ACS Photonics}\ }\textbf {\bibinfo {volume} {1}},\ \bibinfo {pages} {135} (\bibinfo {year} {2014})}\BibitemShut {NoStop}%
\bibitem [{\citenamefont {Zhang}\ \emph {et~al.}(2021)\citenamefont {Zhang}, \citenamefont {Hu}, \citenamefont {Ma}, \citenamefont {Li}, \citenamefont {Krasnok}, \citenamefont {Hillenbrand}, \citenamefont {Al{\`u}},\ and\ \citenamefont {Qiu}}]{zhang2021interface}%
  \BibitemOpen
  \bibfield  {author} {\bibinfo {author} {\bibfnamefont {Q.}~\bibnamefont {Zhang}}, \bibinfo {author} {\bibfnamefont {G.}~\bibnamefont {Hu}}, \bibinfo {author} {\bibfnamefont {W.}~\bibnamefont {Ma}}, \bibinfo {author} {\bibfnamefont {P.}~\bibnamefont {Li}}, \bibinfo {author} {\bibfnamefont {A.}~\bibnamefont {Krasnok}}, \bibinfo {author} {\bibfnamefont {R.}~\bibnamefont {Hillenbrand}}, \bibinfo {author} {\bibfnamefont {A.}~\bibnamefont {Al{\`u}}},\ and\ \bibinfo {author} {\bibfnamefont {C.-W.}\ \bibnamefont {Qiu}},\ }\bibfield  {title} {\bibinfo {title} {Interface nano-optics with van der waals polaritons},\ }\href {https://doi.org/10.1038/s41586-021-03581-5} {\bibfield  {journal} {\bibinfo  {journal} {Nature}\ }\textbf {\bibinfo {volume} {597}},\ \bibinfo {pages} {187} (\bibinfo {year} {2021})}\BibitemShut {NoStop}%
\bibitem [{\citenamefont {Menabde}\ \emph {et~al.}(2022)\citenamefont {Menabde}, \citenamefont {Heiden}, \citenamefont {Cox}, \citenamefont {Mortensen},\ and\ \citenamefont {Jang}}]{menabde2022image}%
  \BibitemOpen
  \bibfield  {author} {\bibinfo {author} {\bibfnamefont {S.~G.}\ \bibnamefont {Menabde}}, \bibinfo {author} {\bibfnamefont {J.~T.}\ \bibnamefont {Heiden}}, \bibinfo {author} {\bibfnamefont {J.~D.}\ \bibnamefont {Cox}}, \bibinfo {author} {\bibfnamefont {N.~A.}\ \bibnamefont {Mortensen}},\ and\ \bibinfo {author} {\bibfnamefont {M.~S.}\ \bibnamefont {Jang}},\ }\bibfield  {title} {\bibinfo {title} {Image polaritons in van der waals crystals},\ }\href {https://doi.org/10.1515/nanoph-2021-0693} {\bibfield  {journal} {\bibinfo  {journal} {Nanophotonics}\ }\textbf {\bibinfo {volume} {11}},\ \bibinfo {pages} {2433} (\bibinfo {year} {2022})}\BibitemShut {NoStop}%
\bibitem [{\citenamefont {Monticone}\ \emph {et~al.}(2025)\citenamefont {Monticone}, \citenamefont {Mortensen}, \citenamefont {Fern\'{a}ndez-Dom\'{i}nguez}, \citenamefont {Luo}, \citenamefont {Zheng}, \citenamefont {Tserkezis}, \citenamefont {Khurgin}, \citenamefont {Shahbazyan}, \citenamefont {Chaves}, \citenamefont {Peres}, \citenamefont {Wegner}, \citenamefont {Busch}, \citenamefont {Hu}, \citenamefont {Sala}, \citenamefont {Zhang}, \citenamefont {Cirac\`{i}}, \citenamefont {Aizpurua}, \citenamefont {Babaze}, \citenamefont {Borisov}, \citenamefont {Chen}, \citenamefont {Christensen}, \citenamefont {Yan}, \citenamefont {Yang}, \citenamefont {Hohenester}, \citenamefont {Huber}, \citenamefont {Wubs}, \citenamefont {Liberato}, \citenamefont {Gon\c{c}alves}, \citenamefont {{Garc\'{\i}a de Abajo}}, \citenamefont {Hess}, \citenamefont {Tarasenko}, \citenamefont {Cox}, \citenamefont {Jelver}, \citenamefont {Dias}, \citenamefont {S\'{a}nchez}, \citenamefont {Margetis}, \citenamefont {G\'{o}mez-Santos},
  \citenamefont {Vasilevskiy}, \citenamefont {Stauber}, \citenamefont {Tretyakov}, \citenamefont {Simovski}, \citenamefont {Pakniyat}, \citenamefont {G\'{o}mez-D\'{i}az}, \citenamefont {Bondarev}, \citenamefont {Biehs}, \citenamefont {Boltasseva}, \citenamefont {Shalaev}, \citenamefont {Krasavin}, \citenamefont {Zayats}, \citenamefont {Al\`{u}}, \citenamefont {Song}, \citenamefont {Brongersma}, \citenamefont {Levy}, \citenamefont {Long}, \citenamefont {Guo}, \citenamefont {Fan}, \citenamefont {Bozhevolnyi}, \citenamefont {Overvig}, \citenamefont {Prud\^{e}ncio}, \citenamefont {Silveirinha}, \citenamefont {Gangaraj}, \citenamefont {Argyropoulos}, \citenamefont {Huidobro}, \citenamefont {Galiffi}, \citenamefont {Yang}, \citenamefont {Pendry},\ and\ \citenamefont {Miller}}]{monticone2025nonlocality}%
  \BibitemOpen
  \bibfield  {author} {\bibinfo {author} {\bibfnamefont {F.}~\bibnamefont {Monticone}}, \bibinfo {author} {\bibfnamefont {N.~A.}\ \bibnamefont {Mortensen}}, \bibinfo {author} {\bibfnamefont {A.~I.}\ \bibnamefont {Fern\'{a}ndez-Dom\'{i}nguez}}, \bibinfo {author} {\bibfnamefont {Y.}~\bibnamefont {Luo}}, \bibinfo {author} {\bibfnamefont {X.}~\bibnamefont {Zheng}}, \bibinfo {author} {\bibfnamefont {C.}~\bibnamefont {Tserkezis}}, \bibinfo {author} {\bibfnamefont {J.~B.}\ \bibnamefont {Khurgin}}, \bibinfo {author} {\bibfnamefont {T.~V.}\ \bibnamefont {Shahbazyan}}, \bibinfo {author} {\bibfnamefont {A.~J.}\ \bibnamefont {Chaves}}, \bibinfo {author} {\bibfnamefont {N.~M.~R.}\ \bibnamefont {Peres}}, \bibinfo {author} {\bibfnamefont {G.}~\bibnamefont {Wegner}}, \bibinfo {author} {\bibfnamefont {K.}~\bibnamefont {Busch}}, \bibinfo {author} {\bibfnamefont {H.}~\bibnamefont {Hu}}, \bibinfo {author} {\bibfnamefont {F.~D.}\ \bibnamefont {Sala}}, \bibinfo {author} {\bibfnamefont {P.}~\bibnamefont {Zhang}}, \bibinfo {author}
  {\bibfnamefont {C.}~\bibnamefont {Cirac\`{i}}}, \bibinfo {author} {\bibfnamefont {J.}~\bibnamefont {Aizpurua}}, \bibinfo {author} {\bibfnamefont {A.}~\bibnamefont {Babaze}}, \bibinfo {author} {\bibfnamefont {A.~G.}\ \bibnamefont {Borisov}}, \bibinfo {author} {\bibfnamefont {X.-W.}\ \bibnamefont {Chen}}, \bibinfo {author} {\bibfnamefont {T.}~\bibnamefont {Christensen}}, \bibinfo {author} {\bibfnamefont {W.}~\bibnamefont {Yan}}, \bibinfo {author} {\bibfnamefont {Y.}~\bibnamefont {Yang}}, \bibinfo {author} {\bibfnamefont {U.}~\bibnamefont {Hohenester}}, \bibinfo {author} {\bibfnamefont {L.}~\bibnamefont {Huber}}, \bibinfo {author} {\bibfnamefont {M.}~\bibnamefont {Wubs}}, \bibinfo {author} {\bibfnamefont {S.~D.}\ \bibnamefont {Liberato}}, \bibinfo {author} {\bibfnamefont {P.~A.~D.}\ \bibnamefont {Gon\c{c}alves}}, \bibinfo {author} {\bibfnamefont {F.~J.}\ \bibnamefont {{Garc\'{\i}a de Abajo}}}, \bibinfo {author} {\bibfnamefont {O.}~\bibnamefont {Hess}}, \bibinfo {author} {\bibfnamefont {I.}~\bibnamefont
  {Tarasenko}}, \bibinfo {author} {\bibfnamefont {J.~D.}\ \bibnamefont {Cox}}, \bibinfo {author} {\bibfnamefont {L.}~\bibnamefont {Jelver}}, \bibinfo {author} {\bibfnamefont {E.~J.~C.}\ \bibnamefont {Dias}}, \bibinfo {author} {\bibfnamefont {M.~S.}\ \bibnamefont {S\'{a}nchez}}, \bibinfo {author} {\bibfnamefont {D.}~\bibnamefont {Margetis}}, \bibinfo {author} {\bibfnamefont {G.}~\bibnamefont {G\'{o}mez-Santos}}, \bibinfo {author} {\bibfnamefont {I.~M.}\ \bibnamefont {Vasilevskiy}}, \bibinfo {author} {\bibfnamefont {T.}~\bibnamefont {Stauber}}, \bibinfo {author} {\bibfnamefont {S.}~\bibnamefont {Tretyakov}}, \bibinfo {author} {\bibfnamefont {C.}~\bibnamefont {Simovski}}, \bibinfo {author} {\bibfnamefont {S.}~\bibnamefont {Pakniyat}}, \bibinfo {author} {\bibfnamefont {J.~S.}\ \bibnamefont {G\'{o}mez-D\'{i}az}}, \bibinfo {author} {\bibfnamefont {I.~V.}\ \bibnamefont {Bondarev}}, \bibinfo {author} {\bibfnamefont {S.-A.}\ \bibnamefont {Biehs}}, \bibinfo {author} {\bibfnamefont {A.}~\bibnamefont {Boltasseva}},
  \bibinfo {author} {\bibfnamefont {V.~M.}\ \bibnamefont {Shalaev}}, \bibinfo {author} {\bibfnamefont {A.~V.}\ \bibnamefont {Krasavin}}, \bibinfo {author} {\bibfnamefont {A.~V.}\ \bibnamefont {Zayats}}, \bibinfo {author} {\bibfnamefont {A.}~\bibnamefont {Al\`{u}}}, \bibinfo {author} {\bibfnamefont {J.-H.}\ \bibnamefont {Song}}, \bibinfo {author} {\bibfnamefont {M.~L.}\ \bibnamefont {Brongersma}}, \bibinfo {author} {\bibfnamefont {U.}~\bibnamefont {Levy}}, \bibinfo {author} {\bibfnamefont {O.~Y.}\ \bibnamefont {Long}}, \bibinfo {author} {\bibfnamefont {C.}~\bibnamefont {Guo}}, \bibinfo {author} {\bibfnamefont {S.}~\bibnamefont {Fan}}, \bibinfo {author} {\bibfnamefont {S.~I.}\ \bibnamefont {Bozhevolnyi}}, \bibinfo {author} {\bibfnamefont {A.}~\bibnamefont {Overvig}}, \bibinfo {author} {\bibfnamefont {F.~R.}\ \bibnamefont {Prud\^{e}ncio}}, \bibinfo {author} {\bibfnamefont {M.~G.}\ \bibnamefont {Silveirinha}}, \bibinfo {author} {\bibfnamefont {S.~A.~H.}\ \bibnamefont {Gangaraj}}, \bibinfo {author} {\bibfnamefont
  {C.}~\bibnamefont {Argyropoulos}}, \bibinfo {author} {\bibfnamefont {P.~A.}\ \bibnamefont {Huidobro}}, \bibinfo {author} {\bibfnamefont {E.}~\bibnamefont {Galiffi}}, \bibinfo {author} {\bibfnamefont {F.}~\bibnamefont {Yang}}, \bibinfo {author} {\bibfnamefont {J.~B.}\ \bibnamefont {Pendry}},\ and\ \bibinfo {author} {\bibfnamefont {D.~A.~B.}\ \bibnamefont {Miller}},\ }\bibfield  {title} {\bibinfo {title} {Nonlocality in photonic materials and metamaterials: roadmap},\ }\href {https://doi.org/10.1364/OME.559374} {\bibfield  {journal} {\bibinfo  {journal} {Opt.\ Mater.\ Express}\ }\textbf {\bibinfo {volume} {15}},\ \bibinfo {pages} {1544} (\bibinfo {year} {2025})}\BibitemShut {NoStop}%
\bibitem [{\citenamefont {Manzoni}\ \emph {et~al.}(2015)\citenamefont {Manzoni}, \citenamefont {Silveiro}, \citenamefont {{Garc\'{\i}a de Abajo}},\ and\ \citenamefont {Chang}}]{manzoni2015second}%
  \BibitemOpen
  \bibfield  {author} {\bibinfo {author} {\bibfnamefont {M.~T.}\ \bibnamefont {Manzoni}}, \bibinfo {author} {\bibfnamefont {I.}~\bibnamefont {Silveiro}}, \bibinfo {author} {\bibfnamefont {F.~J.}\ \bibnamefont {{Garc\'{\i}a de Abajo}}},\ and\ \bibinfo {author} {\bibfnamefont {D.~E.}\ \bibnamefont {Chang}},\ }\bibfield  {title} {\bibinfo {title} {Second-order quantum nonlinear optical processes in single graphene nanostructures and arrays},\ }\href {https://doi.org/10.1088/1367-2630/17/8/083031} {\bibfield  {journal} {\bibinfo  {journal} {New.\ J.\ Phys.}\ }\textbf {\bibinfo {volume} {17}},\ \bibinfo {pages} {083031} (\bibinfo {year} {2015})}\BibitemShut {NoStop}%
\bibitem [{\citenamefont {Constant}\ \emph {et~al.}(2016)\citenamefont {Constant}, \citenamefont {Hornett}, \citenamefont {Chang},\ and\ \citenamefont {Hendry}}]{constant2016alloptical}%
  \BibitemOpen
  \bibfield  {author} {\bibinfo {author} {\bibfnamefont {T.~J.}\ \bibnamefont {Constant}}, \bibinfo {author} {\bibfnamefont {S.~M.}\ \bibnamefont {Hornett}}, \bibinfo {author} {\bibfnamefont {D.~E.}\ \bibnamefont {Chang}},\ and\ \bibinfo {author} {\bibfnamefont {E.}~\bibnamefont {Hendry}},\ }\bibfield  {title} {\bibinfo {title} {All-optical generation of surface plasmons in graphene},\ }\href {https://doi.org/10.1038/nphys3545} {\bibfield  {journal} {\bibinfo  {journal} {Nat.\ Phys.}\ }\textbf {\bibinfo {volume} {12}},\ \bibinfo {pages} {124} (\bibinfo {year} {2016})}\BibitemShut {NoStop}%
\bibitem [{\citenamefont {Cox}\ and\ \citenamefont {{Garc\'{\i}a de Abajo}}(2019)}]{cox2019nonlinear}%
  \BibitemOpen
  \bibfield  {author} {\bibinfo {author} {\bibfnamefont {J.~D.}\ \bibnamefont {Cox}}\ and\ \bibinfo {author} {\bibfnamefont {F.~J.}\ \bibnamefont {{Garc\'{\i}a de Abajo}}},\ }\bibfield  {title} {\bibinfo {title} {Nonlinear graphene nanoplasmonics},\ }\href {https://doi.org/10.1021/acs.accounts.9b00308} {\bibfield  {journal} {\bibinfo  {journal} {Acc.\ Chem.\ Res.}\ }\textbf {\bibinfo {volume} {52}},\ \bibinfo {pages} {2536} (\bibinfo {year} {2019})}\BibitemShut {NoStop}%
\bibitem [{\citenamefont {Christensen}\ \emph {et~al.}(2012)\citenamefont {Christensen}, \citenamefont {Manjavacas}, \citenamefont {Thongrattanasiri}, \citenamefont {Koppens},\ and\ \citenamefont {{Garc\'{\i}a de Abajo}}}]{christensen2012graphene}%
  \BibitemOpen
  \bibfield  {author} {\bibinfo {author} {\bibfnamefont {J.}~\bibnamefont {Christensen}}, \bibinfo {author} {\bibfnamefont {A.}~\bibnamefont {Manjavacas}}, \bibinfo {author} {\bibfnamefont {S.}~\bibnamefont {Thongrattanasiri}}, \bibinfo {author} {\bibfnamefont {F.~H.~L.}\ \bibnamefont {Koppens}},\ and\ \bibinfo {author} {\bibfnamefont {F.~J.}\ \bibnamefont {{Garc\'{\i}a de Abajo}}},\ }\bibfield  {title} {\bibinfo {title} {Graphene plasmon waveguiding and hybridization in individual and paired nanoribbons},\ }\href {https://doi.org/10.1021/nn2037626} {\bibfield  {journal} {\bibinfo  {journal} {ACS Nano}\ }\textbf {\bibinfo {volume} {6}},\ \bibinfo {pages} {431} (\bibinfo {year} {2012})}\BibitemShut {NoStop}%
\bibitem [{\citenamefont {Silveiro}\ \emph {et~al.}(2015)\citenamefont {Silveiro}, \citenamefont {Ortega},\ and\ \citenamefont {{Garc\'{\i}a de Abajo}}}]{silveiro2015quantum}%
  \BibitemOpen
  \bibfield  {author} {\bibinfo {author} {\bibfnamefont {I.}~\bibnamefont {Silveiro}}, \bibinfo {author} {\bibfnamefont {J.~M.~P.}\ \bibnamefont {Ortega}},\ and\ \bibinfo {author} {\bibfnamefont {F.~J.}\ \bibnamefont {{Garc\'{\i}a de Abajo}}},\ }\bibfield  {title} {\bibinfo {title} {Quantum nonlocal effects in individual and interacting graphene nanoribbons},\ }\href {https://doi.org/10.1038/lsa.2015.14} {\bibfield  {journal} {\bibinfo  {journal} {Light Sci.\ Appl.}\ }\textbf {\bibinfo {volume} {4}},\ \bibinfo {pages} {e241} (\bibinfo {year} {2015})}\BibitemShut {NoStop}%
\bibitem [{\citenamefont {Rodrigo}\ \emph {et~al.}(2017)\citenamefont {Rodrigo}, \citenamefont {Tittl}, \citenamefont {Limaj}, \citenamefont {{Garc\'{\i}a de Abajo}}, \citenamefont {Pruneri},\ and\ \citenamefont {Altug}}]{rodrigo2017double}%
  \BibitemOpen
  \bibfield  {author} {\bibinfo {author} {\bibfnamefont {D.}~\bibnamefont {Rodrigo}}, \bibinfo {author} {\bibfnamefont {A.}~\bibnamefont {Tittl}}, \bibinfo {author} {\bibfnamefont {O.}~\bibnamefont {Limaj}}, \bibinfo {author} {\bibfnamefont {F.~J.}\ \bibnamefont {{Garc\'{\i}a de Abajo}}}, \bibinfo {author} {\bibfnamefont {V.}~\bibnamefont {Pruneri}},\ and\ \bibinfo {author} {\bibfnamefont {H.}~\bibnamefont {Altug}},\ }\bibfield  {title} {\bibinfo {title} {Double-layer graphene for enhanced tunable infrared plasmonics},\ }\href {https://doi.org/10.1038/lsa.2016.277} {\bibfield  {journal} {\bibinfo  {journal} {Light Sci.\ Appl.}\ }\textbf {\bibinfo {volume} {6}},\ \bibinfo {pages} {e16277} (\bibinfo {year} {2017})}\BibitemShut {NoStop}%
\bibitem [{\citenamefont {Rasmussen}\ \emph {et~al.}(2023)\citenamefont {Rasmussen}, \citenamefont {{Rodr{\'\i}guez Echarri}}, \citenamefont {{Garc\'{\i}a de Abajo}},\ and\ \citenamefont {Cox}}]{rasmussen2023nonlocal}%
  \BibitemOpen
  \bibfield  {author} {\bibinfo {author} {\bibfnamefont {T.~P.}\ \bibnamefont {Rasmussen}}, \bibinfo {author} {\bibfnamefont {{\'A}.}~\bibnamefont {{Rodr{\'\i}guez Echarri}}}, \bibinfo {author} {\bibfnamefont {F.~J.}\ \bibnamefont {{Garc\'{\i}a de Abajo}}},\ and\ \bibinfo {author} {\bibfnamefont {J.~D.}\ \bibnamefont {Cox}},\ }\bibfield  {title} {\bibinfo {title} {Nonlocal and cascaded effects in nonlinear graphene nanoplasmonics},\ }\href {https://doi.org/10.1039/D2NR06286K} {\bibfield  {journal} {\bibinfo  {journal} {Nanoscale}\ }\textbf {\bibinfo {volume} {15}},\ \bibinfo {pages} {3150} (\bibinfo {year} {2023})}\BibitemShut {NoStop}%
\bibitem [{\citenamefont {Cox}\ \emph {et~al.}(2017)\citenamefont {Cox}, \citenamefont {Marini},\ and\ \citenamefont {{Garc\'{\i}a de Abajo}}}]{cox2017plasmonassisted}%
  \BibitemOpen
  \bibfield  {author} {\bibinfo {author} {\bibfnamefont {J.~D.}\ \bibnamefont {Cox}}, \bibinfo {author} {\bibfnamefont {A.}~\bibnamefont {Marini}},\ and\ \bibinfo {author} {\bibfnamefont {F.~J.}\ \bibnamefont {{Garc\'{\i}a de Abajo}}},\ }\bibfield  {title} {\bibinfo {title} {Plasmon-assisted high-harmonic generation in graphene},\ }\href {https://doi.org/10.1038/ncomms14380} {\bibfield  {journal} {\bibinfo  {journal} {Nat.\ Commun.}\ }\textbf {\bibinfo {volume} {8}},\ \bibinfo {pages} {14380} (\bibinfo {year} {2017})}\BibitemShut {NoStop}%
\bibitem [{\citenamefont {Xia}\ \emph {et~al.}(2014)\citenamefont {Xia}, \citenamefont {Wang},\ and\ \citenamefont {Jia}}]{xia2014rediscovering}%
  \BibitemOpen
  \bibfield  {author} {\bibinfo {author} {\bibfnamefont {F.}~\bibnamefont {Xia}}, \bibinfo {author} {\bibfnamefont {H.}~\bibnamefont {Wang}},\ and\ \bibinfo {author} {\bibfnamefont {Y.}~\bibnamefont {Jia}},\ }\bibfield  {title} {\bibinfo {title} {Rediscovering black phosphorus as an anisotropic layered material for optoelectronics and electronics},\ }\href {https://doi.org/10.1038/ncomms5458} {\bibfield  {journal} {\bibinfo  {journal} {Nat.\ Commun.}\ }\textbf {\bibinfo {volume} {5}},\ \bibinfo {pages} {4458} (\bibinfo {year} {2014})}\BibitemShut {NoStop}%
\bibitem [{\citenamefont {Qiao}\ \emph {et~al.}(2014)\citenamefont {Qiao}, \citenamefont {Kong}, \citenamefont {Hu}, \citenamefont {Yang},\ and\ \citenamefont {Ji}}]{qiao2014highmobility}%
  \BibitemOpen
  \bibfield  {author} {\bibinfo {author} {\bibfnamefont {J.}~\bibnamefont {Qiao}}, \bibinfo {author} {\bibfnamefont {X.}~\bibnamefont {Kong}}, \bibinfo {author} {\bibfnamefont {Z.-X.}\ \bibnamefont {Hu}}, \bibinfo {author} {\bibfnamefont {F.}~\bibnamefont {Yang}},\ and\ \bibinfo {author} {\bibfnamefont {W.}~\bibnamefont {Ji}},\ }\bibfield  {title} {\bibinfo {title} {High-mobility transport anisotropy and linear dichroism in few-layer black phosphorus},\ }\href {https://doi.org/10.1038/ncomms5475} {\bibfield  {journal} {\bibinfo  {journal} {Nat.\ Commun.}\ }\textbf {\bibinfo {volume} {5}},\ \bibinfo {pages} {4475} (\bibinfo {year} {2014})}\BibitemShut {NoStop}%
\bibitem [{\citenamefont {Yang}\ \emph {et~al.}(2015)\citenamefont {Yang}, \citenamefont {Xu}, \citenamefont {Pei}, \citenamefont {Myint}, \citenamefont {Wang}, \citenamefont {Wang}, \citenamefont {Zhang}, \citenamefont {Yu},\ and\ \citenamefont {Lu}}]{yang2015optical}%
  \BibitemOpen
  \bibfield  {author} {\bibinfo {author} {\bibfnamefont {J.}~\bibnamefont {Yang}}, \bibinfo {author} {\bibfnamefont {R.}~\bibnamefont {Xu}}, \bibinfo {author} {\bibfnamefont {J.}~\bibnamefont {Pei}}, \bibinfo {author} {\bibfnamefont {Y.~W.}\ \bibnamefont {Myint}}, \bibinfo {author} {\bibfnamefont {F.}~\bibnamefont {Wang}}, \bibinfo {author} {\bibfnamefont {Z.}~\bibnamefont {Wang}}, \bibinfo {author} {\bibfnamefont {S.}~\bibnamefont {Zhang}}, \bibinfo {author} {\bibfnamefont {Z.}~\bibnamefont {Yu}},\ and\ \bibinfo {author} {\bibfnamefont {Y.}~\bibnamefont {Lu}},\ }\bibfield  {title} {\bibinfo {title} {Optical tuning of exciton and trion emissions in monolayer phosphorene},\ }\href {https://doi.org/10.1038/lsa.2015.85} {\bibfield  {journal} {\bibinfo  {journal} {Light Sci.\ Appl.}\ }\textbf {\bibinfo {volume} {4}},\ \bibinfo {pages} {e312} (\bibinfo {year} {2015})}\BibitemShut {NoStop}%
\bibitem [{\citenamefont {Nourbakhsh}\ and\ \citenamefont {Asgari}(2016)}]{nourbakhsh2016excitons}%
  \BibitemOpen
  \bibfield  {author} {\bibinfo {author} {\bibfnamefont {Z.}~\bibnamefont {Nourbakhsh}}\ and\ \bibinfo {author} {\bibfnamefont {R.}~\bibnamefont {Asgari}},\ }\bibfield  {title} {\bibinfo {title} {Excitons and optical spectra of phosphorene nanoribbons},\ }\href {https://doi.org/10.1103/PhysRevB.94.035437} {\bibfield  {journal} {\bibinfo  {journal} {Phys.\ Rev.\ B}\ }\textbf {\bibinfo {volume} {94}},\ \bibinfo {pages} {035437} (\bibinfo {year} {2016})}\BibitemShut {NoStop}%
\bibitem [{\citenamefont {Carvalho}\ \emph {et~al.}(2016)\citenamefont {Carvalho}, \citenamefont {Wang}, \citenamefont {Zhu}, \citenamefont {Rodin}, \citenamefont {Su},\ and\ \citenamefont {Neto}}]{carvalho2016phosphorene}%
  \BibitemOpen
  \bibfield  {author} {\bibinfo {author} {\bibfnamefont {A.}~\bibnamefont {Carvalho}}, \bibinfo {author} {\bibfnamefont {M.}~\bibnamefont {Wang}}, \bibinfo {author} {\bibfnamefont {X.}~\bibnamefont {Zhu}}, \bibinfo {author} {\bibfnamefont {A.~S.}\ \bibnamefont {Rodin}}, \bibinfo {author} {\bibfnamefont {H.}~\bibnamefont {Su}},\ and\ \bibinfo {author} {\bibfnamefont {A.~H.~C.}\ \bibnamefont {Neto}},\ }\bibfield  {title} {\bibinfo {title} {Phosphorene: from theory to applications},\ }\href {https://doi.org/10.1038/natrevmats.2016.61} {\bibfield  {journal} {\bibinfo  {journal} {Nat.\ Rev.\ Mater.}\ }\textbf {\bibinfo {volume} {1}},\ \bibinfo {pages} {1} (\bibinfo {year} {2016})}\BibitemShut {NoStop}%
\bibitem [{\citenamefont {Neupane}\ \emph {et~al.}(2022)\citenamefont {Neupane}, \citenamefont {Tang}, \citenamefont {Nepal},\ and\ \citenamefont {Ruzsinszky}}]{neupane2022bending}%
  \BibitemOpen
  \bibfield  {author} {\bibinfo {author} {\bibfnamefont {B.}~\bibnamefont {Neupane}}, \bibinfo {author} {\bibfnamefont {H.}~\bibnamefont {Tang}}, \bibinfo {author} {\bibfnamefont {N.~K.}\ \bibnamefont {Nepal}},\ and\ \bibinfo {author} {\bibfnamefont {A.}~\bibnamefont {Ruzsinszky}},\ }\bibfield  {title} {\bibinfo {title} {Bending as a control knob for the electronic and optical properties of phosphorene nanoribbons},\ }\href {https://doi.org/10.1103/PhysRevMaterials.6.014010} {\bibfield  {journal} {\bibinfo  {journal} {Phys.\ Rev.\ Mater.}\ }\textbf {\bibinfo {volume} {6}},\ \bibinfo {pages} {014010} (\bibinfo {year} {2022})}\BibitemShut {NoStop}%
\bibitem [{\citenamefont {Chaves}\ \emph {et~al.}(2025)\citenamefont {Chaves}, \citenamefont {Jelver}, \citenamefont {da~Costa}, \citenamefont {Cox}, \citenamefont {Mortensen},\ and\ \citenamefont {Peres}}]{chaves2025nonlocal}%
  \BibitemOpen
  \bibfield  {author} {\bibinfo {author} {\bibfnamefont {A.~J.}\ \bibnamefont {Chaves}}, \bibinfo {author} {\bibfnamefont {L.}~\bibnamefont {Jelver}}, \bibinfo {author} {\bibfnamefont {D.~R.}\ \bibnamefont {da~Costa}}, \bibinfo {author} {\bibfnamefont {J.~D.}\ \bibnamefont {Cox}}, \bibinfo {author} {\bibfnamefont {N.~A.}\ \bibnamefont {Mortensen}},\ and\ \bibinfo {author} {\bibfnamefont {N.~M.~R.}\ \bibnamefont {Peres}},\ }\bibfield  {title} {\bibinfo {title} {Nonlocal electrodynamics of two-dimensional anisotropic magneto-plasmons},\ }\href {https://doi.org/10.1515/nanoph-2025-0233} {\bibfield  {journal} {\bibinfo  {journal} {Nanophotonics}\ }\textbf {\bibinfo {volume} {14}},\ \bibinfo {pages} {4495} (\bibinfo {year} {2025})}\BibitemShut {NoStop}%
\bibitem [{\citenamefont {Cox}\ \emph {et~al.}(2016)\citenamefont {Cox}, \citenamefont {Silveiro},\ and\ \citenamefont {{Garc\'{\i}a de Abajo}}}]{cox2016quantum}%
  \BibitemOpen
  \bibfield  {author} {\bibinfo {author} {\bibfnamefont {J.~D.}\ \bibnamefont {Cox}}, \bibinfo {author} {\bibfnamefont {I.}~\bibnamefont {Silveiro}},\ and\ \bibinfo {author} {\bibfnamefont {F.~J.}\ \bibnamefont {{Garc\'{\i}a de Abajo}}},\ }\bibfield  {title} {\bibinfo {title} {Quantum effects in the nonlinear response of graphene plasmons},\ }\href {https://doi.org/10.1021/acsnano.5b06110} {\bibfield  {journal} {\bibinfo  {journal} {ACS Nano}\ }\textbf {\bibinfo {volume} {10}},\ \bibinfo {pages} {1995} (\bibinfo {year} {2016})}\BibitemShut {NoStop}%
\bibitem [{\citenamefont {Jelver}\ and\ \citenamefont {Cox}(2023)}]{jelver2023nonlinear}%
  \BibitemOpen
  \bibfield  {author} {\bibinfo {author} {\bibfnamefont {L.}~\bibnamefont {Jelver}}\ and\ \bibinfo {author} {\bibfnamefont {J.~D.}\ \bibnamefont {Cox}},\ }\bibfield  {title} {\bibinfo {title} {Nonlinear plasmonics in nanostructured phosphorene},\ }\href {https://doi.org/10.1021/acsnano.3c05363} {\bibfield  {journal} {\bibinfo  {journal} {ACS Nano}\ }\textbf {\bibinfo {volume} {17}},\ \bibinfo {pages} {20043} (\bibinfo {year} {2023})}\BibitemShut {NoStop}%
\bibitem [{\citenamefont {Jelver}\ and\ \citenamefont {Cox}(2024)}]{jelver2024nonlinear}%
  \BibitemOpen
  \bibfield  {author} {\bibinfo {author} {\bibfnamefont {L.}~\bibnamefont {Jelver}}\ and\ \bibinfo {author} {\bibfnamefont {J.~D.}\ \bibnamefont {Cox}},\ }\bibfield  {title} {\bibinfo {title} {Nonlinear thermoplasmonics in graphene nanostructures},\ }\href {https://doi.org/10.1021/acs.nanolett.4c04048} {\bibfield  {journal} {\bibinfo  {journal} {Nano Lett.}\ }\textbf {\bibinfo {volume} {24}},\ \bibinfo {pages} {13775} (\bibinfo {year} {2024})}\BibitemShut {NoStop}%
\bibitem [{\citenamefont {Boyd}(2020)}]{boyd2020nonlinear}%
  \BibitemOpen
  \bibfield  {author} {\bibinfo {author} {\bibfnamefont {R.~W.}\ \bibnamefont {Boyd}},\ }\href@noop {} {\emph {\bibinfo {title} {Nonlinear optics}}}\ (\bibinfo  {publisher} {Academic press},\ \bibinfo {year} {2020})\BibitemShut {NoStop}%
\bibitem [{\citenamefont {del Coso}\ and\ \citenamefont {Solis}(2004)}]{delcoso2004relation}%
  \BibitemOpen
  \bibfield  {author} {\bibinfo {author} {\bibfnamefont {R.}~\bibnamefont {del Coso}}\ and\ \bibinfo {author} {\bibfnamefont {J.}~\bibnamefont {Solis}},\ }\bibfield  {title} {\bibinfo {title} {Relation between nonlinear refractive index and third-order susceptibility in absorbing media},\ }\href {https://doi.org/10.1364/JOSAB.21.000640} {\bibfield  {journal} {\bibinfo  {journal} {J.\ Opt.\ Soc.\ Am.\ B}\ }\textbf {\bibinfo {volume} {21}},\ \bibinfo {pages} {640} (\bibinfo {year} {2004})}\BibitemShut {NoStop}%
\bibitem [{\citenamefont {Watts}\ \emph {et~al.}(2019)\citenamefont {Watts}, \citenamefont {Picco}, \citenamefont {Russell-Pavier}, \citenamefont {Cullen}, \citenamefont {Miller}, \citenamefont {Bartu{\'s}}, \citenamefont {Payton}, \citenamefont {Skipper}, \citenamefont {Tileli},\ and\ \citenamefont {Howard}}]{watts2019production}%
  \BibitemOpen
  \bibfield  {author} {\bibinfo {author} {\bibfnamefont {M.~C.}\ \bibnamefont {Watts}}, \bibinfo {author} {\bibfnamefont {L.}~\bibnamefont {Picco}}, \bibinfo {author} {\bibfnamefont {F.~S.}\ \bibnamefont {Russell-Pavier}}, \bibinfo {author} {\bibfnamefont {P.~L.}\ \bibnamefont {Cullen}}, \bibinfo {author} {\bibfnamefont {T.~S.}\ \bibnamefont {Miller}}, \bibinfo {author} {\bibfnamefont {S.~P.}\ \bibnamefont {Bartu{\'s}}}, \bibinfo {author} {\bibfnamefont {O.~D.}\ \bibnamefont {Payton}}, \bibinfo {author} {\bibfnamefont {N.~T.}\ \bibnamefont {Skipper}}, \bibinfo {author} {\bibfnamefont {V.}~\bibnamefont {Tileli}},\ and\ \bibinfo {author} {\bibfnamefont {C.~A.}\ \bibnamefont {Howard}},\ }\bibfield  {title} {\bibinfo {title} {Production of phosphorene nanoribbons},\ }\href {https://doi.org/10.1038/s41586-019-1074-x} {\bibfield  {journal} {\bibinfo  {journal} {Nature}\ }\textbf {\bibinfo {volume} {568}},\ \bibinfo {pages} {216} (\bibinfo {year} {2019})}\BibitemShut {NoStop}%
\bibitem [{\citenamefont {Zhao}\ \emph {et~al.}(2020)\citenamefont {Zhao}, \citenamefont {Barin}, \citenamefont {Cao}, \citenamefont {Overbeck}, \citenamefont {Darawish}, \citenamefont {Lyu}, \citenamefont {Drapcho}, \citenamefont {Wang}, \citenamefont {Dumslaff}, \citenamefont {Narita}, \citenamefont {Calame}, \citenamefont {M\"ullen}, \citenamefont {Louie}, \citenamefont {Ruffieux}, \citenamefont {Fasel},\ and\ \citenamefont {Wang}}]{zhao2020optical}%
  \BibitemOpen
  \bibfield  {author} {\bibinfo {author} {\bibfnamefont {S.}~\bibnamefont {Zhao}}, \bibinfo {author} {\bibfnamefont {G.~B.}\ \bibnamefont {Barin}}, \bibinfo {author} {\bibfnamefont {T.}~\bibnamefont {Cao}}, \bibinfo {author} {\bibfnamefont {J.}~\bibnamefont {Overbeck}}, \bibinfo {author} {\bibfnamefont {R.}~\bibnamefont {Darawish}}, \bibinfo {author} {\bibfnamefont {T.}~\bibnamefont {Lyu}}, \bibinfo {author} {\bibfnamefont {S.}~\bibnamefont {Drapcho}}, \bibinfo {author} {\bibfnamefont {S.}~\bibnamefont {Wang}}, \bibinfo {author} {\bibfnamefont {T.}~\bibnamefont {Dumslaff}}, \bibinfo {author} {\bibfnamefont {A.}~\bibnamefont {Narita}}, \bibinfo {author} {\bibfnamefont {M.}~\bibnamefont {Calame}}, \bibinfo {author} {\bibfnamefont {K.}~\bibnamefont {M\"ullen}}, \bibinfo {author} {\bibfnamefont {S.~G.}\ \bibnamefont {Louie}}, \bibinfo {author} {\bibfnamefont {P.}~\bibnamefont {Ruffieux}}, \bibinfo {author} {\bibfnamefont {R.}~\bibnamefont {Fasel}},\ and\ \bibinfo {author} {\bibfnamefont {F.}~\bibnamefont
  {Wang}},\ }\bibfield  {title} {\bibinfo {title} {Optical imaging and spectroscopy of atomically precise armchair graphene nanoribbons},\ }\href {https://doi.org/10.1021/acs.nanolett.9b04497} {\bibfield  {journal} {\bibinfo  {journal} {Nano Lett.}\ }\textbf {\bibinfo {volume} {20}},\ \bibinfo {pages} {1124} (\bibinfo {year} {2020})}\BibitemShut {NoStop}%
\bibitem [{\citenamefont {Lyu}\ \emph {et~al.}(2024)\citenamefont {Lyu}, \citenamefont {Chen}, \citenamefont {Wang}, \citenamefont {Lou}, \citenamefont {Shen}, \citenamefont {Xie}, \citenamefont {Qiu}, \citenamefont {Mitchell}, \citenamefont {Li}, \citenamefont {Hu}, \citenamefont {Zhou}, \citenamefont {Watanabe}, \citenamefont {Taniguchi}, \citenamefont {Wang}, \citenamefont {Jia}, \citenamefont {Liang}, \citenamefont {Chen}, \citenamefont {Li}, \citenamefont {Wang}, \citenamefont {Ouyang}, \citenamefont {Hod}, \citenamefont {Ding}, \citenamefont {Urbakh},\ and\ \citenamefont {Shi}}]{lyu2024graphene}%
  \BibitemOpen
  \bibfield  {author} {\bibinfo {author} {\bibfnamefont {B.}~\bibnamefont {Lyu}}, \bibinfo {author} {\bibfnamefont {J.}~\bibnamefont {Chen}}, \bibinfo {author} {\bibfnamefont {S.}~\bibnamefont {Wang}}, \bibinfo {author} {\bibfnamefont {S.}~\bibnamefont {Lou}}, \bibinfo {author} {\bibfnamefont {P.}~\bibnamefont {Shen}}, \bibinfo {author} {\bibfnamefont {J.}~\bibnamefont {Xie}}, \bibinfo {author} {\bibfnamefont {L.}~\bibnamefont {Qiu}}, \bibinfo {author} {\bibfnamefont {I.}~\bibnamefont {Mitchell}}, \bibinfo {author} {\bibfnamefont {C.}~\bibnamefont {Li}}, \bibinfo {author} {\bibfnamefont {C.}~\bibnamefont {Hu}}, \bibinfo {author} {\bibfnamefont {X.}~\bibnamefont {Zhou}}, \bibinfo {author} {\bibfnamefont {K.}~\bibnamefont {Watanabe}}, \bibinfo {author} {\bibfnamefont {T.}~\bibnamefont {Taniguchi}}, \bibinfo {author} {\bibfnamefont {X.}~\bibnamefont {Wang}}, \bibinfo {author} {\bibfnamefont {J.}~\bibnamefont {Jia}}, \bibinfo {author} {\bibfnamefont {Q.}~\bibnamefont {Liang}}, \bibinfo {author} {\bibfnamefont
  {G.}~\bibnamefont {Chen}}, \bibinfo {author} {\bibfnamefont {T.}~\bibnamefont {Li}}, \bibinfo {author} {\bibfnamefont {S.}~\bibnamefont {Wang}}, \bibinfo {author} {\bibfnamefont {W.}~\bibnamefont {Ouyang}}, \bibinfo {author} {\bibfnamefont {O.}~\bibnamefont {Hod}}, \bibinfo {author} {\bibfnamefont {F.}~\bibnamefont {Ding}}, \bibinfo {author} {\bibfnamefont {M.}~\bibnamefont {Urbakh}},\ and\ \bibinfo {author} {\bibfnamefont {Z.}~\bibnamefont {Shi}},\ }\bibfield  {title} {\bibinfo {title} {Graphene nanoribbons grown in hbn stacks for high-performance electronics},\ }\href {https://doi.org/10.1038/s41586-024-07243-0} {\bibfield  {journal} {\bibinfo  {journal} {Nature}\ }\textbf {\bibinfo {volume} {628}},\ \bibinfo {pages} {758} (\bibinfo {year} {2024})}\BibitemShut {NoStop}%
\bibitem [{\citenamefont {Perdew}\ \emph {et~al.}(1996)\citenamefont {Perdew}, \citenamefont {Burke},\ and\ \citenamefont {Ernzerhof}}]{perdew1996generalized}%
  \BibitemOpen
  \bibfield  {author} {\bibinfo {author} {\bibfnamefont {J.~P.}\ \bibnamefont {Perdew}}, \bibinfo {author} {\bibfnamefont {K.}~\bibnamefont {Burke}},\ and\ \bibinfo {author} {\bibfnamefont {M.}~\bibnamefont {Ernzerhof}},\ }\bibfield  {title} {\bibinfo {title} {Generalized gradient approximation made simple},\ }\href {https://doi.org/10.1103/PhysRevLett.77.3865} {\bibfield  {journal} {\bibinfo  {journal} {Phys.\ Rev.\ Lett.}\ }\textbf {\bibinfo {volume} {77}},\ \bibinfo {pages} {3865} (\bibinfo {year} {1996})}\BibitemShut {NoStop}%
\bibitem [{\citenamefont {Giannozzi}\ \emph {et~al.}(2009)\citenamefont {Giannozzi}, \citenamefont {Baroni}, \citenamefont {Bonini}, \citenamefont {Calandra}, \citenamefont {Car}, \citenamefont {Cavazzoni}, \citenamefont {Ceresoli}, \citenamefont {Chiarotti}, \citenamefont {Cococcioni}, \citenamefont {Dabo}, \citenamefont {Corso1}, \citenamefont {de~Gironcoli}, \citenamefont {Fabris}, \citenamefont {Fratesi}, \citenamefont {Gebauer}, \citenamefont {Gerstmann}, \citenamefont {Gougoussis}, \citenamefont {Kokalj}, \citenamefont {Lazzeri}, \citenamefont {Martin-Samos}, \citenamefont {Marzari}, \citenamefont {Mauri}, \citenamefont {Mazzarello}, \citenamefont {Paolini}, \citenamefont {Pasquarello}, \citenamefont {Paulatto}, \citenamefont {Sbraccia}, \citenamefont {Scandolo}, \citenamefont {Sclauzero}, \citenamefont {Seitsonen}, \citenamefont {Smogunov}, \citenamefont {Umari},\ and\ \citenamefont {Wentzcovitch}}]{giannozzi2009quantum}%
  \BibitemOpen
  \bibfield  {author} {\bibinfo {author} {\bibfnamefont {P.}~\bibnamefont {Giannozzi}}, \bibinfo {author} {\bibfnamefont {S.}~\bibnamefont {Baroni}}, \bibinfo {author} {\bibfnamefont {N.}~\bibnamefont {Bonini}}, \bibinfo {author} {\bibfnamefont {M.}~\bibnamefont {Calandra}}, \bibinfo {author} {\bibfnamefont {R.}~\bibnamefont {Car}}, \bibinfo {author} {\bibfnamefont {C.}~\bibnamefont {Cavazzoni}}, \bibinfo {author} {\bibfnamefont {D.}~\bibnamefont {Ceresoli}}, \bibinfo {author} {\bibfnamefont {G.~L.}\ \bibnamefont {Chiarotti}}, \bibinfo {author} {\bibfnamefont {M.}~\bibnamefont {Cococcioni}}, \bibinfo {author} {\bibfnamefont {I.}~\bibnamefont {Dabo}}, \bibinfo {author} {\bibfnamefont {A.~D.}\ \bibnamefont {Corso1}}, \bibinfo {author} {\bibfnamefont {S.}~\bibnamefont {de~Gironcoli}}, \bibinfo {author} {\bibfnamefont {S.}~\bibnamefont {Fabris}}, \bibinfo {author} {\bibfnamefont {G.}~\bibnamefont {Fratesi}}, \bibinfo {author} {\bibfnamefont {R.}~\bibnamefont {Gebauer}}, \bibinfo {author} {\bibfnamefont
  {U.}~\bibnamefont {Gerstmann}}, \bibinfo {author} {\bibfnamefont {C.}~\bibnamefont {Gougoussis}}, \bibinfo {author} {\bibfnamefont {A.}~\bibnamefont {Kokalj}}, \bibinfo {author} {\bibfnamefont {M.}~\bibnamefont {Lazzeri}}, \bibinfo {author} {\bibfnamefont {L.}~\bibnamefont {Martin-Samos}}, \bibinfo {author} {\bibfnamefont {N.}~\bibnamefont {Marzari}}, \bibinfo {author} {\bibfnamefont {F.}~\bibnamefont {Mauri}}, \bibinfo {author} {\bibfnamefont {R.}~\bibnamefont {Mazzarello}}, \bibinfo {author} {\bibfnamefont {S.}~\bibnamefont {Paolini}}, \bibinfo {author} {\bibfnamefont {A.}~\bibnamefont {Pasquarello}}, \bibinfo {author} {\bibfnamefont {L.}~\bibnamefont {Paulatto}}, \bibinfo {author} {\bibfnamefont {C.}~\bibnamefont {Sbraccia}}, \bibinfo {author} {\bibfnamefont {S.}~\bibnamefont {Scandolo}}, \bibinfo {author} {\bibfnamefont {G.}~\bibnamefont {Sclauzero}}, \bibinfo {author} {\bibfnamefont {A.~P.}\ \bibnamefont {Seitsonen}}, \bibinfo {author} {\bibfnamefont {A.}~\bibnamefont {Smogunov}}, \bibinfo {author}
  {\bibfnamefont {P.}~\bibnamefont {Umari}},\ and\ \bibinfo {author} {\bibfnamefont {R.~M.}\ \bibnamefont {Wentzcovitch}},\ }\bibfield  {title} {\bibinfo {title} {Quantum espresso: a modular and open-source software project for quantum simulations of materials},\ }\href {https://doi.org/10.1088/0953-8984/21/39/395502} {\bibfield  {journal} {\bibinfo  {journal} {J.\ Phys.\ Condens.\ Matter}\ }\textbf {\bibinfo {volume} {21}},\ \bibinfo {pages} {395502} (\bibinfo {year} {2009})}\BibitemShut {NoStop}%
\bibitem [{\citenamefont {Giannozzi}\ \emph {et~al.}(2017)\citenamefont {Giannozzi}, \citenamefont {Andreussi}, \citenamefont {Brumme}, \citenamefont {Bunau}, \citenamefont {Nardelli}, \citenamefont {Calandra}, \citenamefont {Car}, \citenamefont {Cavazzoni}, \citenamefont {Ceresoli}, \citenamefont {Cococcioni}, \citenamefont {Colonna}, \citenamefont {Carnimeo}, \citenamefont {Corso}, \citenamefont {de~Gironcoli}, \citenamefont {Delugas}, \citenamefont {DiStasio}, \citenamefont {Ferretti}, \citenamefont {Floris}, \citenamefont {Fratesi}, \citenamefont {Fugallo}, \citenamefont {Gebauer}, \citenamefont {Gerstmann}, \citenamefont {Giustino}, \citenamefont {Gorni}, \citenamefont {Jia}, \citenamefont {Kawamura}, \citenamefont {Ko}, \citenamefont {Kokalj}, \citenamefont {K{\"u}{\c{c}}{\"u}kbenli}, \citenamefont {Lazzeri}, \citenamefont {Marsili}, \citenamefont {Marzari}, \citenamefont {Mauri}, \citenamefont {Nguyen}, \citenamefont {Nguyen}, \citenamefont {de-la Roza}, \citenamefont {Paulatto}, \citenamefont
  {Ponc{\'e}}, \citenamefont {Rocca}, \citenamefont {Sabatini}, \citenamefont {Santra}, \citenamefont {Schlipf}, \citenamefont {Seitsonen}, \citenamefont {Smogunov}, \citenamefont {Timrov}, \citenamefont {Thonhauser}, \citenamefont {Umari}, \citenamefont {Vast}, \citenamefont {Wu},\ and\ \citenamefont {Baroni}}]{giannozzi2017advanced}%
  \BibitemOpen
  \bibfield  {author} {\bibinfo {author} {\bibfnamefont {P.}~\bibnamefont {Giannozzi}}, \bibinfo {author} {\bibfnamefont {O.}~\bibnamefont {Andreussi}}, \bibinfo {author} {\bibfnamefont {T.}~\bibnamefont {Brumme}}, \bibinfo {author} {\bibfnamefont {O.}~\bibnamefont {Bunau}}, \bibinfo {author} {\bibfnamefont {M.~B.}\ \bibnamefont {Nardelli}}, \bibinfo {author} {\bibfnamefont {M.}~\bibnamefont {Calandra}}, \bibinfo {author} {\bibfnamefont {R.}~\bibnamefont {Car}}, \bibinfo {author} {\bibfnamefont {C.}~\bibnamefont {Cavazzoni}}, \bibinfo {author} {\bibfnamefont {D.}~\bibnamefont {Ceresoli}}, \bibinfo {author} {\bibfnamefont {M.}~\bibnamefont {Cococcioni}}, \bibinfo {author} {\bibfnamefont {N.}~\bibnamefont {Colonna}}, \bibinfo {author} {\bibfnamefont {I.}~\bibnamefont {Carnimeo}}, \bibinfo {author} {\bibfnamefont {A.~D.}\ \bibnamefont {Corso}}, \bibinfo {author} {\bibfnamefont {S.}~\bibnamefont {de~Gironcoli}}, \bibinfo {author} {\bibfnamefont {P.}~\bibnamefont {Delugas}}, \bibinfo {author} {\bibfnamefont
  {R.~A.}\ \bibnamefont {DiStasio}}, \bibinfo {author} {\bibfnamefont {A.}~\bibnamefont {Ferretti}}, \bibinfo {author} {\bibfnamefont {A.}~\bibnamefont {Floris}}, \bibinfo {author} {\bibfnamefont {G.}~\bibnamefont {Fratesi}}, \bibinfo {author} {\bibfnamefont {G.}~\bibnamefont {Fugallo}}, \bibinfo {author} {\bibfnamefont {R.}~\bibnamefont {Gebauer}}, \bibinfo {author} {\bibfnamefont {U.}~\bibnamefont {Gerstmann}}, \bibinfo {author} {\bibfnamefont {F.}~\bibnamefont {Giustino}}, \bibinfo {author} {\bibfnamefont {T.}~\bibnamefont {Gorni}}, \bibinfo {author} {\bibfnamefont {J.}~\bibnamefont {Jia}}, \bibinfo {author} {\bibfnamefont {M.}~\bibnamefont {Kawamura}}, \bibinfo {author} {\bibfnamefont {H.-Y.}\ \bibnamefont {Ko}}, \bibinfo {author} {\bibfnamefont {A.}~\bibnamefont {Kokalj}}, \bibinfo {author} {\bibfnamefont {E.}~\bibnamefont {K{\"u}{\c{c}}{\"u}kbenli}}, \bibinfo {author} {\bibfnamefont {M.}~\bibnamefont {Lazzeri}}, \bibinfo {author} {\bibfnamefont {M.}~\bibnamefont {Marsili}}, \bibinfo {author}
  {\bibfnamefont {N.}~\bibnamefont {Marzari}}, \bibinfo {author} {\bibfnamefont {F.}~\bibnamefont {Mauri}}, \bibinfo {author} {\bibfnamefont {N.~L.}\ \bibnamefont {Nguyen}}, \bibinfo {author} {\bibfnamefont {H.-V.}\ \bibnamefont {Nguyen}}, \bibinfo {author} {\bibfnamefont {A.~O.}\ \bibnamefont {de-la Roza}}, \bibinfo {author} {\bibfnamefont {L.}~\bibnamefont {Paulatto}}, \bibinfo {author} {\bibfnamefont {S.}~\bibnamefont {Ponc{\'e}}}, \bibinfo {author} {\bibfnamefont {D.}~\bibnamefont {Rocca}}, \bibinfo {author} {\bibfnamefont {R.}~\bibnamefont {Sabatini}}, \bibinfo {author} {\bibfnamefont {B.}~\bibnamefont {Santra}}, \bibinfo {author} {\bibfnamefont {M.}~\bibnamefont {Schlipf}}, \bibinfo {author} {\bibfnamefont {A.~P.}\ \bibnamefont {Seitsonen}}, \bibinfo {author} {\bibfnamefont {A.}~\bibnamefont {Smogunov}}, \bibinfo {author} {\bibfnamefont {I.}~\bibnamefont {Timrov}}, \bibinfo {author} {\bibfnamefont {T.}~\bibnamefont {Thonhauser}}, \bibinfo {author} {\bibfnamefont {P.}~\bibnamefont {Umari}}, \bibinfo
  {author} {\bibfnamefont {N.}~\bibnamefont {Vast}}, \bibinfo {author} {\bibfnamefont {X.}~\bibnamefont {Wu}},\ and\ \bibinfo {author} {\bibfnamefont {S.}~\bibnamefont {Baroni}},\ }\bibfield  {title} {\bibinfo {title} {Advanced capabilities for materials modelling with quantum espresso},\ }\href {https://doi.org/10.1088/1361-648X/aa8f79} {\bibfield  {journal} {\bibinfo  {journal} {J.\ Phys.\ Condens.\ Matter}\ }\textbf {\bibinfo {volume} {29}},\ \bibinfo {pages} {465901} (\bibinfo {year} {2017})}\BibitemShut {NoStop}%
\bibitem [{\citenamefont {van Setten}\ \emph {et~al.}(2018)\citenamefont {van Setten}, \citenamefont {Giantomassi}, \citenamefont {Bousquet}, \citenamefont {Verstraete}, \citenamefont {Hamann}, \citenamefont {Gonze},\ and\ \citenamefont {Rignanese}}]{vansetten2018pseudodojo}%
  \BibitemOpen
  \bibfield  {author} {\bibinfo {author} {\bibfnamefont {M.~J.}\ \bibnamefont {van Setten}}, \bibinfo {author} {\bibfnamefont {M.}~\bibnamefont {Giantomassi}}, \bibinfo {author} {\bibfnamefont {E.}~\bibnamefont {Bousquet}}, \bibinfo {author} {\bibfnamefont {M.~J.}\ \bibnamefont {Verstraete}}, \bibinfo {author} {\bibfnamefont {D.~R.}\ \bibnamefont {Hamann}}, \bibinfo {author} {\bibfnamefont {X.}~\bibnamefont {Gonze}},\ and\ \bibinfo {author} {\bibfnamefont {G.-M.}\ \bibnamefont {Rignanese}},\ }\bibfield  {title} {\bibinfo {title} {The pseudodojo: Training and grading a 85 element optimized norm-conserving pseudopotential table},\ }\href {https://doi.org/10.1016/j.cpc.2018.01.012} {\bibfield  {journal} {\bibinfo  {journal} {Comput.\ Phys.\ Commun.}\ }\textbf {\bibinfo {volume} {226}},\ \bibinfo {pages} {39} (\bibinfo {year} {2018})}\BibitemShut {NoStop}%
\bibitem [{\citenamefont {Pizzi}\ \emph {et~al.}(2020)\citenamefont {Pizzi}, \citenamefont {Vitale}, \citenamefont {Arita}, \citenamefont {Bl{\"u}gel}, \citenamefont {Freimuth}, \citenamefont {G{\'e}ranton}, \citenamefont {Gibertini}, \citenamefont {Gresch}, \citenamefont {Johnson}, \citenamefont {Koretsune}, \citenamefont {Iba{\~{n}}ez-Azpiroz}, \citenamefont {Lee}, \citenamefont {Lihm}, \citenamefont {Marchand}, \citenamefont {Marrazzo}, \citenamefont {Mokrousov}, \citenamefont {Mustafa}, \citenamefont {Nohara}, \citenamefont {Nomura}, \citenamefont {Paulatto}, \citenamefont {Ponc{\'{e}}}, \citenamefont {Ponweiser}, \citenamefont {Qiao}, \citenamefont {Thöle}, \citenamefont {Tsirkin}, \citenamefont {Wierzbowska}, \citenamefont {Marzari}, \citenamefont {Vanderbilt}, \citenamefont {Souza}, \citenamefont {Mostofi},\ and\ \citenamefont {Yates}}]{pizzi2020wannier90}%
  \BibitemOpen
  \bibfield  {author} {\bibinfo {author} {\bibfnamefont {G.}~\bibnamefont {Pizzi}}, \bibinfo {author} {\bibfnamefont {V.}~\bibnamefont {Vitale}}, \bibinfo {author} {\bibfnamefont {R.}~\bibnamefont {Arita}}, \bibinfo {author} {\bibfnamefont {S.}~\bibnamefont {Bl{\"u}gel}}, \bibinfo {author} {\bibfnamefont {F.}~\bibnamefont {Freimuth}}, \bibinfo {author} {\bibfnamefont {G.}~\bibnamefont {G{\'e}ranton}}, \bibinfo {author} {\bibfnamefont {M.}~\bibnamefont {Gibertini}}, \bibinfo {author} {\bibfnamefont {D.}~\bibnamefont {Gresch}}, \bibinfo {author} {\bibfnamefont {C.}~\bibnamefont {Johnson}}, \bibinfo {author} {\bibfnamefont {T.}~\bibnamefont {Koretsune}}, \bibinfo {author} {\bibfnamefont {J.}~\bibnamefont {Iba{\~{n}}ez-Azpiroz}}, \bibinfo {author} {\bibfnamefont {H.}~\bibnamefont {Lee}}, \bibinfo {author} {\bibfnamefont {J.-M.}\ \bibnamefont {Lihm}}, \bibinfo {author} {\bibfnamefont {D.}~\bibnamefont {Marchand}}, \bibinfo {author} {\bibfnamefont {A.}~\bibnamefont {Marrazzo}}, \bibinfo {author} {\bibfnamefont
  {Y.}~\bibnamefont {Mokrousov}}, \bibinfo {author} {\bibfnamefont {J.~I.}\ \bibnamefont {Mustafa}}, \bibinfo {author} {\bibfnamefont {Y.}~\bibnamefont {Nohara}}, \bibinfo {author} {\bibfnamefont {Y.}~\bibnamefont {Nomura}}, \bibinfo {author} {\bibfnamefont {L.}~\bibnamefont {Paulatto}}, \bibinfo {author} {\bibfnamefont {S.}~\bibnamefont {Ponc{\'{e}}}}, \bibinfo {author} {\bibfnamefont {T.}~\bibnamefont {Ponweiser}}, \bibinfo {author} {\bibfnamefont {J.}~\bibnamefont {Qiao}}, \bibinfo {author} {\bibfnamefont {F.}~\bibnamefont {Thöle}}, \bibinfo {author} {\bibfnamefont {S.~S.}\ \bibnamefont {Tsirkin}}, \bibinfo {author} {\bibfnamefont {M.}~\bibnamefont {Wierzbowska}}, \bibinfo {author} {\bibfnamefont {N.}~\bibnamefont {Marzari}}, \bibinfo {author} {\bibfnamefont {D.}~\bibnamefont {Vanderbilt}}, \bibinfo {author} {\bibfnamefont {I.}~\bibnamefont {Souza}}, \bibinfo {author} {\bibfnamefont {A.~A.}\ \bibnamefont {Mostofi}},\ and\ \bibinfo {author} {\bibfnamefont {J.~R.}\ \bibnamefont {Yates}},\ }\bibfield
  {title} {\bibinfo {title} {Wannier90 as a community code: new features and applications},\ }\href {https://doi.org/10.1088/1361-648X/ab51ff} {\bibfield  {journal} {\bibinfo  {journal} {J.\ Phys.\ Condens.\ Matter}\ }\textbf {\bibinfo {volume} {32}},\ \bibinfo {pages} {165902} (\bibinfo {year} {2020})}\BibitemShut {NoStop}%
\bibitem [{\citenamefont {de~Vega}\ \emph {et~al.}(2020)\citenamefont {de~Vega}, \citenamefont {Cox}, \citenamefont {Sols},\ and\ \citenamefont {{Garc\'{\i}a de Abajo}}}]{devega2020strong}%
  \BibitemOpen
  \bibfield  {author} {\bibinfo {author} {\bibfnamefont {S.}~\bibnamefont {de~Vega}}, \bibinfo {author} {\bibfnamefont {J.~D.}\ \bibnamefont {Cox}}, \bibinfo {author} {\bibfnamefont {F.}~\bibnamefont {Sols}},\ and\ \bibinfo {author} {\bibfnamefont {F.~J.}\ \bibnamefont {{Garc\'{\i}a de Abajo}}},\ }\bibfield  {title} {\bibinfo {title} {Strong-field-driven dynamics and high-harmonic generation in interacting one dimensional systems},\ }\href {https://doi.org/10.1103/PhysRevResearch.2.013313} {\bibfield  {journal} {\bibinfo  {journal} {Phys.\ Rev.\ Res.}\ }\textbf {\bibinfo {volume} {2}},\ \bibinfo {pages} {013313} (\bibinfo {year} {2020})}\BibitemShut {NoStop}%
\end{thebibliography}

%

\end{document}


\title{Nonlocal and nonlinear plasmonics in atomically thin heterostructures \\ {\color{gray} \small -- SUPPLEMENTAL MATERIAL --}}

\author{Line Jelver\,\orcidlink{0000-0001-5503-5604}}
\affiliation{POLIMA---Center for Polariton-driven Light--Matter Interactions, University of Southern Denmark, Campusvej 55, DK-5230 Odense M, Denmark}

\author{Joel~D.~Cox\,\orcidlink{0000-0002-5954-6038}}
\email[Corresponding author: ]{cox@mci.sdu.dk}
\affiliation{POLIMA---Center for Polariton-driven Light--Matter Interactions, University of Southern Denmark, Campusvej 55, DK-5230 Odense M, Denmark}
\affiliation{Danish Institute for Advanced Study, University of Southern Denmark, Campusvej 55, DK-5230 Odense M, Denmark}

\begin{abstract}
We provide additional details on the computational parameters used to simulate the electronic properties of graphene and phosphorene nanoribbons, along with a comparison of the electronic band structures obtained from first-principles calculations and their tight-binding counterparts based on maximally localized Wannier functions (MLWFs) for all nanoribbons studied in this work. We then showcase the tunability of the linear optical response, along with the switching behavior of the second-order susceptibility, in both armchair and zigzag graphene nanoribbon dimers of $\approx10$\,nm width across various stacking configurations. Furthermore, we explore the tunability of the linear, second-order, and third-order responses in heterostructures of $\approx10$\,nm and $\approx5$\,nm wide graphene ribbons.
\end{abstract}

\date{\today}
\maketitle


\begin{table}
\centering
\label{tab:ribbon_parameters}
\begin{tabular}{l@{\hskip 0.5cm}l@{\hskip 0.5cm}l@{\hskip 0.5cm}l@{\hskip 0.5cm}l@{\hskip 0.5cm}l@{\hskip 0.5cm}l}
\toprule
Type & Width & Length of unit cell & Wannier orbitals & K-points \\
\midrule
AC-GNR & 5.1 nm & 4.27 \AA & 81  & 401 \\
AC-GNR & 10.2 nm & 4.27 \AA & 162 & 401\\
ZZ-GNR & 5.1 nm & 2.47 \AA & 48  & 801 \\
ZZ-GNR & 9.9 nm & 2.47 \AA & 96  & 801\\
ZZ-PNR & 5.5 nm & 3.31 \AA & 192 & 501 \\
\bottomrule
\end{tabular}
\caption{\textbf{Computational parameters used for graphene and phosphorene nanoribbon optical response simulations.} The electronic ground state is computed using a plane-wave cutoff energy of 48 Ry for phosphorene and 86 Ry for graphene. A one-dimensional $k$-point grid is used along the direction of translational invariance, with 16 points for phosphorene, and 12 and 24 points for armchair (AC) and zigzag (ZZ) graphene nanoribbons (GNRs), respectively. Gamma-point sampling is applied in the two directions perpendicular to the ribbons. Graphene ribbon edges are hydrogen-passivated, while phosphorene ribbons feature bare edges.
    }
\end{table}

\begin{figure*}[h!]
    \centering
    \includegraphics[width=0.98\textwidth]{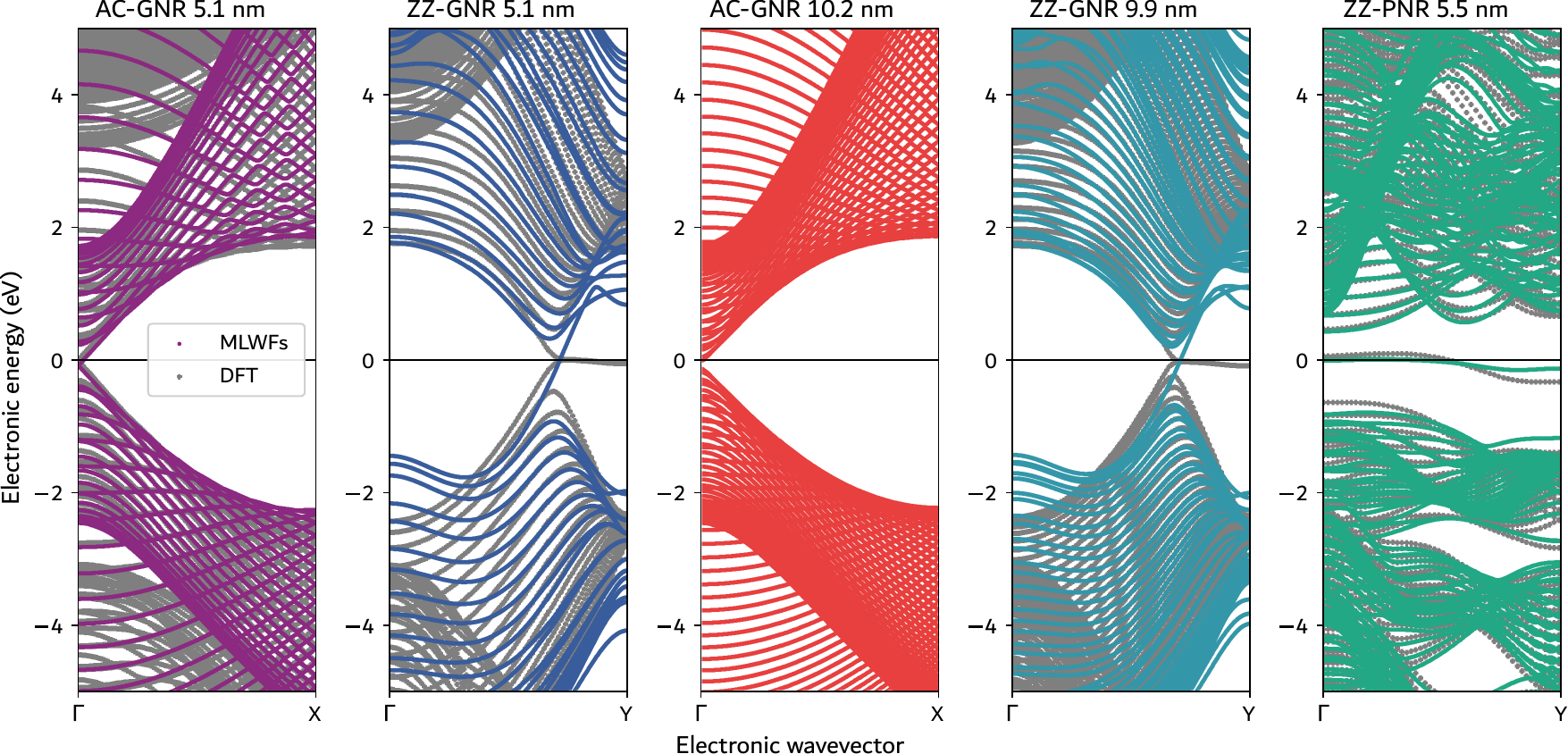}
    \caption{\textbf{Electronic states in graphene and phosphorene nanoribbons.} The electronic bands of all armchair (AC) and zigzag (ZZ) edge-terminated nanoribbons considered in this study. Band structures obtained from Quantum ESPRESSO calculations (grey lines) are compared to those derived from the tight-binding Hamiltonian produced via the Wannierization procedure outlined in the main text (colored lines). NB: the ab-initio dispersion for the largest AC terminated graphene ribbon is not included due to computational limitations.
    }
\end{figure*}

\begin{figure*}[h!]
    \centering
    \includegraphics[width=0.75\textwidth]{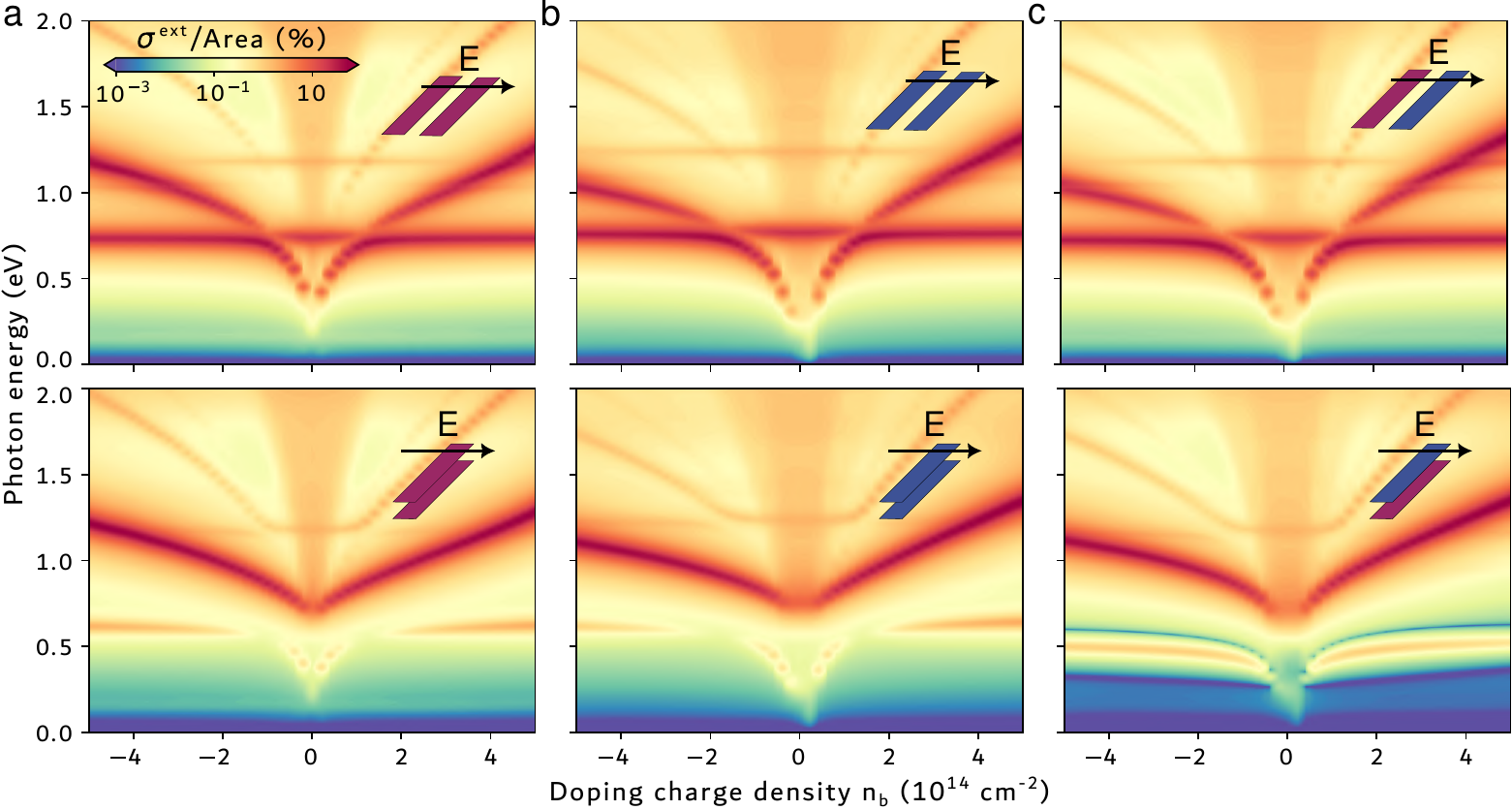}
    \caption{\textbf{Linear extinction spectra of heterostructured $\approx$\,10 nm wide graphene nanoribbons.} The extinction cross-sections $\sigma^{\rm ext}$ per unit area for dimers formed by (a) two coplanar (upper panel) or vertically stacked (lower panel) graphene nanoribbons (GNRs) with armchair (AC) edges and 2\,nm spacing. Panels (b) and (c) feature ZZ and mixed AC-ZZ GNR dimers, respectively. In the upper (lower) panel, we consider that the leftmost (lower) GNR has a fixed electron doping corresponding to a Fermi level of $E_\mathrm{F} = 1.0$\,eV while the charge carrier density in the rightmost (upper) GNR is varied. A colorbar indicating the scale of all contour plots is shown in (a). Strong hybridization of plasmonic modes in the vertically stacked systems leads to quenching of the response from the lower branch. 
    }
\end{figure*}

\begin{figure*}[h!]
    \centering
    \includegraphics[width=0.6\textwidth]{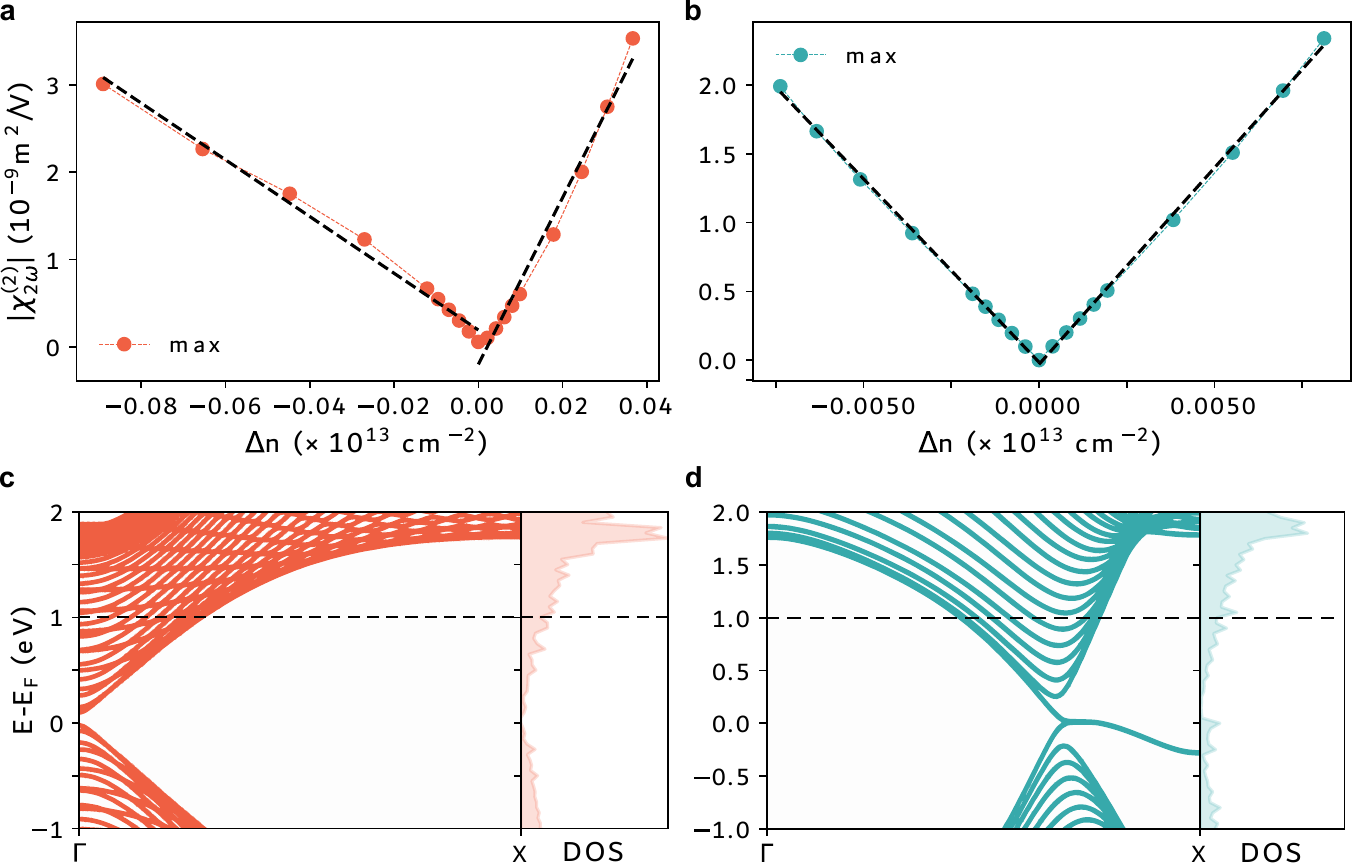}
    \caption{\textbf{Switching of second-harmonic generation (SHG) in $\approx$\,10 nm wide graphene nanoribbons.} Panels (a) and (b) show the second-order susceptibility in a system of two coplanar graphene nanoribbons separated by 2\,nm and terminated by armchair or zigzag edges, respectively, as the doping level in the first ribbon is tuned near that of the second ribbon. Panels (c) and (d) show the electronic band structure and density of states near the Fermi level of the left ribbon, which is held at a fixed doping level corresponding to $E_\mathrm{F} = 1.0$ eV. The steep dispersion in the zigzag ribbon results in significantly greater sensitivity to changes in charge density.
    }
\end{figure*}

\begin{figure*}[h!]
    \centering
    \includegraphics[width=0.98\textwidth]{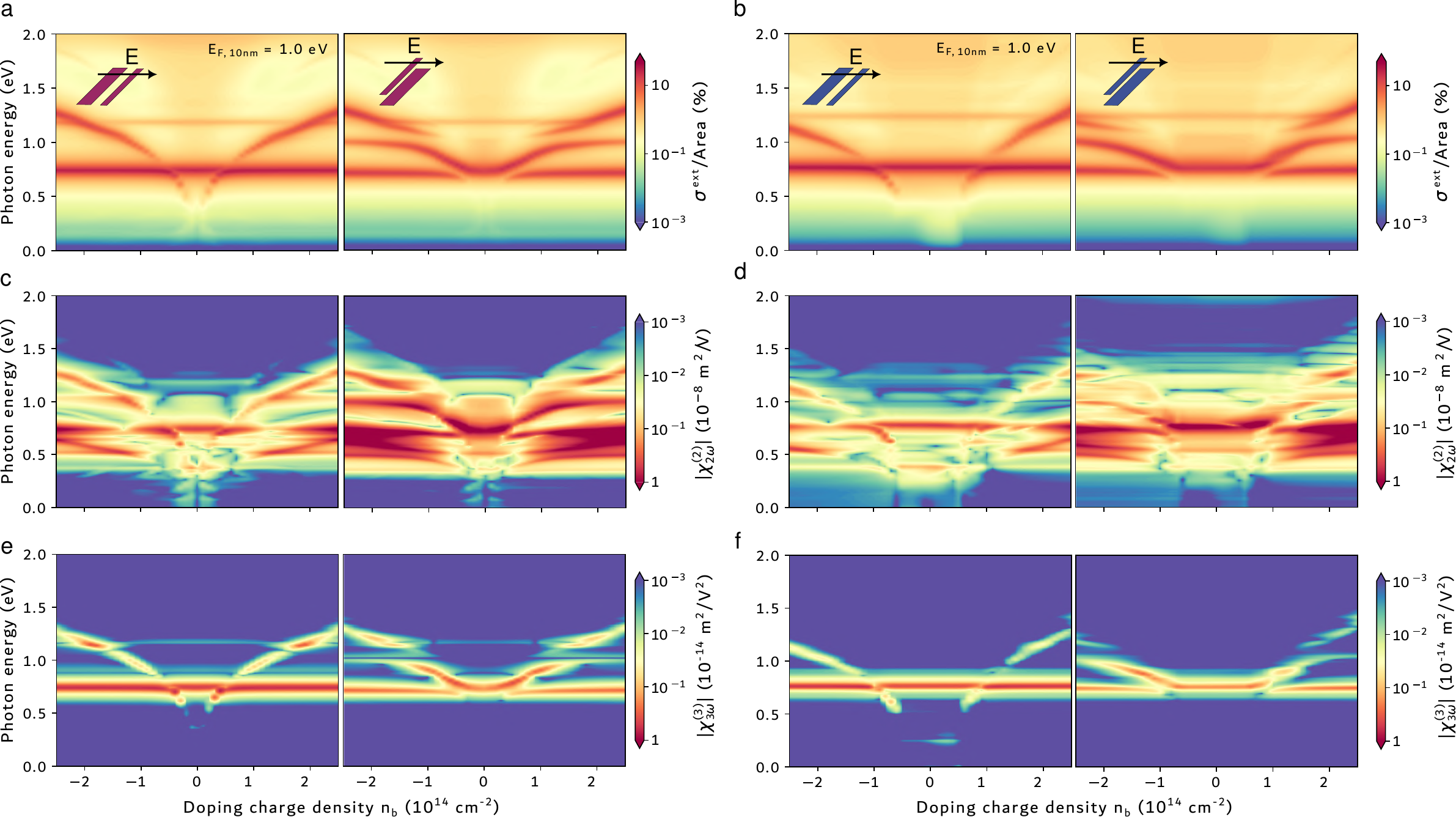}
    \caption{\textbf{Nonlinear susceptibilities of heterostructured graphene nanoribbons.} Absorption cross section per area (top), second-order susceptibility associated with SHG (middle), and third-order susceptibility associated with THG (bottom) for coplanar (left) or vertically stacked (right) $\approx10$ and $\approx5$\,nm wide graphene nanoribbons with (a) armchair and (b) zigzag edge terminations, in each case separated by 2\,nm. In all cases, the wider ribbon is doped to a Fermi level of $E_\mathrm{F} = 1.0$\,eV, and the charge carrier density in the narrower ribbon is varied. Radiating dark modes in the linear response and SHG enhancement are evident in the vertical configurations, while the third-order response is optimized in the coplanar systems. 
    }
\end{figure*}